\documentclass[fleqn,usenatbib]{mnras}

\usepackage{natbib}
\usepackage[T1]{fontenc}

\usepackage{soul}

\DeclareRobustCommand{\VAN}[3]{#2}
\let\VANthebibliography\thebibliography
\def\thebibliography{\DeclareRobustCommand{\VAN}[3]{##3}\VANthebibliography}

\usepackage{graphicx}	% Including figure files
\usepackage{amsmath}	% Advanced maths commands
\usepackage{amssymb}	% Extra maths symbols
\usepackage{xspace}
\usepackage{calrsfs}
\usepackage[para,online,flushleft]{threeparttable}
\usepackage{siunitx}
\usepackage[table]{xcolor}
\usepackage{comment}
\usepackage{newtxtext,newtxmath}
\usepackage{pdflscape}
\usepackage{placeins}
\usepackage{longtable}
\usepackage{multirow}   % To allow multi-row items in tables
\usepackage{subcaption} % Put this in your preamble
\usepackage{threeparttable} % in preamble

\newcommand{\kima}{\texttt{kima}\xspace}

\newcommand{\rearth}{{R$_\oplus$}\xspace}
\newcommand{\mearth}{{M$_\oplus$}\xspace}
\DeclareMathAlphabet{\pazocal}{OMS}{zplm}{m}{n}
\definecolor{lightgray}{gray}{0.9}

\newcommand{\degs}{$^{\circ}$}
\newcommand{\vsini}{$v\,\textrm{sin}\,i$}{}
{}
\newcommand{\kms}{km\,s$^{-1}$}{}

\title[DMPP - Occurence rates]{The Dispersed Matter Planet Project Sample - Detection limits, Occurrence Rates and New Planets}

\author[M. R. Standing et al.]{
Matthew R. Standing,$^{1,2}$\thanks{E-mail: matthew.standing@esa.int}
John R. Barnes,$^{2}$ %
Carole A. Haswell,$^{2}$ %
Adam T. Stevenson,$^{2,3}$ %
Jo\~ao P. Faria,$^{4}$ \newauthor %
Erwan Quintin,$^{1}$ %
Zachary O. B. Ross,$^{2}$ %
Luca Fossati,$^{5}$ %
James S. Jenkins,$^{6,7}$ %
Douglas Alves,$^{8}$ %
Daniel Staab$^{2}$
\\
$^{1}$European Space Agency (ESA), European Space Astronomy Centre (ESAC), Camino Bajo del Castillo s/n, E-28692 Villanueva de la Ca\~{n}ada, Madrid, Spain\\
$^{2}$School of Physical Sciences, The Open University, Milton Keynes MK7 6AA, UK\\
$^{3}$School of Physics \& Astronomy, University of Birmingham, Edgbaston, Birmingham B15 2TT, UK\\
$^{4}$Observatoire Astronomique de l'Universit\'{e} de Gen\`{e}ve, Chemin Pegasi 51, CH-1290 Versoix,
Switzerland\\
$^{5}$Space Research Institute, Austrian Academy of Sciences, Schmiedlstra{\ss}e~6, 8042 Graz, Austria \\
$^{6}$Instituto de Estudios Astrof\'{i}sicos, Facultad de Ingenier\'{i}a y Ciencias, Universidad Diego Portales, Av. Ej\'{e}rcito 441, Santiago, Chile \\
$^{7}$Centro de Astrof\'isica y Tecnolog\'ias Afines (CATA), Casilla 36-D, Santiago, Chile\\
$^{8}$Departamento de Astronom\'ia, Universidad de Chile, Casilla 36-D, Santiago, Chile\\
}

\date{Accepted 2026 February 18. Received 2026 February 18; in original form 2026 January 08}

\pubyear{2026}

\begin{document}
\label{firstpage}
\pagerange{\pageref{firstpage}--\pageref{lastpage}}
\maketitle

\begin{abstract}
DMPP is a radial-velocity survey that aims to detect planets around stars exhibiting anomalous activity signatures, consistent with the presence of close-in evaporating planets.
Here, we report the discovery of 7 new planetary signals in 5 different systems: DMPP-2\,c \& d, HD\,67200/DMPP-6\,b \& c, HD\,118006/DMPP-7\,b, HD\,191122/DMPP-8\,b, and HD\,200133/DMPP-9\,b. We update the orbital parameters of the DMPP-1, DMPP-2, and DMPP-3 systems, along with those of the planetary systems orbiting HD\,181433, HD\,39194, and HD\,89839.
We derive detection limits for all 24 targets in our sample with adequate observational coverage, and test the DMPP hypothesis by calculating the occurrence rates for planets in this configuration. 
We find that the occurrence rates of planets in our sample with orbital periods shorter than $50~\mathrm{d}$ and masses in the range $3$--$10$~\mearth are $83.0^{+27.1}_{-24.4}~\%$, for $10$--$30$~\mearth are $27.0^{+15.0}_{-11.2}~\%$, and for $30$--$100$~\mearth are $13.9^{+11.8}_{-7.5}~\%$.
This is significantly higher than the occurrence rates reported by other radial velocity surveys, providing strong support for the DMPP hypothesis.
\end{abstract}


%%%%%%%%%%%%%%%%%%%%%%%%%%%%%%%%%%%%%%%%%%%%%%%%%%

%%%%%%%%%%%%%%%%% BODY OF PAPER %%%%%%%%%%%%%%%%%%

\section{Introduction}\label{sec:intro}

The two most prolific exoplanet discovery methods to date, radial velocities (RVs) and transits, are generally applied with no significant {\it{a priori}} knowledge of the properties of the planets the target stars are likely to host. RVs were initially collected for FGK stars in the hope of detecting planets akin to the Solar System's giant planets on orbits exceeding 1\,au \citep{CochranHatzes1994}. The first RV planet detection, 51\,Peg\,b with a 4.23\,d period and 0.05\,au orbit was surprising \citep{MayorQueloz1995}. Since both the transit and RV methods are particularly sensitive to close-in orbits \citep[e.g.][]{Haswell2010}, the known exoplanet population is skewed to short period, hot planets. For transit surveys the geometric selection effect for short orbital periods can be analytically corrected, and the sensitivity as a function of planet radius can be estimated with injection-recovery tests. \citet{petigura2013} consequently showed that the fraction of stars hosting planets with radii exceeding $1\,$\rearth rises quite steeply with orbital period, being only $\sim 1\%$ at $P = 5\,$d. Despite this property of the Galaxy's population of exoplanets, selection effects meant early discoveries were dominated by close-in planets, particularly transiting planets found by large scale ground-based transit searches such as WASP \citep{Pollacco2006} and the deeper survey carried out by \textit{Kepler} \citep{Borucki2010}.

These hot planets exhibit some of the most extreme processes in planetary system evolution, including the sublimation of rocky planet surfaces due to intense irradiation \citep{Rappaport2012}. Mass loss from the closest-orbiting planets is common, and was first dramatically demonstrated in the Lyman\,$\alpha$ transits of HD\,209458\,b \citep{Vidal-Madjar2003}.

The inflated ultra-hot Jupiter WASP-12\, b, with $P_\textrm{orb} = 1.09$\,d, shows an enhanced transit depth in the near-UV compared to the optical \citep{fossati2010,haswell2012}, and an anomalous lack of the expected chromospheric emission cores in  the  Mg {\sc ii} h\&k lines. This is consistent with absorption due to mass lost from Roche-lobe overflow \citep{haswell2012}. Along with other hot Jupiter hosts, \cite{fossati2013} showed that WASP-12 also shows anomalously low chromospheric emission cores in the Ca {\sc ii} H\&K resonance lines. These planet hosts lie below the stellar chromospheric basal flux level \citep{schrijver1987,rutten1991}, which is  a lower limit exhibited by extremely inactive stars. The limit has been shown to be consistent with acoustic heating in the chromospheres of main sequence stars \citep{buchholz1998} where magnetic activity is absent.
This minimum level for main sequence FGK stars lies at $\log$ R$^\prime_\textrm{HK} = -5.1$ \citep{Henry1996,Wright2004}.

Main sequence stars with $\log R^\prime_\textrm{HK} < -5.1$ therefore have impossibly low Ca {\sc ii} H\&K line core emission, indicating an upper stellar atmospheric structure different from other stars, or alternatively their intrinsic line core emission has been absorbed by circumstellar material. Since the most extreme example of the sub-basal $\log$ R$^\prime_\textrm{HK}$ phenomenon was WASP-12 \citep{fossati2013}, which hosts one of the hottest known giant planets, Occam's razor strongly suggests the two phenomena are connected. Material dispersed from mass loss from close-orbiting planets can naturally give rise to circumstellar material \citep{haswell2012}, and this material will naturally absorb starlight most strongly in the resonance line cores of abundant elements. These lines are precisely those which exhibit prominent stellar chromospheric emission. If we conclude planetary mass loss depresses a star's apparent $\log$ R$^\prime_\textrm{HK}$ value,  this leads to the corollary that main sequence stars exhibiting sub-basal $\log$ R$^\prime_\textrm{HK}$ can be identified {\it a priori} as the probable hosts of hot, mass-losing planets. This potentially provides an efficient short-cut to finding particularly rare and interesting exoplanets orbiting bright, nearby stars. The Dispersed Matter Planet Project aims to test and exploit this hypothesis.

\citet{Staabthesis} details a painstaking analysis of extant $\log$ R$^\prime_\textrm{HK}$ measurements to identify these probable hosts of hot, mass-losing planets. The steps taken to identify 39 main sequence stars with $\log$ R$^\prime_\textrm{HK} < -5.1$ from a supersample of 7864 stars catalogued by \citet{pace2013} included re-calibrating the conversions from S-index to $\log$ R$^\prime_\textrm{HK}$, selecting only stars with 0.4 < $B$-$V$ < 1.2 where the $\log$ R$^\prime_\textrm{HK}$ is well-calibrated, and using $B$-$V$ and $M_{\rm V}$ to identify a main sequence sample of  2716 stars. The 39 sub-basal objects became the targets of the Dispersed Matter Planet Project \citep[DMPP,][]{Haswell2020}; they might be expected to harbour mass-losing planets in short period orbits.

RV observations of the DMPP sample has identified diverse planet systems, including DMPP-1, a compact multiplanet system \citep{Staab2020};  DMPP-2 a hot Saturn-mass planet \citep{Haswell2020};  DMPP-3Ab, a hot terrestrial, S-type circumprimary planet in a compact eccentric binary star system \citep{Barnes2020, Stevenson2023-DMPP3};  and DMPP-4, a naked eye star hosting either one or two sub-Neptune mass planets \citep{barnes2023}. 
A summary of target observations for additional candidates was given by \cite{Haswell2020} with a comparison of the detection probabilities of the reported systems compared with a random selection of main sequence stars. Here, with further observations and a larger sample, we present new planet discoveries, a more complete, Bayesian analysis of detection sensitivities for each system and occurrence rates for the DMPP sample. 

This paper comprises Section~\ref{sec:Observations}, outlining target selection and DMPP observational protocol; Section~\ref{sec:Methods} describes our RV analysis using \texttt{kima} \citep{Faria2018}, and outlines the detection limit calculation methodology; Section~\ref{sec:results_dis} announces new planetary detections with associated activity analysis, and updated planet parameters, and presents occurrence rates for the survey, comparing them to occurrence rates from other RV surveys; we conclude in Section~\ref{sec:conclusion}.

\section{Observational and data protocols}\label{sec:Observations}

Observations were made on a number of observing runs, with the first observations as part of the DMPP programme made with HARPS \citep{Pepe2002b} in late 2015 and the most recent in early 2023. Motivated by the premise that the host stars harbor exoplanets in short-period orbits, individual observing runs initially intensively observed a handful of targets. By cycling through a small sub-sample of the DMPP targets, several observations of each chosen star could be obtained during a single night of observing, ensuring sensitivity to short-period orbits. Signals became generally evident on the timescales $>1$\,d, with the result that the survey found low-mass planets on the timescales of a few days, i.e. the duration of the observing campaigns. The ESO P110 campaign in 2023, spanning 12 nights focused on long-term verification of already detected signals along with first epoch RVs for a number of previously unobserved targets. The main DMPP observations were made during ESO runs 096.C-0499(A), 097.C-0390(B), 098.C-0269(B), 099.C-0798(A) and 0100.C-0836(A), 107.22UN.001, 110.248C.001 and are supplemented for some targets by further archival HARPS observations (programme IDs detailed in the Data Availbility statement). DMPP-3 was also observed in the ESPRESSO \citep{Pepe2021} run 112.25LZ.001. DMPP-4 was observed with SOPHIE \citep{Perruchot2008} and HARPS-N \citep{Cosentino2012} in programmes supported by the OPTICON Trans-National Access Programme (FP7II, 2013–2016).

\subsection{Observational protocol}
The majority of the observations were carried out in visitor mode or delegated visitor mode at the ESO 3.6m telescope at La Silla with the High Accuracy Radial velocity Planet Searcher (HARPS).  We used HARPS in High Accuracy Mode with our target spectra recorded on Fibre A and Fabry--P\'erot (FP) reference spectrum on Fibre B to track the small instrument drifts relative to the nightly master reference spectrum. Before 2023, we used ThAr reference lamps, but switched to using the nightly master laser frequency comb and simultaneous Fabry--P\'erot reference in P110. Service observations were also made with the \'{E}chelle SPectrograph for Rocky Exoplanets and Stable Spectroscopic Observations instrument (ESPRESSO) in singleHR mode ($R \sim 140,000$) at the ESO VLT telescopes.

\subsection{Data reduction}

To search for periodic signals in our data, we elected to re-reduce the RVs from the HARPS spectra, rather than using the automatically generated, cross-correlation function (CCF) derived Data Reduction Software (\textsc{DRS}) RVs. DRS data products derived from the CCFs, specifically the Bisector Inverse Span (BIS) and CCF full width at half maximum (FWHM) measurements, were used for the activity correlation analyses. Archival pre-fibre and DMPP post-fibre upgrade spectra were reduced respectively with DRS versions 3.5 and 3.8\footnote{\url{https://www.eso.org/sci/software/pipelines/harps/harps-pipe-recipes.html}}. We used the recently developed \textsc{s-BART}, which re-casts a template matching approach into a Bayesian framework to provide a more robust estimate of the RVs \citep{S-BARTpaper}. \textsc{s-BART} creates a high S/N stellar template and removes activity-sensitive lines. It also includes an involved treatment of telluric features and removes areas near telluric lines where transmittance drops below 99\% of the continuum. Uncertainties (from flux measurement) in the template are propagated for use in estimating the error in the final RV value. Deriving independent RVs for each \'{e}chelle order and combining them is common practice. However, this approach can be subject to chromatic effects due to stellar activity or instrument and observing practicalities. Instead, to mitigate these factors, \cite{S-BARTpaper} assume a single RV shift, appropriate for a dynamically induced Doppler shift, between each spectrum and the master template. The Bayesian statistical framework allows characterisation of the RV posterior probability for any observation, and includes the possibility of marginalisation with respect to parameters of polynomials used to adjust continuum levels between spectrum and template. This ensures that the only difference is a radial velocity shift. \citet{S-BARTpaper} demonstrate improvements in RV scatter over the standard DRS, \textsc{harps-terra} \citep{TERRApaper} and \textsc{serval} \citep{SERVALpaper} pipelines. \citet{Stevenson2025-hd28471} evaluated the use of \textsc{s-BART} for the multiplanet system HD\,28471, finding an improvement in RV precision of over 19\% compared with the DRS. We therefore use \textsc{s-BART} RVs for all targets in this work, including our ESPRESSO dataset, with the exception of the SOPHIE and HARPS-N observations on DMPP-4, which are detailed in \cite{barnes2023}.

\subsection{Choice of targets \& observation summary}

The 39 FGK stars on the DMPP target list (see Section~\ref{sec:intro}) were divided into objects observable from La Silla and objects only observable from the north, and further divided into groups which could sensibly be observed in succession over a single night. We proposed to observe such groups and then, where necessary, modified our plans to accommodate variance between the requested and awarded observing time. We typically observed a target three or four times a night during the earlier runs, but analysis began to suggest that there was much more variability on timescales of a day or longer. Consequently, we observed targets only once or occasionally twice a night in the later runs. Our empirical findings, and the lower cadence we subsequently adopted are consistent with more recent results showing RVs are affected by super-granulation noise which is correlated on a timescale of 1-2\,d \citep[e.g.][]{Klein+2024}. We consider herein only DMPP targets with 10 or more RV epochs; there are 24 targets satisfying this, they are listed in Table~\ref{tab:System_parameters}-\ref{tab:System_parameters_4}.
Observation metrics such as number of observation on each target, SNR, observation period can be found in Table~\ref{tab:obs_summary}.

\section{Methods}\label{sec:Methods}

Here we describe how we calculate detection limits for the DMPP survey.
We analyse our RV timeseries with \texttt{kima} \citep{Faria2018}, a diffusive nested sampling algorithm which fits a sum of ${N_{\rm p}}$ Keplerian signals to the data. \kima treats the number of planetary signals ${N_{\rm p}}$ as a free parameter, which enables Bayesian model comparison to determine the number of planetary signals present \citep{Standing2022}. During fitting, we use a Students' T distribution for the likelihood as in \cite{Agol2021} \& \cite{Standing2023} which allows for any large RV outliers without strongly influencing the fit.

\subsection{Prior distributions}\label{sec:priors}
We set up our model with priors similar to those used in \cite{Standing2022} with minor changes detailed below.

The $\beta$-prior distribution described in \cite{Kipping2013} favours lower eccentricities but allows exploration of more eccentric orbits where the data requires \citep{Standing2022}. In \kima, we use the more flexible Kumaraswamy distribution \citep{Kumaraswamy1980} as a prior, with $\alpha=0.881$ and $\beta=2.878$ \citep{Stevenson2025} as a proxy for the \cite{Kipping2013} $\beta$-distribution. 

We limit the upper range of our $K$ prior following an initial run on the data. An initial run with loose priors provides us with information on the signals which could explain the data and motivates pragmatic upper limits for the $K$ prior without leaving the parameter space too large. We still allow room for variation and use a median upper limit of $15~\rm{m\,s^{-1}}$ for our targets.

The lower period limit is chosen as 1-day to avoid the 1-day period alias and because shorter intervals are affected by correlated super-granulation noise; the upper period limit was chosen depending on the individual time-span of each dataset. For well sampled datasets we set the upper limit to double the timespan of the data. Whereas, for poorly sampled datasets with long times between observations we define an ``effective timespan'' based on the time periods covered by the sampling, e.g. if two observing runs each with a duration of 1 week are spaced 1 year apart, we take an effective timespan of 14 days, rather than 1 year. The majority of our least-sampled datasets have time-spans $<14$~days, in this case we set the upper period limit to $20$~days. 

For each system, we use Angular Momentum Deficit (AMD) enforcement within \kima. This option computes the AMD \citep{Laskar1997, Laskar2000} for each posterior sample obtained and ensures that the solution proposed is stable, and thus a realistic solution. Samples which fail the AMD stability test are discarded, which increases the efficiency of the sampling.
We used the priors in Table~\ref{tab:prior distributions}, but implemented minor changes to the limits where required, always using the same distributions. DMPP-3 is an eccentric binary; consequently, the AMD stability criterion could not be effectively applied to this system. For HD\,181433, achieving an adequate fit with the AMD stability criterion also proved challenging, likely due to the presence of eccentric, Jovian-mass planets with long orbital periods. We therefore deactivated the criterion for this system as well.

\kima\ fits for additive offsets between different instruments using a uniform prior, $\mathcal{U}(\min(\mathrm{all\ data}), \max(\mathrm{all\ data}))$. When our observations span epochs in which an instrument is known to have undergone a zero-point change (e.g. due to a fibre upgrade), we treat the pre- and post-change data as originating from separate instruments and fit independent offsets; examples include the HARPS fibre upgrade in May 2015 \citep{LoCurto2015, Trifonov2020} and the COVID shutdown (see e.g. \citealt{Faria2022, Standing2023, Stevenson2023-DMPP3, Liebing2024, Stevenson2025-hd28471}).

We do not incorporate Gaussian Processes (GPs) in our fits because our targets all have low activity  and the observation sampling is generally unsuitable for effective GP modeling.

\renewcommand{\arraystretch}{1.4}

\begin{table}
	\centering
	\caption{Prior distributions used in RV model planetary signals in \kima. Initial (wider) priors are used for a first-fit of the data. The narrower priors are used once an initial fit has been established.}
	\begin{tabular}{c c c c}
		\hline
		\hline
	    Parameter & Unit & \multicolumn{2}{c}{Prior distribution}\\
		\hline
		    &   & Initial & Narrower\\
		$N_{\rm p}$ &  & $\pazocal{U}(0,4)$ & $\pazocal{U}(0,3)$\\
		$P$ & days & $\pazocal{LU}(1,100)$ & $\pazocal{LU}(1,20)$\\
		$K$ & $\rm{m\,s^{-1}}$ & $\pazocal{MLU}(0.1, 100)$ & $\pazocal{MLU}(0.1, \lesssim15)$\\
		$e$ & & \multicolumn{2}{c}{$\pazocal{K}(0.881, 2.878)$}\\
		$\phi$ & & \multicolumn{2}{c}{$\pazocal{U}(0,2\pi)$}\\
		$\omega$ & & \multicolumn{2}{c}{$\pazocal{U}(0,2\pi)$}\\
		$\sigma_{\rm jit}$ & $\rm{m\,s^{-1}}$ & $\pazocal{LU}(0.001,100)$ & $\pazocal{LU}(0.001,10\times \rm{rms})$\\
		$\gamma$ & $\rm{m\,s^{-1}}$ & $\pazocal{U}(\rm{RV_{min}, RV_{max}})$ & $\pazocal{U}(V_{sys}\pm{50})$\\
		\hline
	\end{tabular}
	\label{tab:prior distributions}
    \begin{tablenotes}
    \small
    \item \textbf{Notes:} $N_{\rm p}$ - Number of Keplerian signals. $\pazocal{U}$ -  Uniform prior with upper and lower limit. $\pazocal{LU}$ - Log-uniform (Jeffreys) prior with upper and lower limit. $\pazocal{MLU}$ - Modified log-uniform prior with knee and upper limit. $\pazocal{K}$ - Kumaraswamy prior \citep{Kumaraswamy1980} with two shape parameters.
    \end{tablenotes}
\end{table}

\subsection{Detection limit method}\label{subsec:Detlim_method}

We follow the method laid out in \cite{Standing2022} and tested with injection recovery tests. This method is used in e.g. \cite{Grieves2022, Standing2023, John2023, Balsalobre-Ruza2025} \& \cite{Figueira2025arXiv}:
\begin{enumerate}
    \item For each system, we begin by running \kima using wide-uninformative priors similar to those used in \cite{Standing2022} to establish the presence and parameters of any signals. 
    \item Priors for each target are then adjusted individually to sample the parameter space as efficiently as possible. For example, should there be no evidence supporting $>2~N_{\rm p}$ signals then $N_{\rm p}$ would only be sampled uniformly from $0-2$; should there be no evidence of signals with semi-amplitudes $>15~\rm{m\,s^{-1}}$ then the upper limit on our semi-amplitude prior was lowered to $<15~\rm{m\,s^{-1}}$, and so on. A second longer run would then be carried out on each dataset ensuring the most appropriate solution is obtained.
    \item If the system has no signals which pass a Bayes Factor (BF) threshold of $150$ to claim a detection \citep{Jeffreys1961, Standing2022}, then we proceed to the next step. However, if $\ge 1$ planetary signal is detected, then the posterior sample with the highest likelihood is subtracted from the data, and \kima is run on the subtracted dataset to ensure no signals remain.
    \item \kima is then run on the data set (or the signal-subtracted data set) with $N_{\rm p}$ fixed to a value of $1$. This step yields the posterior samples which can be used to calculate our detection limits as detailed in \cite{Standing2022}. We run \kima until a minimum of $200,000$ posterior samples have been obtained on each system to ensure robust detection limits can be calculated\footnote{On average we have $\approx400,000$ samples for each of our systems.}.
    \item Once the posterior samples are obtained, we calculate the detection limits by binning the samples in log-period space, ${P_\mathrm{p}}$, and obtaining the (upper) $99\%$ confidence levels of the semi-amplitudes, ${K_\mathrm{p}}$, as in \cite{Standing2022}. The uncertainty ($E$) on the detection limits are calculated as in \cite{Standing2023} using the equation $E = \frac{1.97}{\sqrt{N}}$. \footnote{We note that the incorrect scaling value was stated in \cite{Standing2023} and should be $a = 1.97 \pm 0.12$. This does not affect the results of \cite{Standing2023}, only the uncertainty plotted in the detection limit in Figure~4 of the paper.}
\end{enumerate}

\begin{figure*}
	\includegraphics[width=\textwidth]{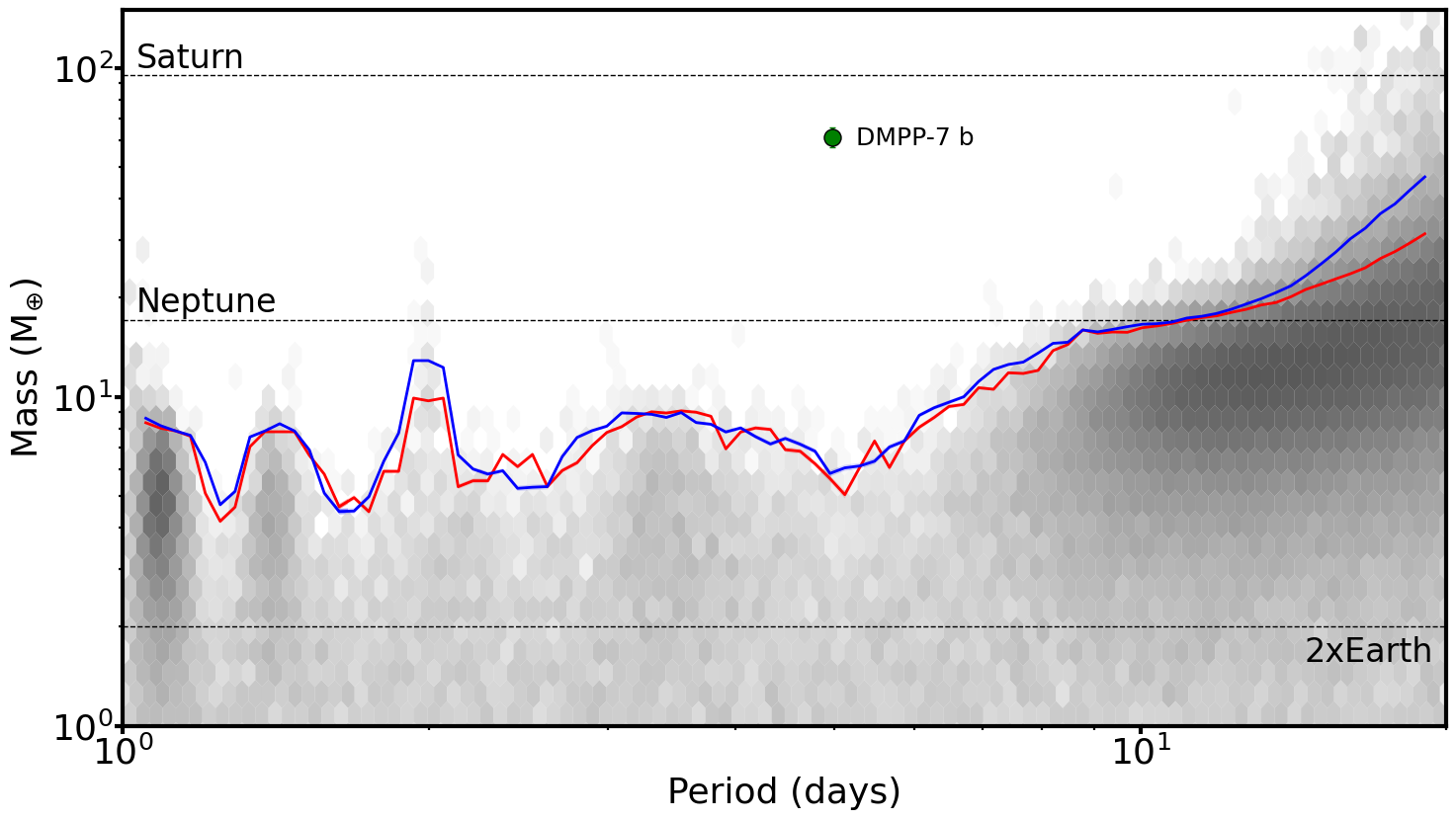}
    \caption{Detection limit plot for HD118006 / DMPP-7. The grey hexbin plot denotes the density of posterior samples obtained from \kima runs on our HD118006 / DMPP-7 dataset, the darker the region the more samples are present. The blue line is the $99\%$ upper detection limit, with uncertainties denoted by the blue shaded region. The red line is the $99\%$ upper limit calculated on posterior samples with $e<0.1$, with associated uncertainty denoted by the red shaded region. Shaded uncertainties for the detection limits are present but small. The green point represents the detected planetary signal of HD118006 / DMPP-7\,b.}
    \label{fig:HD118006_Det_lim}
\end{figure*}

An example detection limit calculated following the above steps can be seen in Figure~\ref{fig:HD118006_Det_lim} for HD\,118006 / DMPP-7. We perform the above steps for our sample of 24 DMPP targets (see Table~\ref{tab:System_parameters}). Detection limit uncertainties, calculated as described above, are plotted for each detection limit shown here and in Appendix~\ref{app:det-lims}. In some cases, the uncertainties are too small to be visible.
For DMPP-1 and the previously unreported system HD\,67200 / DMPP-6, which have more extensive datasets, we investigate the solutions further, including the application of GPs in \cite{Barnes2025multiplanet}. For the purposes of this study, we assume all RV signals are of dynamically induced origin, resulting from orbiting exoplanet candidates.

\subsection{Survey completeness}\label{subsec:Completeness}

\begin{figure*}
\includegraphics[width=\textwidth]{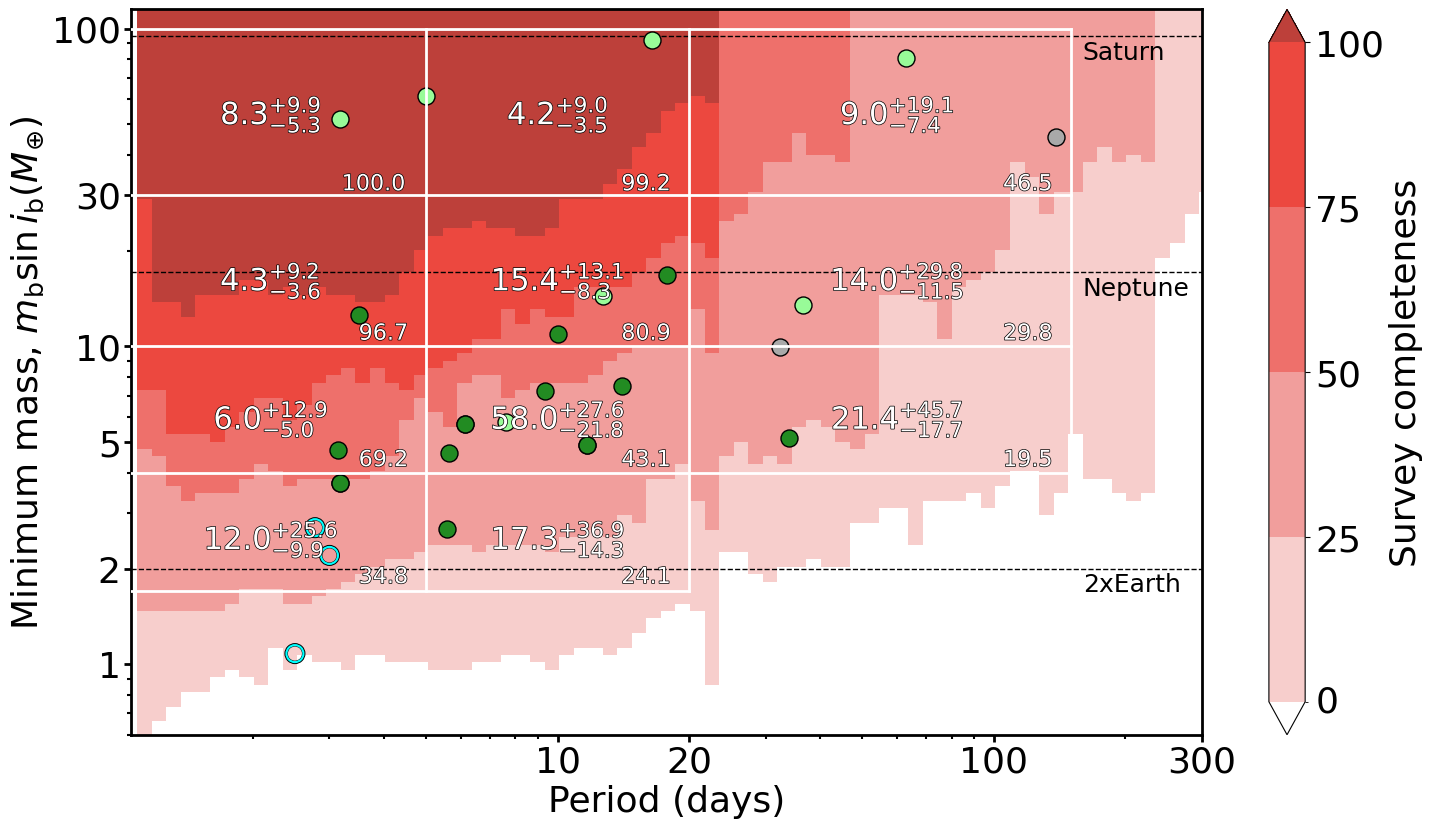}
    \caption{Completeness and planet abundance plot for the DMPP radial velocity survey of 24 targets with sub-basal activity, as a function of minimum mass and orbital period. The red contours indicate the completeness of the survey from $0\%$ (white) to $100\%$ (dark red) as in Figure 10 in \citet{Martin2019}. The dark green circles show previously published planet signals present in our data, while the light green show signals found to pass our detection threshold in this work. Cyan open circles are the three candidate signals we identified with moderate statistical evidence, while the grey circles represent signals which we suspect arise from activity. The white lines delineate 11 boxes enclosing regions of parameter space within which we are able to constrain our planetary abundance. Box limits were chosen to be roughly equal in $\log$ space for short periods within our completeness, with mass limits on integer masses. The number in the centre of each box is a percentage indicating the planet abundance with its $1\sigma$ uncertainty, or the $2\sigma$ ($95\%$) upper limit on abundance where no planet detection was made. The white values in the lower right of each box is its mean completeness as a percentage.}
    \label{fig:DMPP_Occ-rates}
\end{figure*}

With detection limits calculated for each of the systems in the DMPP sample, it then becomes possible to calculate the survey's completeness. We follow the approach of \cite{Martin2019}, \cite{cumming2008}, and \cite{Mayor2011}, combining the 99\% detection limits across all individual targets and averaging them within mass–period bins.
This completeness is defined as \( C(P_{b}, m_{b} \sin i_{b}) \), i.e., the minimum mass our survey is sensitive to at a given period. It is illustrated by the red shaded regions in Figure~\ref{fig:DMPP_Occ-rates}, where white indicates 0\% completeness (no sensitivity in any system), and dark red indicates 100\% completeness—regions of parameter space to which we are sensitive in all 24 systems. The 100\% completeness region does not extend past 20~days in orbital period as some of our datasets have timespans $<14$~days as discussed in Section~\ref{sec:priors}.

\subsection{Occurrence rates}\label{subsec:Occurrence_rate_calculation}

In regions of the mass-period space where no planetary signals are formally detected we are only able to place upper limits on planet occurrence rates. We follow \cite{He2017}; the upper limits on planet abundance is calculated \citep{Martin2019} using:

\begin{equation}\label{eq:abundance_up_lim}
  \mathcal{N}_{\rm{p}}=\frac{1-(1-\kappa)^{1/n_{\rm{stars}}}}{C(P_{\rm{b}},m_{\rm{b}}\sin i_{\rm{b}})}  
\end{equation}
Here, $\mathcal{N}_{\rm{p}}$ is the upper limit on the number of planets, $\kappa$ is the confidence interval (i.e. for $2\sigma$, $95\%$ interval $\kappa=0.95$); $n_{\rm{stars}} = 24$ herein and $C(P_{\rm{b}},m_{\rm{b}}\sin i_{\rm{b}})$ is the mean completeness within the region of parameter space in question.

In regions where there are planetary signals detected the planetary abundance is calculated as in \cite{Mayor2011} using \citep{Martin2019, Bennett2021}:
\begin{equation}\label{eq:abundance}
\mathcal{N}_{\rm{p}}=\frac{1}{n_{\rm{stars}}}\sum_{i=1}^{n_{\rm{det}}}\frac{1}{C(P_{\rm{b}},m_{\rm{b}}\sin i_{\rm{b}})}
\end{equation}
where $n_{\rm{det}}$ is the number of planet detections in the region. 

The uncertainties on the occurrence rates are computed using the Clopper-Pearson confidence interval \citep{ClopperPearson1934}. This method allows retrieval of 
conservative and realistic uncertainties even for a low number of observed successes, solving the issues arising in the standard Poisson error $\sigma=2\mathcal{N}_{\rm{p}}/\sqrt{n_{\rm{det}}}$ \citep[e.g.][]{Martin2019}. In practice, we compute the Clopper-Pearson 1$\sigma$ confidence interval for the observed occurrence rate in each $\{P_{\rm{b}},m_{\rm{b}}\sin i_{\rm{b}}\}$ bin, and then correct for completeness by dividing both the upper and lower values by the corresponding $C(P_{\rm{b}},m_{\rm{b}}\sin i_{\rm{b}})$.

\section{Results and Discussion}\label{sec:results_dis}

\subsection{Detection limit results}\label{sec:det_lim_results}
Individual detection limits for each target in our sample can be found in Appendix~\ref{app:det-lims}.  
Detection limits for the survey reveal that our most sensitively probed systems are DMPP-1, DMPP-3, DMPP-4, HD\,39194, HD\,67200, and HD\,181433. We are  sensitive to Earth mass planets orbiting DMPP-3, HD39194, HD67200, and HD181433 at periods $<5$~days (see Figures~\ref{fig:Det-lim_DMPP-3}, \ref{fig:Det-lim_HD39194}, \ref{fig:Det-lim_HD67200}, and \ref{fig:Det-lim_HD181433}).
In all six of the systems listed above, our data allow us to detect planets with masses smaller than $2$\,\mearth\ on orbital periods ranging from less than 5~days (DMPP-1 and DMPP-4) to less than 50~days (HD\,181433). This is attributable to a combination of sufficient well-sampled observations (between 71 and 288 data points, including 20 ESPRESSO spectra for DMPP-3) and the targets' brightnesses ($V$ magnitudes as high as 5.7).

In contrast, in our least-sensitive system, HD\,58489 (with 40 data points and a $V$~magnitude of 9.74), we are still sensitive to sub-Neptune mass planets out to $\approx5$~days, and sub-Saturn mass planets out to $\approx1000$~days (Figure~\ref{fig:Det-lim_HD58489}).

\subsection{Detection limits in the presence of planetary signals}\label{sec:det_lim_revcovery}
As stated in Section~\ref{subsec:Detlim_method}, if a system hosts a candidate signal that passes our detection threshold, a Bayes Factor (BF) of 150, this signal is subtracted prior to calculating the detection limit. The subtracted signals, corresponding to the posterior samples with the highest likelihood, are listed in Table~\ref{tab:Subtracted_signals}.

If, on the other hand, a candidate signal does not exceed the detection threshold, it is not removed before computing the detection limit. In such cases, if there remains significant evidence for the signal, the majority of posterior samples obtained during the detection limit run (with \( N_{\rm p} = 1 \)) tend to cluster around the orbital period of the candidate.

\begin{figure*}
\includegraphics[width=\textwidth]{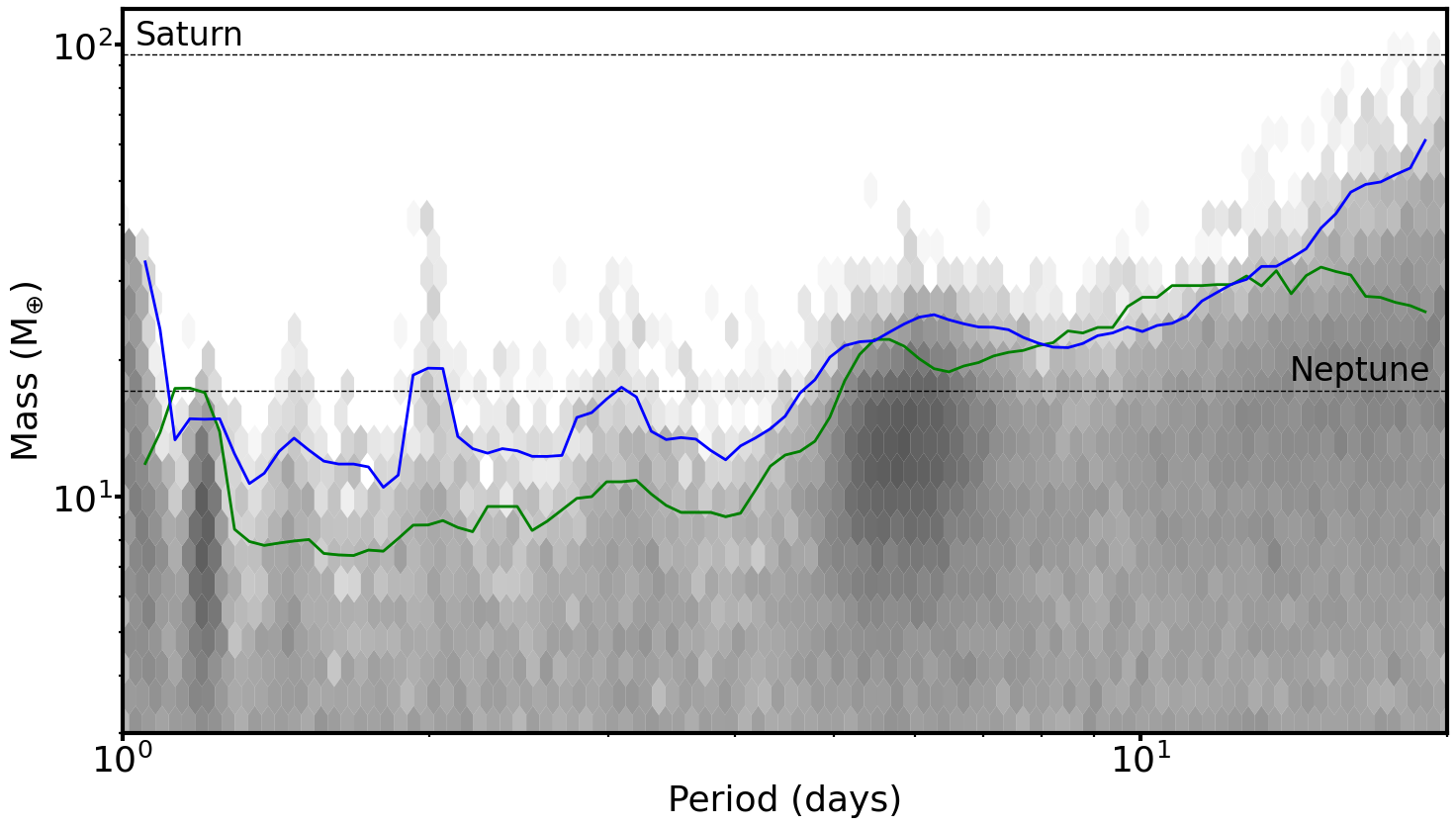}
    \caption{Detection limit plot for BD+03580. The grey hexbin plot denotes the density of 222,000 posterior samples obtained from \kima runs. The blue line is the $99\%$ upper detection limit, with uncertainties denoted by the blue shaded region. The green line is the same $99\%$ upper limit calculated on posterior samples obtained from separate runs where $N_{\rm p}=2$, with associated uncertainty denoted by the green shaded region. Shaded uncertainties for the detection limits are present but small}
    \label{fig:ForcingNp_plus_two}
\end{figure*}

For these systems, we investigated the effect of forcing \( N_{\rm p} = 2 \) (to account for both the candidate and additional signals), rather than relying solely on a single-planet model. The results of this test are shown for the target BD+03580 in Figure~\ref{fig:ForcingNp_plus_two}. The green detection limit is derived from posterior samples obtained with \( N_{\rm p} = 2 \), while the blue curve corresponds to the detection limit from a higher number of posterior samples with \( N_{\rm p} = 1 \). We find that forcing \( N_{\rm p} = 2 \) results in a lower detection limit than simply increasing the number of samples under the single-planet assumption.

From this analysis, we conclude that when calculating detection limits in the presence of strong candidate signals that are not subtracted (as in step (iii) of Section~\ref{subsec:Detlim_method}), best practice is to obtain more than 200,000 posterior samples to adequately explore the parameter space. If this is not feasible, and the returned samples still cluster at the orbital period of the candidate, this provides strong evidence in favour of the signal. In such cases, it becomes reasonable to force \( N_{\rm p} = 2 \) to account for both the candidate and any additional signals.

This situation applies to only one target in our sample: HD\,67200 / DMPP-6 (see Section~\ref{sec:HD67200}).

\subsection{Targets with previously published solutions}\label{sec:prev_published}

In this section we present the orbital parameters obtained from our fits to the data, updating planetary parameters that have been previously published in our target systems. 
We obtain parameters from our posterior samples as in \cite{Standing2022, Standing2023}, by clustering the samples in period, semi-amplitude and eccentricity using \texttt{fast\_HDBSCAN}\footnote{\texttt{fast\_HDBSCAN} is an updated version of the original algorithm by \cite{McInnes2017} to allow running on multiple cores \url{https://github.com/TutteInstitute/fast_hdbscan}}. The clusters are then plotted using the \texttt{corner} package \citep{corner}, the 50th percentile provides the orbital parameters, and the cluster's 14th and 84th percentile provide the $1\sigma$ uncertainty. Orbital parameters for these systems can be found in Table~\ref{tab:Prev_published}, phase plots for each of the signals mentioned here in Appendix~\ref{app:phase_plots}, and corner plots for the same signals here can be found in Appendix~\ref{app:corners}. 

\textbf{\underline{DMPP-1:}}
DMPP-1 was previously proposed to host a system of four planets \citep{Staab2020}. However, simultaneous fitting of multiple signals with \kima yields results that differ from the periodogram-based recursive search employed by \citet{Staab2020}.
The new solution with the additional observations taken in 2023 has a less constrained eccentricity prior (i.e. beta distribution compared vs the previously adopted half-Gaussian width of 0.05) and uses the AMD constraint to reject non-stable orbits. This combination of factors leads to a less prior-driven posterior distribution of eccentricities. The Bayesian interpretation thus favours a three-planet solution with slightly more eccentric, but stable, orbits. The BF exceeds 15,000.
Our new maximum posterior solution includes three significant signals with orbital periods of 3.14, 9.79, and 17.83~days. These findings are consistent with the default and alternative periodicities reported in \citet{Staab2020} and are investigated further in a forthcoming publication.

\begin{figure}
    \centering
    \begin{subfigure}[b]{0.48\textwidth}
        \centering
        \includegraphics[width=\textwidth]{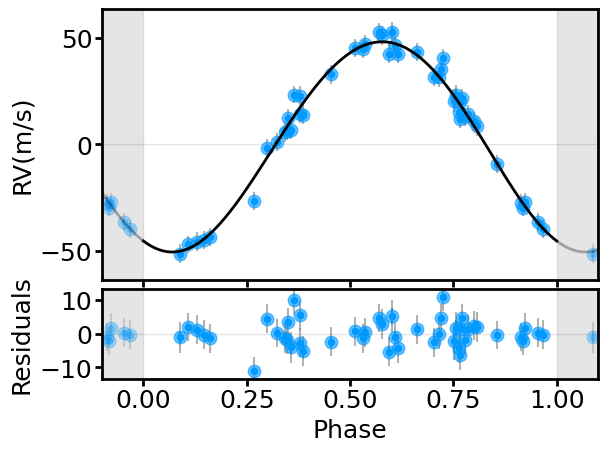}
    \end{subfigure}
    \hfill
    \begin{subfigure}[b]{0.48\textwidth}
        \centering
        \includegraphics[width=\textwidth]{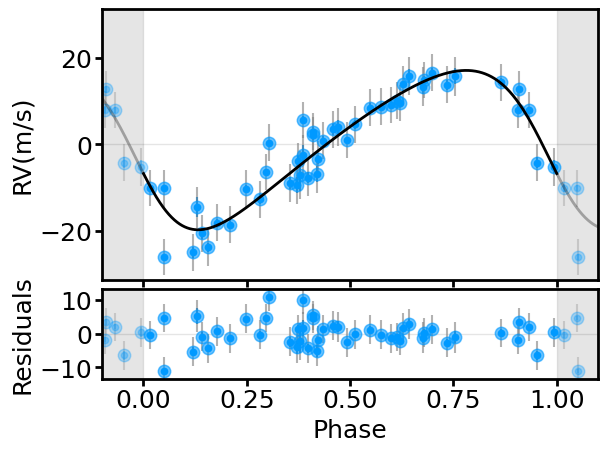}
    \end{subfigure}
    \begin{subfigure}[b]{0.48\textwidth}
        \centering
        \includegraphics[width=\textwidth]{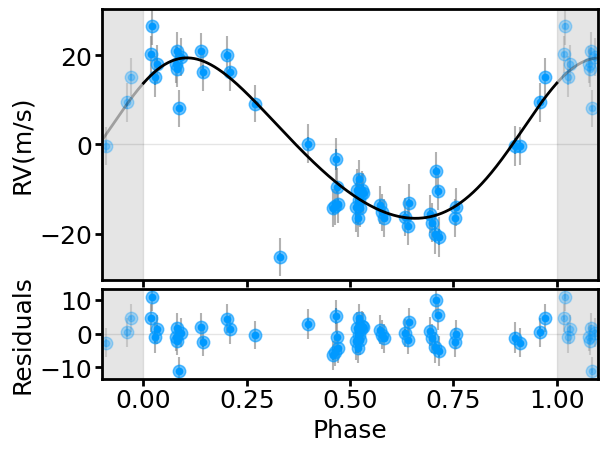}
    \end{subfigure}    
    \caption{Phased Keplerian RV models of DMPP-2\,b (top), DMPP-2\,c (middle), and DMPP-2\,d (bottom) with HARPS data post-fibre upgrade (blue) and its residuals after removing the planetary signals. The additional RV jitter term has been added to the plotted uncertainties and is shown by the grey error bars. No random samples are shown for these signals as their eccentricity is unconstrained and consistent with 0. The shaded regions display the repeating signal.}
    \label{fig:DMPP-2_combined_phase}
\end{figure}

\textbf{\underline{DMPP-2:}}
DMPP-2 was previously analysed by \citet{Haswell2020}, who identified a single planet with an orbital period of 5.203~days. Using \texttt{kima}, our analysis favors the presence of three planetary signals, with BF of 390 supporting this model over simpler alternatives. The posterior sample with the highest likelihood reveals orbital periods of 3.168, 5.203, and 16.491~days. The second of these is the period reported by \citet{Haswell2020}. 
Figure~\ref{fig:DMPP-2_combined_phase} shows the three signals detected in our analysis. We present two additional planets: DMPP-2\,c is a 51.8~\mearth planet on a 3.168~day orbit, and DMPP-2\,d is a 91.8~\mearth planet on a 16.5~d orbit. This solution yields an increase in mass for the previously detected planet DMPP-2\,b at 171.5~\mearth compared to the 138.9~\mearth originally reported in \citet{Haswell2020}. Figure~\ref{fig:DMPP-2_combined_phase} shows the three signals corresponding to DMPP-2\,b, c, and d from the posterior sample with the highest likelihood for our DMPP-2 analysis. A full activity analysis on this target was carried out in \citet{Haswell2020} and no correlations or significant periodogram power were identified at the above periods. DMPP-2 was observed in TESS Sector 30, but is too close to the edge of the field to enable photometric extraction from the full frame images.
This system is discussed in greater detail in \cite{Barnes2025multiplanet}.

\textbf{\underline{DMPP-3:}}
DMPP-3AB is an eccentric ($e \sim 0.6$) binary system with an orbital period of 507~days and a semi-major axis of 1.2~au \citep{Stevenson2023-DMPP3}. A circumprimary (S-type) planet, DMPP-3Ab, was previously reported in the system with an orbital period of 6.67~days \citep{Barnes2020, Stevenson2023-DMPP3}.

Since then, new radial velocity data from ESPRESSO \citep{Pepe2021} have become available. Including this new data, and modelling without constraining the planet’s eccentricity, we find revised orbital parameters for DMPP-3Ab which differ from those reported by \citet{Stevenson2023-DMPP3}. Specifically, we now find an orbital period of approximately 5.6~days and an eccentricity of $e \approx 0.42$, although the planet’s minimum mass remains consistent at $2.67~\rm{M_{\oplus}}$. The updated parameters are summarised in Table~\ref{tab:Prev_published}.

We note that the 5.6~day signal is not well-supported by the ESPRESSO data alone (see the top plot in Figure~\ref{fig:DMPP-3_combined_phase}), the signal's power appears to originate primarily from our HARPS observations of the system. When allowing for an additional planetary signal in the model, we continue to observe power near a period of $\sim2.26$~days with a mass of $\sim1.08~\rm{M_{\oplus}}$, as previously discussed in \citet{Stevenson2023-DMPP3}. However, additional observations are required to confirm the presence of this potential second planet.

\textbf{\underline{DMPP-4:}}
\citet{barnes2023} identified stellar activity signals which they modelled using a Gaussian Process (GP), along with a candidate Neptune-mass planet, DMPP-4b. 
Here, for the purposes of calculating the detection limits, we subtracted the preferred solution from \citet{barnes2023}, consisting of the 3.498~day planetary signal and the GP model fit to both the radial velocity and bisector span time series. Consequently, in the absence of additional data, we do not update the planetary parameters for this system.

\textbf{\underline{HD\,28471 (DMPP-5):}} 
\citet{Stevenson2025-hd28471} identified three planetary signals in this system, HD\,28471 (DMPP-5)\,b, c and d. 
Here, for the purposes of calculating the detection limits, we subtracted a preferred solution corresponding to the 3.16, 6.12, and 11.68~day planetary signals from \citet{Stevenson2025-hd28471}. Consequently, in the absence of additional data, we also do not update the planetary parameters for this system.

\textbf{\underline{HD\,181433:}}
The planetary system HD\,181433 was initially reported by \citet{Bouchy2009} to host three planets with orbital periods of 9.37, 962, and 2172~days. These orbital solutions were subsequently refined by \citet{Campanella2013} using dynamical simulations, yielding revised periods of 9.37, 975, and 2468~days. With the inclusion of additional data, \citet{Horner2019} further updated the system architecture, increasing the orbital separation of planet\,d and deriving best-fit periods of 9.37, 1014.5, and 7012~days.

For HD\,181433, we have obtained ten additional HARPS spectra over a timespan of ten days, with the aim of detecting any planets interior to HD\,181433\,b. However, incorporating this new dataset into our analysis reveals no evidence for additional planets in the system. We include the \citet{Horner2019} solution as known objects\footnote{Setting a known object in \kima allows tight individual priors to be set on a specific signal, and allows the samples to explore the known signal along with any additional signals that may be present in the data.} in our \texttt{kima} model, and recover parameters consistent with their reported uncertainties.

\textbf{\underline{HD\,39194:}}
The HD\,39194 multiplanet system was reported by \citet{Unger2021} to host three super-Earth planets. We obtained 12 additional HARPS spectra over a 9-day timespan in an effort to search for planets interior to the 5.6~day period HD\,39194\,b. Our analysis reveals no evidence for additional planets beyond those identified by \citet{Unger2021}, although we do detect the 1-day alias of the 14 and 5.6-day signals previously reported. Excluding 
these aliases, our preferred \kima solution is consistent with that of \citet{Unger2021} (see Table~\ref{tab:Prev_published}). However, we note that the stellar mass used in \citet{Unger2021} of $0.67~\rm{M_\odot}$ is very low for a K0V type star, and instead we use the mass derived from \textit{Gaia} FLAME of $0.86\pm0.04$ \citep{Gaia2020}.

\textbf{\underline{HD\,89839}:}
is known to host a giant planet with an orbital period of 3440~d, as reported by \citet{Xiao2023}. We include this known signal in our \kima analysis by treating it as a known object, using the parameters provided in their study. However, since the orbital period lies well beyond the range of interest in our current search, we do not subtract the signal from the data. To probe for additional short-period companions, we contribute 22 new HARPS spectra spanning ten days. Unfortunately, with the current dataset, we find no evidence for additional short-period planets in the system. Our recovered parameters for HD\,89839~b can be found in Table~\ref{tab:Prev_published}.

\renewcommand{\thefootnote}{\fnsymbol{footnote}}
\begin{table*}
    \centering
	\caption{Orbital parameter comparison table for previously published planetary signals. Upper rows for each planet are previously published solutions, lower rows are from this work. For eccentricities with posterior distributions consistent with zero we provide the 1-sigma upper limit of the distribution.}
	\label{tab:Prev_published}

	\begin{tabular}{llllllll}
		\hline
		\hline
		 System name & Period [days] & K [m\,s$^{-1}$] & $e$ & $\omega$ [rad] & $m_{\rm{p}}\sin i~[\rm{M_{\oplus}}]$ & T$_0$ [BJD-2400000] & a [AU] \\

		\hline
		DMPP-1\,b$^1$ & $18.57^{+0.01}_{-0.01}$ & $5.162^{+0.19}_{-0.31}$ & $<0.083$ & - & $24.27^{+1.16}_{-1.59}$ & $57375.58981035$ & $0.1462 ^{+0.0012}_{-0.0012}$\\
& $17.812^{+0.025}_{-0.019}$ & $3.63^{+0.60}_{-0.54}$ & $<0.2$ & $3.21^{+2.42}_{-2.31}$ & $16.80^{+2.5}_{-2.5}$ & $58679.85^{+7.38}_{-5.23}$ & $0.1423^{+0.0013}_{-0.0013}$ \\
		  DMPP-1\,c & $6.584^{+0.003}_{-0.002}$ [10.57] & $2.884 ^{+0.16}_{-0.47}$ & $<0.057$ & - & $9.6^{+0.53}_{-1.58}$ & - & $0.0733 ^{+0.0006}_{-0.0007}$\\
& $10.03^{+0.53}_{-0.24}$ & $2.89^{+0.16}_{-0.15}$ & $<0.22$ & $2.11^{+1.53}_{-1.52}$ & $10.94^{+0.61}_{-0.59}$ & $58684.72^{+3.87}_{-3.29}$ & $0.0973^{+0.0031}_{-0.0016}$ \\
		  DMPP-1\,d & $2.882^{+0.001}_{-0.001}$ [3.146] & $1.326 ^{+0.149}_{-0.130}$ & $<0.07$ & - & $3.35^{+0.38}_{-0.34}$ & - & $0.0422 ^{+0.0004}_{-0.0003}$\\
& $3.1417^{+0.0020}_{-0.0248}$ & $1.88^{+0.33}_{-0.30}$ & $<0.36$ & $4.06^{+0.60}_{-0.65}$ & $4.72^{+0.81}_{-0.73}$ & $58688.54^{+0.84}_{-1.12}$ & $0.04467^{+0.00045}_{-0.00040}$ \\
		  DMPP-1\,e\footnotemark[1] & $5.516^{+0.002}_{-0.004}$ & $1.316 ^{+0.358}_{-0.206}$ & $<0.07$ & - & $4.13^{+0.66}_{-1.14}$ & - & $0.0651 ^{+0.0005}_{-0.0005}$\\
& N/A & N/A & N/A & N/A & N/A & N/A & N/A \\

            \hline
		DMPP-2\,b$^2$ & $5.2072^{+0.0002}_{-0.0055}$ & $40.26^{+2.69}_{-5.40}$ & $<0.082$ & - & $138.89^{+9.53}_{-18.75}$ & $53254.84461$ & $0.0664^{+0.0005}_{-0.0005}$ \\
        & $5.2030^{+0.0012}_{-0.0012}$ & $49.60^{+3.12}_{-2.19}$ & $<0.06$ & $2.41^{+1.68}_{-1.23}$ & $171.48^{+10.99}_{-8.34}$ & $57521.43^{+1.47}_{-0.10}$ & $0.06638 \pm 0.00046$ \\
		  DMPP-2\,c\footnotemark[2] & N/A & N/A & N/A & N/A & N/A & N/A & N/A\\
         & $3.168^{+0.355}_{-0.002}$ & $17.99^{+1.76}_{-2.02}$ & $<0.28$ & $2.00^{+0.62}_{-0.48}$ & $51.79^{+5.43}_{-5.49}$ & $57524.30^{+0.22}_{-0.75}$ & $0.04789^{+0.00327}_{-0.00049}$ \\
 		  DMPP-2\,d & N/A & N/A & N/A & N/A & N/A & N/A & N/A\\
         & $16.491^{+0.067}_{-0.053}$ & $18.21^{+1.32}_{-1.38}$ & $<0.1$ & $4.66^{+1.01}_{-3.61}$ & $91.79^{+6.11}_{-7.32}$ & $57516.66^{+3.69}_{-4.02}$ & $0.1432^{+0.0012}_{-0.0013}$ \\
            \hline
		DMPP-3A\,b$^3$ & $6.67^{+0.03}_{-0.01}$ & $0.82^{+0.20}_{-0.07}$ & $0.174^{+0.032}_{-0.084}$ & $0.9189^{+0.0017}_{-0.0080}$ & $2.22^{+0.5}_{-0.28}$ & - & $0.0670^{+0.0003}_{-0.0002}$\\
                     & $5.590^{+0.169}_{-0.002}$ & $1.17\pm0.27$ & $0.42\pm0.17$ & $1.96^{+0.48}_{-0.52}$ & $2.67^{+0.50}_{-0.56}$ & $57432.1^{+2.8}_{-2.1}$ & $0.05958^{+0.00119}_{-0.00043}$\\
		DMPP-3B & $506.89\pm0.01$ & $2657.31^{+0.33}_{-0.02}$ & $2.773\pm0.001$ & $0.596\pm0.001$ & $82.52\pm0.53\rm{M_J}$ & - & $1.139\pm0.004$ \\
                     & $506.93^{+0.01}_{-0.05}$ & $2671.1^{+3.5}_{-9.8}$ & $0.59811^{+0.00036}_{-0.00078}$ & $2.77356^{+0.00113}_{-0.00041}$ & $78.25\pm0.56\rm{M_J}$ & $56936.288^{+0.254}_{-0.027}$ & $1.2336^{+0.0039}_{-0.0040}$\\
		DMPP-3+ & $809.38^{+0.20}_{-0.34}$ & $3.52^{+0.20}_{-0.34}$ & $0.0$(Fixed) & $5.0027^{+0.0049}_{-0.0019}$ & $0.156\pm0.007\rm{M_J}$ & - & $1.641^{+0.006}_{-0.005}$ \\
                    & $809.29^{+0.93}_{-1.01}$ & $2.8^{+0.60}_{-0.28}$ & $0.0$(Fixed) & $3.36^{+0.80}_{-2.46}$ & $0.124^{+0.024}_{-0.013}\rm{M_J}$ & $57090.0^{+212.3}_{-247.8}$ & $1.6409^{+0.0056}_{-0.0055}$ \\
            \hline
		DMPP-4\,b$^4$ & $3.4982^{+0.0015}_{-0.0027}$ & $4.58^{+0.59}_{-0.67}$ & $<0.063$ & $3.6^{+1.8}_{-2.5}$ & $12.6^{+1.6}_{-1.8}$ & $57139.8^{+1.3}_{-1.3}$ & $0.04854^{+0.00033}_{-0.00054}$\\
            \hline
		HD\,28471 / DMPP-5\,b$^5$ & $3.1649^{+0.0002}_{-0.0003}$ & $1.679^{+0.180}_{-0.197}$ & $0.195^{+0.061}_{-0.073}$ & $2.66^{+0.40}_{-0.45}$ & $3.72^{+0.40}_{-0.43}$ & $52945.806$ & $0.042^{+0.001}_{-0.001}$\\
        HD\,28471 / DMPP-5\,c & $6.1245^{+0.0856}_{-0.0009}$ & $2.045^{+0.190}_{-0.257}$ & $0.088^{+0.058}_{-0.054}$ & $1.32^{+3.88}_{-0.78}$ & $5.72^{+0.57}_{-0.72}$ & - & $0.065^{+0.001}_{-0.001}$\\
		HD\,28471 / DMPP-5\,d & $11.6810^{+0.0042}_{-0.0055}$ & $1.411^{+0.233}_{-0.218}$ & $0.093^{+0.064}_{-0.058}$& $1.67^{+3.56}_{-1.08}$ & $4.91^{+0.82}_{-0.77}$ & - & $0.100^{+0.002}_{-0.002}$\\
            \hline
		HD\,181433\,b$^6$ & $9.3745\pm0.0002$ & $2.7\pm0.1$ & $0.336\pm0.014$ & $3.672\pm0.044$ & $7.09\pm0.10$ & $52939.16\pm0.06$ & $0.0801\pm0.0001$\\
                         & $9.37450^{+0.00025}_{-0.00038}$ & $2.646^{+0.089}_{-0.124}$ & $0.396^{+0.015}_{-0.024}$ & $3.523^{+0.073}_{-0.138}$ & $7.22^{+0.44}_{-0.45}$ & $56470.15^{+0.10}_{-0.19}$ & $0.0828^{+0.0019}_{-0.0020}$\\
		HD\,181433\,c & $1014.5\pm0.6$ & $16.55\pm0.07$ & $0.235\pm0.003$ & $0.150\pm0.012$ & $214.29\pm0.95$ & $52184.3\pm1.9$ & $1.819\pm0.001$\\
                         & $1018.50^{+1.47}_{-0.78}$ & $17.05^{+0.19}_{-0.18}$ & $0.2402^{+0.0093}_{-0.0080}$ & $0.177^{+0.013}_{-0.026}$ & $235.50^{+10.80}_{-11.13}$ & $55711.3^{+3.4}_{-7.1}$ & $1.885^{+0.042}_{-0.044}$\\
            HD\,181433\,d & $7012\pm276$ & $8.7\pm0.1$ & $0.469\pm0.013$ & $4.212\pm0.042$ & $194.63\pm1.27$ & $46915\pm239$ & $6.60\pm0.22$\\
                         & $6896.1^{+102.7}_{123.7}$ & $8.58^{+0.15}_{-0.13}$ & $0.444^{+0.020}_{-0.018}$ & $4.365^{+0.059}_{-0.030}$ & $206.91^{+10.49}_{-10.49}$ & $50647.2^{+155.4}_{-87.9}$ & $6.74^{+0.18}_{-0.18}$\\
            \hline
		HD\,39194\,b$^7$ & $5.6368\pm0.0004$ & $1.86\pm0.13$ & $<0.105; <0.207$ & - & $4.0\pm0.3$ & $55503.8\pm1.4$ & $0.056\pm0.001$\\
                        & $5.63692\pm0.00032$ & $1.84\pm0.11$ & $<0.101$ & $3.1 \pm 1.7$ & $4.62\pm0.32$ & $52942.3 \pm 2.5$ & $0.05895\pm0.00092$\\
		HD\,39194\,c & $14.030\pm0.003$ & $2.19\pm0.14$ & $<0.078; <0.154$ & - & $6.3\pm0.5$ & $55499\pm3$ & $0.103\pm0.002$\\
                        & $14.0309\pm0.0022$ & $2.19\pm0.13$ & $<0.078$ & $2.2^{+2.8}_{-1.4}$ & $7.49\pm0.50$ & $52937.9 \pm 5.8$ & $0.1083\pm0.0017$\\
		HD\,39194\,d & $33.91\pm0.03$ & $1.04\pm0.14$ & $<0.174; <0.333$ & - & $4.0\pm0.6$ & $55516\pm7$ & $0.185\pm0.0033$\\
                        & $33.874\pm0.024$ & $1.13\pm0.13$ & $<0.159$ & $4.2^{+1.4}_{-3.3}$ & $5.16\pm0.59$ & $52932.3^{+8.8}_{-16.8}$ & $0.1949\pm0.0031$\\
            \hline
		HD\,89839\,b$^8$ & $3441\pm18$ & - & $0.187\pm0.013$ & $2.818\pm0.061$ & $1211.3\pm24.5$ & $56682.0\pm35.0$ & $4.761\pm0.042$\\
                        & $3391.2^{+26.1}_{-24.6}$ & $45.65^{+0.63}_{-0.60}$ & $0.185\pm0.014$ & $2.757 \pm 0.081$ & $1196.9 \pm 26.1$ & $49842.2 \pm 60.2$ & $4.713 \pm 0.045$\\
            \hline
    \multicolumn{8}{l}{\textbf{Notes:} 1 - \cite{Staab2020}, 2 - \cite{Haswell2020}, 
    3 - \cite{Stevenson2023-DMPP3}, 4 - \cite{barnes2023} (parameters not updated here),} \\
    \multicolumn{8}{l}{5 - \cite{Stevenson2025-hd28471} (parameters not updated here), 6 - \cite{Horner2019}, 7 - \cite{Unger2021}, 8 - \cite{Xiao2023}}\\
    \multicolumn{8}{l}{\footnotemark[1]We only find evidence of three signals present in our DMPP-1 data.}\\
    \multicolumn{8}{l}{\footnotemark[2]In \cite{Haswell2020} only a single planetary signal was detected in the DMPP-2 data.}\\
    \end{tabular}
\end{table*}
\renewcommand{\thefootnote}{\arabic{footnote}}

\subsection{The Dangers of Assuming Circular Orbits}\label{subsec:assume_circular}
As shown in \cite{Standing2022}, fixing $e_{\rm{p}}=0$ when calculating detection limits can lead to overly optimistic results, by as much as 40\%. In this study, we extend their analysis by calculating detection limits for 24 target systems, an eightfold increase in sample size. For each system, we plot the standard $3\sigma$ detection limit in blue and the corresponding $3\sigma$ limit calculated using only posterior samples with $e_{\rm{p}}<0.1$ in red. These plots are available in Appendix~\ref{app:det-lims}.

\begin{figure}
    \centering
	\includegraphics[width=0.48\textwidth]{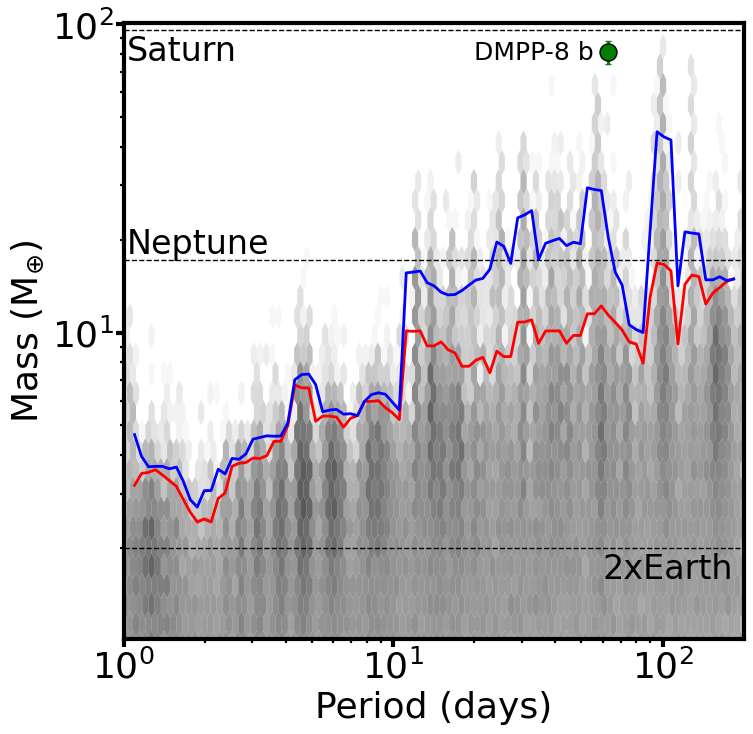}
    \caption{Detection limit plot as in Figure~\ref{fig:HD118006_Det_lim} for target HD191122 / DMPP-8 demonstrating the difference between the circular (red) detection limit and the standard (blue) detection limit.}
    \label{fig:e_cut_limit}
\end{figure}

Across all targets, the circular (red) detection limits are consistently lower than the standard (blue) limits, with a median reduction of $7.99\%$ and a maximum difference of $38.7\%$, equivalent to $5.1~\rm{M_\oplus}$ for HD191122 (Figure~\ref{fig:e_cut_limit}).

Our findings reinforce the conclusion of \cite{Standing2022} that using a diffusive nested sampler where no simplifying assumptions are necessary yields more robust detection limits. Moreover, enforcing circular orbits can lead to an apparent oversensitivity of up to 8\%, potentially biasing both detection limits and the resulting occurrence rates.

\subsection{Planet discoveries}\label{sec:planet_params}
We now present new planetary signals discovered in our analysis with BF\,$ >150$.
The new planetary parameters are listed in Table~\ref{tab:Detected_planets}, while Table~\ref{tab:Subtracted_signals} lists the maximum likelihood parameters of all signals used to derive detection limits, i.e. it also includes signals discussed in Section~\ref{sec:prev_published}.

Although the DMPP sample comprises low activity targets, we cannot rule out the presence of intrinsic stellar activity above the $\log$~$R^\prime_\textrm{HK} = -5.1$ basal cutoff. Thus, while we have modeled all signals as Keplerians, we also examined common stellar activity metrics for targets with sufficient RV measurements or with candidate planet signals. For each target spectrum, the Ca {\sc ii} H\&K indices along with the bisector inverse span (BIS) and full width at half maximum (FWHM), derived from the HARPS DRS cross-correlation functions (CCFs), were tested for periodicities and correlations with the derived \textsc{s-BART} RVs. The mean BIS and FWHM variation for each target can be found in Table~\ref{tab:obs_summary}. In addition, we tested the third line moment, $M_3$ of the CCFs \citep{Barnes2024}, which also returns a similar, but potentially more sensitive measure than the BIS of photospheric activity components. We also examined the \textit{TESS} photometry for targets with planet signals. The \textit{TESS} SPOC Simple Aperture Photometry (SAP) fluxes contain very large systematics with discontinuities and large amplitude sector-length trends, meaning they are not suitable for stellar periodicity searches for low activity stars. By contrast, the Pre-Search Data Conditioned Simple Aperture Photometry (PDCSAP) fluxes only show low level variability for low activity stars, since they are optimised for transit searches\footnote{see https://archive.stsci.edu/missions-and-data/tess and \textit{TESS} archive manual at https://outerspace.stsci.edu/display/TESS/TESS+Archive+Manual}. The PDCSAP fluxes often show peak period power at around $\lesssim 10$~d, using single and multiple \textit{TESS} sectors; however, the corresponding signal amplitudes of $10 - 100$~ppm are very low and likely arise from remaining systematics after pre-conditioning for transit searches. For comparison, we note that \citet{claytor2024} have employed \textit{TESS} full frame images (FFIs) and machine learning to obtain rotation periods for over 70000 stars in the \textit{TESS} southern continuous viewing zone. They are sensitive to periods up to $35$~d for GK stars and $80$~d for M stars and generally recover periodicities for targets with $300 \lesssim S_\textrm{ph} \lesssim 10000$ ppm (where $S_\textrm{ph}$ is a measure of variability amplitude from a mean of standard deviations of lightcurve segments of length $5 \times P_\textrm{rot}$). We used {\sc unpopular} \citep{hattori2022} to detrend FFIs for individual sectors and check for stellar periodicities. Below, we give a brief summary for those targets where we find new candidate planets.

\textbf{\underline{HD\,2134:}} Our data for HD\,2134 provides strong evidence, BF\,$ >$ 25,000, for a single signal with a period of 32.3\,d and a minimum mass of $9.93~\rm{M_\oplus}$.

Two sets of observations were acquired for this target. However, the most recent set was affected by significant issues, leaving only two usable data points. Accounting for an additional RV offset and white-noise jitter parameter renders these two points uninformative, as they contribute negligible statistical weight to the model fit. We therefore exclude these most recent observations from our final analysis. We note, however, that their inclusion does not significantly alter the recovered 32-day signal.

The system also shows moderate evidence for a second planetary signal, with BF of 73.5 in favour of a two-planet model. The highest-likelihood two-planet solution includes a 2.78\,d, $2.69~\rm{M_\oplus}$ planet on a near-circular orbit ($e = 0.15$), and a more eccentric ($e = 0.66$), 21.18\,d, $6.8~\rm{M_\oplus}$ signal. However, given that our current dataset robustly supports only a single-signal interpretation, we  subtract this single-planet model from the data when computing detection limits.

With the exception of FWHM, the activity indices show only weak indications of a correlation, with Pearson's {$|r| \lesssim 0.24$}. Although the FWHM correlation is weak, $r = 0.31 \pm 0.10$, the probability of no correlation is excluded via the p-statistic, $p=0.004$. With \vsini{} $= 2.61 \pm 0.55$~\kms{} and $R_* = 1.16 \pm 0.01$~R$_\odot$ (values from \texttt{SPECIES}; \citealt{Soto2018}), and assuming randomly distributed stellar axial inclinations, $i$, (i.e. observed probability density $P(i) \propto \textrm{sin}\,i$) we predict a most probable $P_\textrm{rot} = 18.6$\,d. By fixing the inclinations to the observed mean of $\int_{0}^{\frac{\pi}{2}} i \sin(i) \, \mathrm{d}i = 1$\,rad~$=57.3$\degs{} and $i=90$\degs{}, we find respectively $P_\textrm{rot} = 17.5$\,d and $20.8$\,d. A periodogram of FWHM shows marginal evidence for periodicities at $24.5, 21.1$ and $31.6$~d ($\Delta\log{L}$ = $15.5, 13.5 $ and $13.4$), which are distinct from window function peaks. HD\,2134 was observed in three sets of consecutive \textit{TESS} sectors: Sectors 27 and 28, Sectors 67 and 68 and Sectors 94 and 95. We performed periodogram analysis of the photometry using likelihood periodogram analysis \citep{Anglada2013,Anglada2016}, which allows for offsets between sectors. We assumed a model with a simple sinusoidal variability. The photometric data were obtained with {\sc unpopular} \citep{hattori2022}, which implements casual pixel model detrending to retain stellar variability.

Using all six sectors, {a} period peak of $26.4$
is found at $\Delta \log L = 26.6$ with a sinusoidal flux (half) amplitude of 0.0105\% ($105$\,ppm). 
Treating all normalised sectors without offsets in the periodogram search again yields $26.4$\,d but with reduced significance ($\Delta \log L = 14.6$) and slightly reduced amplitude 0.009\% ($89$\,ppm) variability, suggesting the modulation is present across the Sector timescale, and not a windowing effect due to allowing for offsets. Further investigation of only Sectors 67 and 68, which show the highest photometric amplitude among the six sectors, returns a peak period of $27.2$\,d and sinusoidal amplitude of 0.115\% (115 ppm). Sectors 27 and 28 return peak period of $13.4$\,d ($99$\,ppm), but with a second peak at $27.5$\,d (84\,ppm), whereas Sectors 94 and 95 return peaks at $6.2$\,d ($110$\,ppm), $18.0$\,d ($113$\,ppm) and an indistinct broad peak / inflection at $25.7$\,d ($114$\,ppm). The shorter period maximum peaks in Sectors 27 and 28 and Sectors 94 and 95 appear to be harmonics of the $\sim 27$\,d rotation period, implying changing activity. Importantly, while $27$\,d is close to the FWHM periodicities, we also note that it essentially matches the \textit{TESS} sector length, with the harmonics matching TESS momentum dumps and the spacecraft orbital period of $\sim 13.5$\,d.

The Keplerian solution for HD\,2134 with \kima is multi-modal, with samples also predominantly appearing at $26.3$ and $21.5$\,d in addition to the 32-day signal. Thus, while the candidate planet signal is coincident with the 32-day FWHM signal and close to some \textit{TESS} photometric periodicities, it is distinct from the predicted rotation periods. Taking this into account, we refrain from announcing that this signal is planetary in nature here, and refer to it only as a planet candidate signal. As such the signal is not considered in the occurrence rate calculations, and is shown as a grey point in Figure~\ref{fig:DMPP_Occ-rates}.
Further data are required to determine the significance of the other RV and FWHM periodicities, since they are also closer to the predicted stellar rotation period. A more detailed analysis of this system will be presented in \cite{Barnes2025multiplanet} (in prep).

\begin{figure}
    \centering
    
    \begin{subfigure}[b]{0.48\textwidth}
        \centering
        \includegraphics[width=\textwidth]{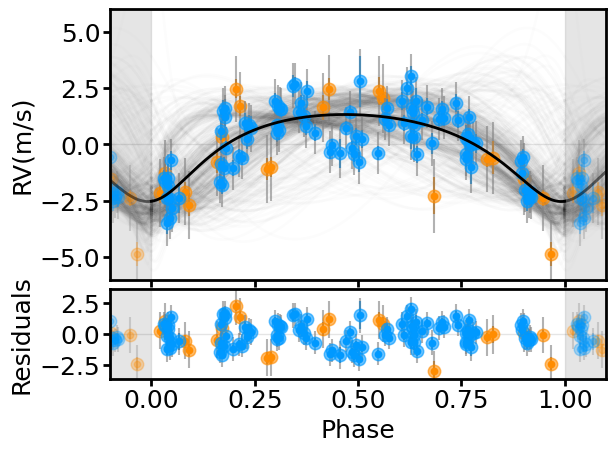}
    \end{subfigure}
    \hfill
    \begin{subfigure}[b]{0.48\textwidth}
        \centering
        \includegraphics[width=\textwidth]{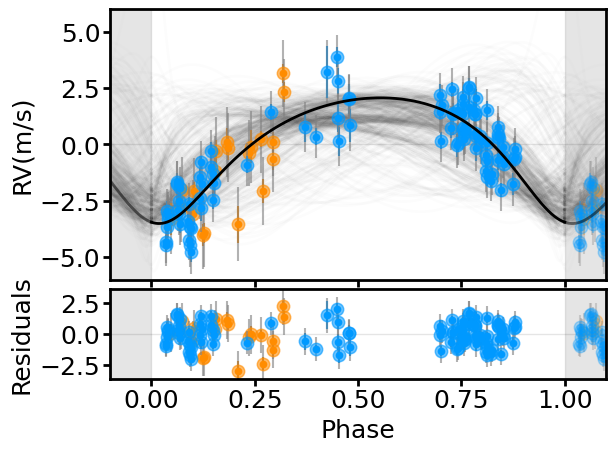}
    \end{subfigure}
    
    \caption{Phased Keplerian RV models of HD\,67200 / DMPP-6\,b (top) HD\,67200 / DMPP-6\,c (bottom) with HARPS data post-fibre upgrade (blue) and HARPS data post-COVID warm-up (orange) and their respective residuals. The additional RV jitter term has been added to the plotted uncertainties and is shown by the grey error bars. Faded Keplerian models are based on 500 randomly drawn posterior samples from a \kima run. The shaded regions display the repeating signal.}
    \label{fig:HD67200b+c_combined}
\end{figure}

\textbf{\underline{HD\,67200 (DMPP-6):}}\label{sec:HD67200} Our RVs for HD\,67200 provide strong evidence (BF~$> 31,000$) in favour of two Keplerians with orbital periods of 7.6 and 36.4\,d, corresponding to minimum masses of 5.8 and $13.5~\rm{M_\oplus}$, respectively. We designate these planets as DMPP-6\,b and DMPP-6\,c (Figure~\ref{fig:HD67200b+c_combined}), and subtract their signals from the dataset to compute detection limits.

There is also moderate evidence (BF = 36) for a third periodic signal in the RV data, with a period of $\sim3$\,d and minimum mass of $2.2~\rm{M_\oplus}$ (indication of this signal can be seen in Figure~\ref{fig:Det-lim_HD67200}. This signal appears alongside the 7.6 and 36.4\,d periods in the highest-likelihood three-planet solution.  

We find \vsini{}~$= 2.4 \pm 0.9$~\kms{} and $R_* = 1.27 \pm 0.01$~R$_\odot$ using \texttt{SPECIES}. A most probable $P_\textrm{rot} = 19.2$\,d is found, while for fixed mean and maximum inclinations (see HD\,2134 above) of $i=57.3$\degs{} and $i=90$\degs{}, the respective $P_\textrm{rot} = 18.9$ and $22.7$\,d. The pre-COVID data set shows an activity-RV Pearson's linear correlation of $|r| < 0.28$. The $M_3$, BIS, FWHM and S-index post-COVID correlations are weak to moderate with, respectively, $r=-0.19, 0.26, -0.43$ and $0.32$, but large uncertainties in $r$ are a reflection of the small (20 observations) data set. Only S-index shows a significant periodicity at $15.99$~d, with other peaks at longer periods, including $29.8$\,d. While this may be related to the expected rotation period, it is not coincident with any of the candidate Keplerian periodicities. With a declination of $-70.0$\degs{}, HD\,67200 has been observed fairly continuously by \textit{TESS} in 42 sectors during Years 1, 3, 5 and 7. We split the data into odd and even sectors for Years 1 and 3 since they contain respectively 12 and 13 sectors with usable full frame images (sectors $7$ and $95$, were rejected due to extraction issues with the full frame images). The semi-amplitudes of the sinusoidal fits arising from periodogram analysis are between $30$\,ppm and $49$\,ppm. However, we do not find consistent periods for the data sets considered. For Year 1 (odd), Year 1(even), Year 3 (odd), Year 3(even), Year 5 and Year 7, we find respectively $P=10.4, 6.23, 13.7, 10.7, 4.52~\textrm{and}~3.65$\,d with semi-amplitudes of $48, 30, 47, 46, 38, 49$\,ppm and $\Delta \log L = 24.3, 10.2, 28.8, 26.1, 17.9, 36.4$\,d. We note that the Year 1 (even) data set does not recover any significant periods, while the Year 1 (odd) data set finds significant periodicity. Additional peaks at around $10-15$\,d are common, even when the peak power lies at lower period, but these often appear with  $\Delta \log L < 15$. The sinusoidal amplitudes for HD\,67200, one of our brightest targets ($V=7.7$) are relatively low. It is likely that systematics related to the \textit{TESS} orbit and photometric extraction remain at these amplitudes and thus preclude reliable identification of periodicities at constant amplitude with periodogram analyses.

Despite this, we note that the periodicities for HD\,67200 are not inconsistent with the S-index periodicities derived from our spectra. A more detailed examination of HD\,67200 will be presented in \cite{Barnes2025multiplanet}.

When calculating the detection limits for HD\,67200/DMPP-6, we initially forced \texttt{kima} to fit $N_{\rm{p}} = 1$ signal to the residual (subtracted) dataset. However, the resulting posterior samples predominantly clustered around the 3.2\,d signal. This is expected, as the 3.2\,d candidate was not subtracted from the data. Notably, when following the detection-limit recovery methodology described in Section~\ref{sec:det_lim_revcovery}, we found that the sampler did not explore other regions of the period space. In this case, it was necessary to set $N_{\rm{p}} = 2$ to enable adequate exploration of the entire period range. This was the only system in our sample for which this adjustment was required, and supports the reality of the 3.2-day signal, despite the merely moderate statistical evidence in the initial model comparison.

\textbf{\underline{HD\,118006 (DMPP-7):}} We find strong evidence for a planetary signal in the radial velocity (RV) data of HD\,118006, with BF $>$20,700 in favour of a single-planet model. The planet, henceforth designated DMPP-7\,b, has an orbital period of $4.98$\,d and an RV semi-amplitude of $21.8~\rm{m\,s^{-1}}$, corresponding to a minimum mass of $61.4$\,\mearth. Although there is weak evidence for a second signal (BF = 9.2), the current data set of 19 observations is insufficient to confirm the presence of additional planets.  

{HD\,118006 was classified as a G1V star by \citep{Houk1999}. Our \texttt{SPECIES} age estimate of $3.87^{+0.28}_{-0.31}$\,Gyr and \hbox{$\log\,g = 4.17 \pm 0.09$} confirm main sequence status, but are at tension with the corresponding radius estimate of $R_* = 1.49 \pm 0.02$~R$_\odot$, which implies evolved status. Further, an independent radius estimate using parameters derived from the Gaia mission also returns $R_* = 1.64\pm0.04$~R$_\odot$.} We find \vsini{} $= 3.04 \pm 0.74$~\kms{} from \texttt{SPECIES}, yielding a most probable $P_\textrm{rot} = 19.8$\,d. The method for obtaining rotational velocities described by \cite{Murphy2016} simultaneously fits for the macroturbulent velocity, giving \vsini{} $= 2.6^{+2.3}_{-2.6}$~\kms{}. This shifts the most probable $P_\textrm{rot} = 12.6$\,d, indicating sensitivity of the Monte Carlo estimate to the \vsini{} value and its uncertainty. Nevertheless, neither estimate is coincident with the large RV signal. RV-activity correlations for $M_3$, BIS, FWHM and S-index are respectively $r = -0.04, 0.02, -0.13$ and $0.32$ ($p=0.88, 0.94, 0.59$ and $0.19$) indicating weak or no correlations. There are no significant periodicities evident in the activity indicators. Likelihood periodogram analysis of \textit{TESS} Sectors 23, 46, 50 and 91 indicates significant periodicity, with peak power of $\Delta \log L = 15.0$ at $21.5$\,d and with a sinusoidal semi-amplitude of 111 ppm. A low significance peak cluster at $9.6$\,d ($\Delta \log L = 10.4$ is also present) and may be a harmonic of the $21.5$\,d peak. There is no significant power at the $4.98$\,d period seen in the RVs. 

\begin{figure}
    \centering
	\includegraphics[width=0.48\textwidth]{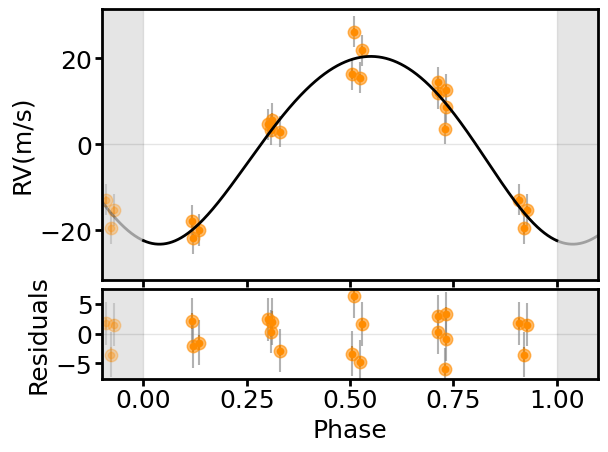}
    \caption{Phased Keplerian Radial-Velocity (RV) model of HD\,118006 / DMPP-7\,b with HARPS data post-COVID warm-up (orange) and associated residuals after removing the planetary signal. The additional RV jitter term has been added to the plotted uncertainties and is shown by the grey error bars. No random samples are shown for this signal as the eccentricity is unconstrained and consistent with 0. The shaded regions display the repeating signal.}
    \label{fig:HD118006b_phase}
\end{figure}

Figure~\ref{fig:HD118006b_phase} shows the Keplerian fit for HD\,118006/DMPP-7\,b. The observational cadence results in clustered data points, as the planet's orbital period is near an integer number of days, causing periodic observational gaps from the Earth’s diurnal cycle. Figure~\ref{fig:HD118006_Det_lim} presents the detection limits for the system, after subtracting the most likely posterior sample, with parameters as stated in Table~\ref{tab:Subtracted_signals}.

\textbf{\underline{HD\,122640:}}
For this system we find a BF $>14,000$ in favour of a single planet solution with a period of 138.5\,d on an eccentric orbit with $e\sim 0.3 $. 
However, for this target we have two datasets, one pre HARPS fibre upgrade and another post HARPS COVID warm-up with a large separation of $>$1500\,d between them. Our sampling within each of these datasets is limited to three tightly sampled periods of $\sim2$, $\sim4$, and $\sim10$\,d each, yielding an effective timespan of $\sim16$\,d. The offset between the two datasets introduces uncertainty about the signal: we suspect this signal is not planetary in nature,  arising instead from poor sampling. Therefore, the signal is not included in any of our occurrence rate calculations, and is represented by a grey circle in Figure~\ref{fig:DMPP_Occ-rates}.

We do not see any strong evidence for short-period signals.
When limiting the period search to $<$20\,d, highly eccentric solutions are found with a large number of posterior samples near the upper end of the period prior.

Therefore, we subtract the 138.5\,d signal from the dataset for the detection limit analysis, but restrict our analysis to orbital periods $<$20\,d.

\begin{figure}
    \centering
	\includegraphics[width=0.48\textwidth]{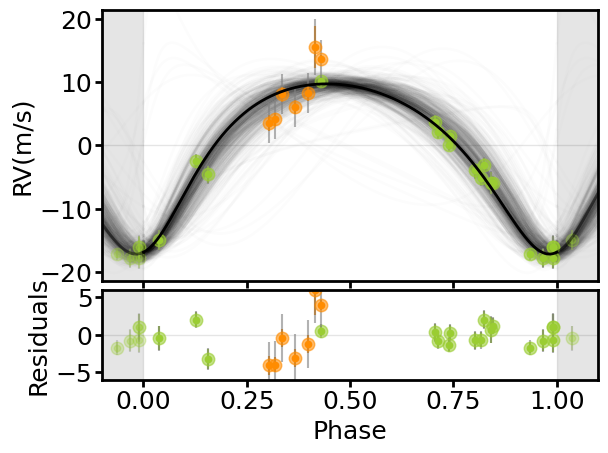}
    \caption{Phased Keplerian Radial-Velocity (RV) models of HD\,191122 / DMPP-8\,b with HARPS data pre-fibre upgrade (green) and HARPS data post-COVID warm-up (orange) along with associated residuals after removing the planetary signal. The additional RV jitter term has been added to the plotted uncertainties and is shown by the grey error bars. Faded Keplerian models are based on 500 randomly drawn posterior samples from a \kima run. The shaded regions display the repeating signal.}
    \label{fig:HD191122b_phase}
\end{figure}

\textbf{\underline{HD\,191122 (DMPP-8):}}
hosts a very clear $62.92$\,d, $81.1$\,\mearth signal which we henceforth refer to as DMPP-8\,b, with weak evidence for additional signals.
See Figure~\ref{fig:HD191122b_phase} for a phase plot of the radial velocity signal of DMPP-8\,b. 
Inspection of the detection limit plot for this target reveals a concentration of posterior samples near an orbital period of $\approx 4.8$~days. This feature is suggestive of an additional planetary signal consistent with the DMPP hypothesis \citep{Haswell2020}; however, the Bayes factor in favour of a two-planet model is only 2.3. 
The penultimate post-COVID warmup RV data point is derived from a spectrum with low counts (S/N in order 50 = 31). Further observations are required to confirm the source of this additional signal. 

With \texttt{SPECIES}, we find \vsini{} $= 2.13 \pm 0.83$~\kms{} and $R_* = 1.24 \pm 0.01$~R$_\odot$, yielding a most probable $P_\textrm{rot} = 24.2$\,d. There are too-few data in the post-COVID set for useful correlation statistics. The pre-fibre upgrade data RV-activity Pearson's correlations for $M_3$, BIS, FWHM and S-index are respectively $r = -0.39, -0.42, -0.13$ and $0.35$  indicating moderate-weak correlations, with low confidence for rejecting the null hypothesis of no correlation ($p=0.11, 0.08, 0.60$ and $0.15$). Periodicities close to the most probable $P_\textrm{rot} = 23.8$\,d are notable in $M_3$ at $P = 23.1$\,d and BIS at $P = 27.7$\,d and S-index at $P = 21.5$\,d and $25.4$\,d, but the respective significances of these periods of $\Delta \log L = 9.3, 7.2, 7.4$ and $6.0$ are low. There is no evidence that the $\sim$63~d signal we identify as DMPP-8\,b is due to stellar activity. More data are needed to assess these periodicities further in conjunction with the tentative $4.8$~d signal. Periodogram analysis of photometry from seven TESS sectors yield a low significance peak ($\Delta \log L = 11.6$) at $9.7$\,d with corresponding $55$\,ppm sinusoidal flux semi-amplitude. 
There are no significant periodicities at $>$10\,d.

The posterior sample with the highest likelihood subtracted from our data to calculate the detection limit for the system can be found in Table~\ref{tab:Subtracted_signals}.

\begin{figure}
    \centering
	\includegraphics[width=0.48\textwidth]{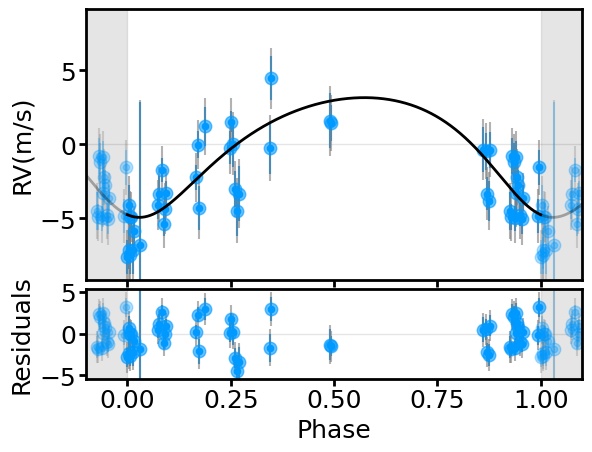}
    \caption{Phased Keplerian Radial-Velocity (RV) models of HD\,200133 / DMPP-9\,b with HARPS data post-fibre upgrade (blue) along with associated residuals after removing the planetary signal. The additional RV jitter term has been added to the plotted uncertainties and is shown by the grey error bars. No random samples are shown for this signal as the eccentricity is unconstrained and consistent with 0. The shaded regions display the repeating signal.}
    \label{fig:HD200133b_phase}
\end{figure}

\textbf{\underline{HD\,200133 (DMPP-9):}} For this target, a single planet solution is preferred with BF\,$ > 26,140$. The signal (Figure~\ref{fig:HD200133b_phase}) has an orbital period of 12.73\,d, and a RV semi-amplitude of $3.8~\rm{m\,s^{-1}}$ corresponding to a 14.4~$\rm{M_\oplus}$ planet (see Figure~\ref{fig:corner_HD200133b}). 
\texttt{SPECIES} yields \vsini{}~$= 3.22 \pm 0.68$~\kms{} and $R_* = 1.26 \pm 0.01$~R$_\odot$, from which we find a most probable $P_\textrm{rot} = 16.2$\,d. HD\,200133 was observed in the post-fibre upgrade interval only. There are no indications of significant periodicity in $M_3$, BIS or S-index, with Pearson's correlations of $0.07 < |r| < 0.14$. The FWHM periodogram peak at $13.8$\,d is close to the Markov Chain (MC) derived most probable period and recovered Keplerian periodicity, but shows low significance of $\Delta \log L = 9.5$ or 5\% FAP.

HD\,200133 was observed in \textit{TESS} sectors 13, 27, 67 and 94. The dominant period peak is found at $10.9$\,d with $\Delta \log L = 26.3$ with $89$\,ppm semi-amplitude. Weaker peaks at $4.1$\,d and $24.2$\,d are seen with marginal or low significances of $\Delta \log L = 15.4$ and $8.9$.  With four sectors from different cycles, there are likely not sufficient data on the timescales of the longer low-significance candidate rotation period. 

As there is no significant evidence for activity periodicities at the RV period, and we did not obtain any posterior samples at this period, we accept the RV signal planet candidate and henceforth refer to this signal as DMPP-9\,b. For the HD\,200133/DMPP-9 detection limit calculation, the posterior sample with the highest likelihood with eccentricity $ < 0.2$ was subtracted as the posterior samples corresponding to the signal are consistent with a circular orbit. The subtracted signal parameters can be found in Table~\ref{tab:Subtracted_signals}.

\begin{table*}
    \centering
	\caption{Orbital parameters for planetary signals identified in this work. For eccentricities with posterior distributions consistent with zero we provide the 1-sigma upper limit of the distribution.}
	\label{tab:Detected_planets}

	\begin{tabular}{llllllll}
		\hline
		\hline
		 Planet name & Period [days] & K [m\,s$^{-1}$] & $e$ & $\omega$ [rad] & $m_{\rm{p}}\sin i~[\rm{M_{\oplus}}]$ & T$_0$ [BJD-2400000] & a [AU] \\

		\hline
DMPP-2\,c & $3.168^{+0.355}_{-0.002}$ & $17.99^{+1.76}_{-2.02}$ & $<0.28$ & $2.00^{+0.62}_{-0.48}$ & $51.79^{+5.43}_{-5.49}$ & $57524.30^{+0.22}_{-0.75}$ & $0.04789^{+0.00327}_{-0.00049}$ \\
DMPP-2\,d & $16.491^{+0.067}_{-0.053}$ & $18.21^{+1.32}_{-1.38}$ & $<0.1$ & $4.66^{+1.01}_{-3.61}$ & $91.79^{+6.11}_{-7.32}$ & $57516.66^{+3.69}_{-4.02}$ & $0.1432^{+0.0012}_{-0.0013}$ \\

HD\,67200 / DMPP-6\,b & $7.6027^{+0.0261}_{-0.0016}$ & $1.93^{0.30}_{-0.27}$ & $0.34^{+0.11}_{-0.14}$ & $3.06^{+0.39}_{-0.26}$ & $5.80^{+0.94}_{-0.80}$ & $58862.32^{+0.42}_{-3.71}$ & $0.0779^{+0.0016}_{-0.0012}$\\
          
HD\,67200 / DMPP-6\,c & $36.441^{+0.077}_{-0.070}$ & $2.54^{+0.26}_{-0.30}$ & $<0.31$ & $3.11^{+0.66}_{-0.49}$ & $13.47^{+1.60}_{-1.97}$ & $58842.50^{+19.79}_{-13.59}$ & $0.2206^{+0.0033}_{-0.0039}$\\

HD\,118006 / DMPP-7\,b & $4.984^{+0.069}_{-0.061}$ & $21.82^{+1.40}_{-1.41}$ & $<0.1$ & $2.97^{+1.22}_{-1.27}$ & $61.39^{+4.33}_{-4.06}$ & $59996.46^{+1.45}_{-0.81}$ & $0.05873^{+0.00086}_{-0.00089}$ \\

HD\,191122 / DMPP-8\,b & $62.915^{+0.041}_{-0.033}$ & $13.58^{+0.64}_{-0.69}$ & $0.275^{+0.049}_{-0.049}$ & $3.26^{+0.21}_{-0.22}$ & $81.09^{+7.09}_{-7.10}$ & $54205.20^{+2.34}_{-2.12}$ & $0.3098^{+0.0098}_{-0.0108}$ \\
          
HD\,200133 / DMPP-9\,b & $12.733^{+0.584}_{-0.524}$ & $3.80^{+0.66}_{-0.59}$ & $<0.25$ & $2.665^{+1.367}_{-1.373}$ & $14.43^{+2.57}_{-2.36}$ & $57586.981^{+2.458}_{-2.279}$ & $0.1097^{+0.0041}_{-0.0037}$ \\
        
\end{tabular}
\end{table*}

\subsection{Planet abundance for the DMPP survey}
Figure~\ref{fig:DMPP_Occ-rates} presents the completeness of the DMPP survey alongside the corresponding planet occurrence rates.
Our survey achieves near-complete sensitivity ($\sim100\%$) to planets with masses greater than $30\,\mathrm{M_{\oplus}}$ and orbital periods up to 20\,d. We also reach $\sim95\%$ completeness for planets with masses above $10\,\mathrm{M_{\oplus}}$ and periods shorter than 5\,d.

In our most sensitively searched systems, HD\,39194 and HD\,181433, we are capable of detecting Earth-mass planets with orbital periods $\lesssim$5\,d, based on 288 and 208 spectra, respectively (see Figures~\ref{fig:Det-lim_HD39194} \& \ref{fig:Det-lim_HD181433}). Survey completeness declines rapidly for periods exceeding 20 days, consistent with our focus on short-period planet detection.

The occurrence rates plotted in Figure~\ref{fig:DMPP_Occ-rates} suggest a high abundance of terrestrial-mass planets ($ < 10\,\mathrm{M_{\oplus}}$) within orbital periods of 20\,d. Our survey sensitivity declines with decreasing mass as expected. The observational strategy means we are formally most sensitive at the shortest periods, however the methodology does not take into account the correlated astrophysical systematic effects which arise due to plasma flows in the stellar photospheres. The effects on measured RVs of super-granulation are correlated on 1-2\,d timescales, and reduce the effectiveness of our sub-day sampling to resolve RVs due to low mass planets in short period orbits. We have clearly detected a $2.67\,\mathrm{M_{\oplus}}$ planet in a 5.6\,d orbit (DMPP-3A\,b) but due to super-granulation we only have a single planet signal detection $<4\,\mathrm{M_{\oplus}}$ for $P < 5$\,d (HD\,28471 / DMPP-5\,b; \citealt{Stevenson2025-hd28471}). Despite this we do have some moderate evidence for planets in/below this bin: the $\sim2.5\,$d $1.08\,\mathrm{M_{\oplus}}$ signal in DMPP-3, the 2.78\,d, $2.69\,\mathrm{M_{\oplus}}$ signal in HD\,2134, and the $\sim3\,$d $2.2\,\mathrm{M_{\oplus}}$ signal in HD\,67200 / DMPP-6. Further observations are urgently required to confirm these detections.

As demonstrated for HD\,39194 and HD\,181433, additional observations would allow us to detect terrestrial-mass planet signals in our sample, thereby increasing survey completeness and improving the accuracy of our occurrence rates.

\subsubsection{Comparison with other RV planet occurrence rates}

The DMPP project aimed to pre-select stars hosting hot, mass-losing planets using the signatures of circumstellar absorption in their spectra before embarking on RV observations to search for them. If the hypothesis underlying DMPP is correct, we should expect significantly higher occurrence rates of short period planets for the DMPP targets than for the targets of other RV planet searches. To examine this, in Table~\ref{tab:occ_rate_comparison} we compare our results with occurrence rates reported by other RV surveys involving FGK-type stars in \citet{Howard2010}, \citet{Mayor2011}, and \citet{Bashi2020}. We see the DMPP sample exhibits significantly higher planetary occurrence rates, supporting the DMPP hypothesis laid out in \cite{Haswell2020}.

The prior works enumerated in Table~\ref{tab:occ_rate_comparison} assumed circular planetary orbits when calculating occurrence rates, citing \citet{Endl2002} as a justification. In contrast, our analysis demonstrates that this assumption is no longer best practice for deriving detection limits and, consequently, occurrence rates. For a more detailed critique of the continued use of \citet{Endl2002} to justify circular orbit assumptions, we refer the reader to \citet{Standing2022}.

To allow a more direct comparison of our results with those of \citet{Howard2010}, \citet{Mayor2011}, and \citet{Bashi2020}, we apply a correction to account for their circular orbit assumption. Specifically, we apply a uniform correction of $8\%$ to our survey completeness (see Section~\ref{subsec:assume_circular}) and recalculate our occurrence rates. The corrected values are presented in Table~\ref{tab:occ_rate_comparison}.
Even after this correction, the DMPP survey demonstrates a far higher planetary occurrence rate compared to large samples of FGK stars. We note that increasing the completeness within a given parameter space region results in a decrease in the inferred occurrence rates, as described by Equation~\ref{eq:abundance}. This occurs because the number of detected planets remains constant, whereas the total number of systems capable of detecting such planets, yet showing no detection, increases.

Additional comparisons with \textit{Kepler} occurrence rates are not feasible without making strong and uncertain assumptions about the planet mass--radius relation in this specific population.

\begin{table}
    \centering
    \begin{threeparttable}
        \caption{Comparison of occurrence rates from the DMPP survey with other RV works for periods shorter than 50 days. Also included are the DMPP rates corrected to assume circular orbits.}
        \label{tab:occ_rate_comparison}
        \begin{tabular}{c|c|c|c}
            \hline
                &  $3-10~\mathrm{M_{\oplus}}$ & $10-30~\mathrm{M_{\oplus}}$ & $30-100~\mathrm{M_{\oplus}}$    \\
            \hline
            \cite{Howard2010}   &   $11.8 \pm 4.3$\% & $6.5 \pm 3$\%  &   $1.6 \pm 1.2$\%  \\
            \cite{Mayor2011}    &   $16.6 \pm 4.4$\% & $11.1 \pm 2.4$\%  &   $1.2 \pm 0.5$\%  \\
            \cite{Bashi2020}    &   $10.0 \pm 2.0$\% & $6.0 \pm 1.0$\%   &    --   \\
            \hline
            This work   &   $83.0^{+27.1}_{-24.4}$\% &  $27.0^{+15.0}_{-11.2}$\%  &   $13.9^{+11.8}_{-7.5}$\%\tnote{1}  \\
            This work (circular)  &   $70.5^{+23.0}_{-20.7}$\% &  $24.5^{+13.6}_{-10.2}$\%  &   $12.8^{+10.9}_{-6.9}$\%  \\

            \hline
        \end{tabular}
        \begin{tablenotes}
            \footnotesize
            \item[1] HD\,118006/DMPP-7 could be an evolved star (see section~\ref{sec:planet_params}).  
            When removing this planet from the calculations we obtain  
            $9.3^{+11.0}_{-6.0}$\% and $8.5^{+10.1}_{-5.5}$\% respectively.
        \end{tablenotes}
    \end{threeparttable}
\end{table}

Based on these occurrence rates, we estimate the number of sub-basal stars expected in a representative local stellar population. Following \citet{Staabthesis}, approximately $1.4$\% of main-sequence (MS) stars exhibit sub-basal activity. From \textit{Gaia} DR3 \citep{GaiaDR32023}, we selected all MS stars with $V < 12$ within $500~\mathrm{pc}$, yielding $16{,}776$ stars and implying $\sim241$ sub-basal objects.
This calculation is intended as a population-level estimate only; the DMPP survey targets a substantially smaller and brighter subset of stars ($V<10$, $d<100~\mathrm{pc}$), and therefore a much lower absolute number of sub-basal stars is expected in that sample.

Using our occurrence rate statistics for planets with orbital periods $P <$ 50\,d from Table~\ref{tab:occ_rate_comparison}, we estimate the potential number of planets as follows:

\begin{itemize}
    \item Planets with masses $3$--$10~M_\oplus$: $200^{+65}_{-59}$%$142 \pm 58$
    \item Planets with masses $10$--$30~M_\oplus$: $65^{+36}_{-27}$
    \item Planets with masses $30$--$100~M_\oplus$: $22.4^{+27}_{-14}$
\end{itemize}

These values provide a first-order expectation for the number of planets detectable in the selected sample of nearby, bright MS stars with sub-basal activity.

We also estimate the observational efficiency of the DMPP survey purely in terms of the number of observations (OBs) per planet detection. 
The DMPP program obtained a total of 1,336 OBs on the 29 targets contributing to 16 detected planets (excluding previously known systems such as HD\,181433, HD\,39194, and HD\,89839), corresponding to approximately $93$~OBs per planet. 
For comparison, the Eta\,Earth survey \citep{Howard2010} had 5,259 OBs for 34 detections ($\sim$ 155\,OBs per planet), while \citet{Bashi2020} reported $>8,316$ OBs for 55 planets ($>$151\,OBs per planet). 
Thus, the DMPP survey is roughly 55\% more efficient in terms of OBs per detection relative to these previous programs. However, we note (i) that this is calculated assuming a standard integration time for each of the observations in the aforementioned surveys and (ii) the effectiveness of the DMPP observations was hampered by correlated noise due to stellar super-granulation, so the current DMPP observing protocols have an even more pronounced efficiency gain.

\subsection{Demographics of the DMPP sample}\label{subsec:demographics}
\begin{figure*}
    \begin{tabular}{cc}
	      \includegraphics[trim=0mm 0mm -1mm 0mm, height=0.63\columnwidth]{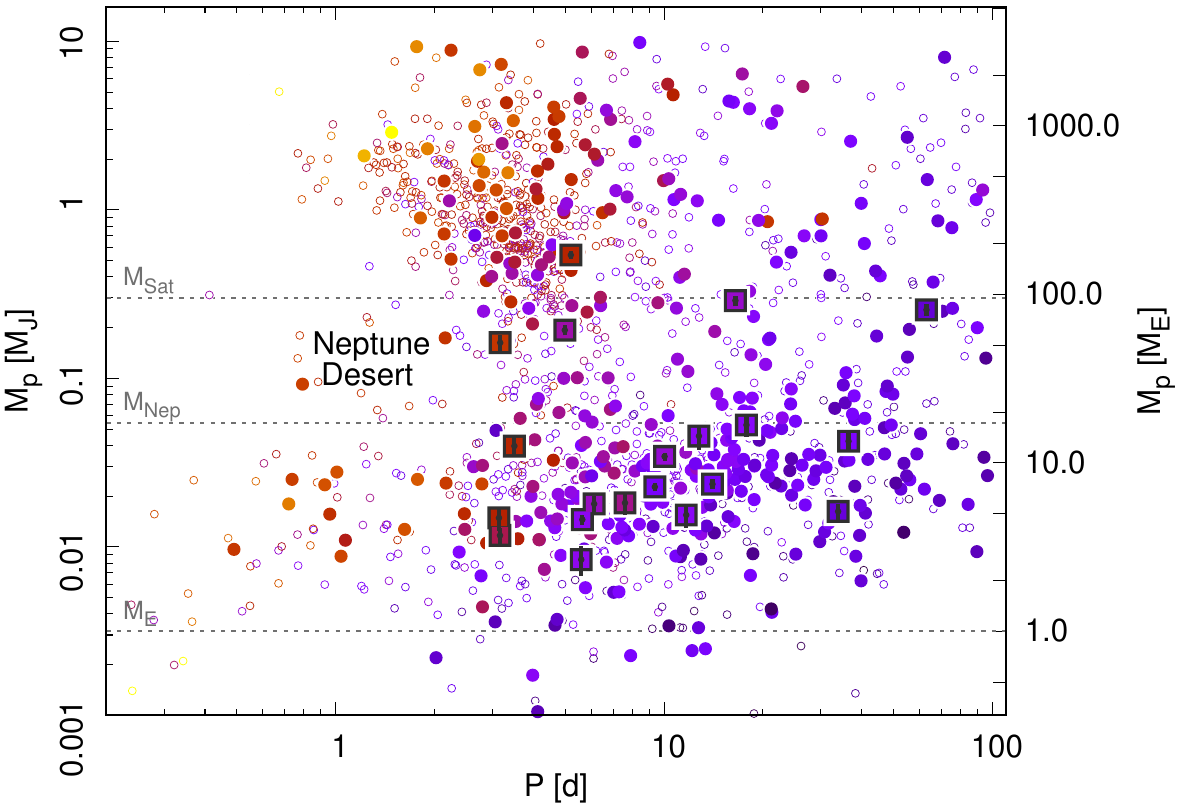} &
        \includegraphics[trim=-1mm 0mm 0mm 0mm, height=0.63\columnwidth]{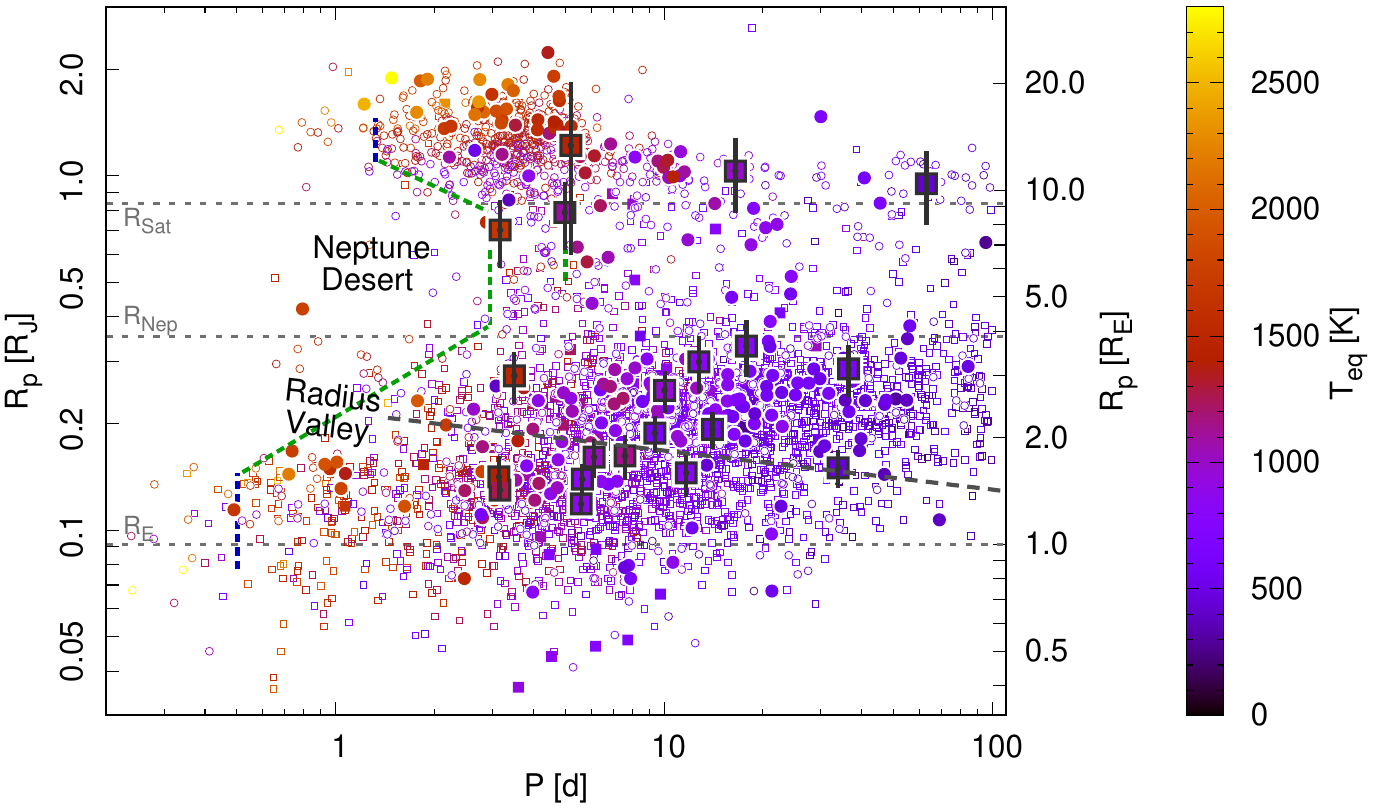} \\
    \end{tabular}
    \caption{Mass-Period and Radius-Period diagrams for the detected DMPP planets (large filled squares) and other known planets from the NASA Exoplanet Archive (\url{https://exoplanetarchive.ipac.caltech.edu} on 2025/11/14). The DMPP radii and ranges (vertical bars) are predicted from their minimum masses using the relationships from \citep{muller2024}. The archival planets with mass or minimum mass determinations are plotted with circles in both panels, while planets with only radius measurements are plotted with squares in the Radius-Period panel. Archival planets are divided into those with $V<10$ (larger filled symbols) and those with $V \geq 10$ (smaller open symbols). The Neptune Desert boundaries (green and blue dashed) are taken from \citet{Castro-Gonzalez2024}. The Radius Valley (grey dashed curve) is taken from the hydrodynamic model prediction of \citet{affolter2023}, and is here defined as $R_\textrm{p}(P) = (1.88/11.2)(P/10)^{-0.11}$ $M_\textrm{J}$.}
    \label{fig:mass_radius_period_v2}
\end{figure*}

Fig~\ref{fig:mass_radius_period_v2} shows our planets compared to the population of other known planets. 
In mass-period space, the DMPP planets are broadly distributed in a similar way to the known population, with some clear differences.
It is immediately clear 
that we have not detected any planets with orbital periods $\sim 1\,$d. The reasons for this may be twofold:
(i) correlated RV noise due to stellar supergranulation makes such periods difficult to detect via RVs, even if the data sampling is designed to do so.
   (ii)
    Such close-in planets experience profound irradiation, leading to equilibrium temperatures $\gtrapprox 2000\,$K. They may have lost a lot of mass, and hence be below our RV detection threshold.
There are also no discoveries of DMPP planets with masses exceeding $0.6\,\rm{M_{J}}$. There are two possible reasons for this: 
(i) many of the known hot Jupiters are fairly distant, and were discovered because their transits are relatively deep and thus easy to detect. We limited our sample to main sequence FGK stars within $\sim 100\,$pc, so it is unsurprising we do not find examples of an intrinsically rare population. Lower mass planets are much more plentiful, and thus well represented within our volume-limited sample. 
(ii)
Our targets were selected based on the signature of planetary mass loss, and mass loss is inhibited by the strong gravity of a massive planet.
Four of the planet discoveries lie along the upper boundary of the Neptune Savannah.

We searched the TESS light curves for transiting planets orbiting our targets, but the shallow ($\sim 100\,$ ppm) transits of terrestrial planets across FGK stars are challenging to reach with TESS. As more sectors of TESS data accumulate and the detrending techniques for TESS light curves improve, such transits may be found. Furthermore, the forthcoming PLATO mission may reveal transits of the short period, edge-on planets originally anticipated by DMPP \citep{Haswell2020}.

In mass-radius space, the DMPP planets below the Neptune Savannah appear to straddle the radius valley, but we note that we have only model-derived radii from \citet{muller2024}, so this is merely indicative. Nonetheless, it is clear that among the sub-Neptune and lower mass DMPP discoveries that the shorter period planets are more likely to be below the radius valley, while those with longer periods lie above it. This is consistent with the hypothesis that the DMPP systems have recent or ongoing planetary mass loss: instellation and mass loss are more likely for the closer-in planets. However, it is also true to say that the broader population of known exoplanets also exhibit this property, which could also be attributed to the simple fact that the stronger instellation at shorter periods leads the locus of the radius valley to be inclined to larger radii at shorter periods.

\section{Conclusions}\label{sec:conclusion}

We conducted a blind search of 24 systems in the DMPP radial-velocity survey, designed to test the hypothesis of \citet{Haswell2020} that stars exhibiting sub-basal chromospheric activity are preferentially orbited by short-period, mass-losing planets. Our analysis yields the following main results:

\begin{enumerate}
    \item We detect a total of 24 planetary signals in the sample, including 7 newly discovered planets: DMPP-2\,c \& d, DMPP-6\,b \& c, DMPP-7\,b, DMPP-8\,b, and DMPP-9\,b along with updated solutions in previously published systems.
    \item For six systems with previously published planets --- DMPP-1, DMPP-2, DMPP-3, HD~181433, HD~39194, and HD~89839 --- we update the orbital parameters of 16 planetary signals using homogeneous \kima fits, with expanded datasets.
    \item We compute robust 99\% confidence detection limits for each of our systems, taking into account eccentric orbits and using the methodology of \citet{Standing2022}. From these, we derive survey completeness as a function of planet mass and orbital period.
    \item Using the completeness maps, we calculate planet occurrence rates for the DMPP sample in well-sampled regions of parameter space (Figure~\ref{fig:DMPP_Occ-rates}). For periods $<50$~d, the occurrence rates are $83.0^{+27.1}_{-24.4}\%$ for $3$--$10\,M_{\oplus}$ planets, $27.0^{+15.0}_{-11.2}\%$ for $10$--$30\,M_{\oplus}$, and $13.9^{+11.8}_{-7.5}\%$ for $30$--$100\,M_{\oplus}$.
    \item Comparing with other radial-velocity surveys (Table~\ref{tab:occ_rate_comparison}), we find the DMPP sample exhibits significantly higher planet occurrence rates than a random sample of FGK-type stars, strongly supporting the DMPP hypothesis.
    \item Based on $\sim 241$ nearby main-sequence stars with sub-basal activity and our measured occurrence rates, we estimate that roughly
    $200$, $65$, and $22$
    planets with masses $3$--$10$, $10$--$30$, and $30$--$100~M_\oplus$, respectively, and short orbital periods ($P < 50~\mathrm{days}$) could be waiting to be discovered in such systems.
    \item The DMPP survey made $\sim 93$~observations per discovered planet, compared to $\sim 155$ for Eta\,Earth \citep{Howard2010} and $>151$ for \citet{Bashi2020}. DMPP is approximately 40\% more efficient purely in terms of observations per planet detection. However, we note that this is calculated assuming a standard integration time for each of the observations in the aforementioned surveys.
    \item Further systematic observations, e.g. 1 epoch per night over an extended period, would ameliorate the affect of ~1-2\,d correlations in RVs due to stellar photospheric flows (super-granulation) and should yield additional low mass hot planet discoveries.
\end{enumerate}

Angular momentum considerations imply the circumstellar material lost from hot planets will be concentrated in the orbital plane of the mass-losing planet. This means the DMPP planets are particularly likely to be in edge-on orientations. Despite this, no transits have been repeatedly detected. The hot planets are likely to be small as they will have lost their gaseous envelopes, and $\sim 1\, \mathrm{M_{\oplus}}$ planets orbiting FGK stars produce transit depths $\lesssim \,100$~ppm.

The DMPP planets exhibit demographics rather similar to the overall population (Fig.~\ref{fig:mass_radius_period_v2}). This may imply that planetary mass loss is ubiquitous for the known short period exoplanet systems. In this interpretation, the special property of the DMPP sample is most likely that their host stars are intrinsically particularly chromospherically quiet, so any circumstellar absorption depresses the apparent activity below the basal value. For a moderately active system, depressing the apparent activity simply makes the star appear less active. This has important implications for the selection of targets hosting short period planets for intensive transit spectroscopy: their host stars may be more active than their $\log$ R$^\prime_\textrm{HK}$ values indicate.

\section*{Acknowledgements}
M.R.S acknowledges support from the European Space Agency as an ESA Research Fellow, and was supported by grant ST/T000295/1 from the Science and Technology Facilities Council (STFC) for the early part of this work. 
JRB and CAH were supported by grants ST/T000295/1, and ST/X001164/1 from STFC. ATS and ZOBR were supported by STFC studentships. 

We thank the anonymous referee for their comments to improve the quality and clarity of this manuscript.
We thank Thomas Baycroft for comments which helped to improve the manuscript.
Some data products used in this publication were taken on observing runs where funding was assisted by travel grants from the Royal Astronomical Society (RAS), allowing postgraduate students the opportunity to get in-person experience of observational astronomy that had previously been hindered by the COVID-19 pandemic.
\\

This work used Astropy, numpy, pandas, scipy, corner, and matplotlib packages.

%%%%%%%%%%%%%%%%%%%%%%%%%%%%%%%%%%%%%%%%%%%%%%%%%%
\section*{Data Availability} \label{sec:data_availability}

The data used in this work is available for download at the ESO public archive\footnote{\url{http://archive.eso.org/eso/eso_archive_main.html}}.
%The inclusion of a Data Availability Statement is a requirement for articles published in MNRAS. Data Availability Statements provide a standardised format for readers to understand the availability of data underlying the research results described in the article. The statement may refer to original data generated in the course of the study or to third-party data analysed in the article. The statement should describe and provide means of access, where possible, by linking to the data or providing the required accession numbers for the relevant databases or DOIs.

Programme IDs for the HARPS data, for stars in Table A1-A3 \\
072.C-0488(E), 077.C-0364(E), 079.C-0927(B,C), 081.C-0148(A), 085.C-0019(A), 087.C-0368(A,B), 087.C-0531(A), 088.C-0662(A,B), 089.C-0497(A,B), 089.C-0732(A), 090.C-0421(A), 091.C-0866(A,B,C), 091.C-0936(A), 092.C-0721(A), 093.C-0409(A), 095.C-0551(A), 095.C-0799(A), 096.C-0460(A), 096.C-0499(A), 096.C-0876(A), 097.C-0390(B), 098.C-0269(A,B), 098.C-0366(A), 098.C-0499(A), 099.C-0458(A), 099.C-0798(A), 0100.C-0097(A) 0100.C-0836(A),  0102.C-0558(A), 0103.C-0432(A), 0106.C-1067(A), 0108.C-0879(A), 0110.C-4334(A), 183.C-0972(A), 192.C-0836(A), 192.0852(A,M), 196.C-1006(A), 2107.C-5066(A).



\bibliographystyle{mnras}
\bibliography{DMPP_Occ_rates} % if your bibtex file is called example.bib

%\begin{thebibliography}{99}
%\end{thebibliography}

%%%%%%%%%%%%%%%%%%%%%%%%%%%%%%%%%%%%%%%%%%%%%%%%%%

%%%%%%%%%%%%%%%%% APPENDICES %%%%%%%%%%%%%%%%%%%%%

\appendix

\section{Target tables}

Here we present the targets of the survey along with relevant properties and parameters.

\begin{landscape}
\begin{table}
    \centering
	\caption{DMPP targets with $\geq10$ observations studied in this analysis. Uncertainties are given in the brackets, as the last two significant figures, except for $\sigma_{\rm jit}$ where uncertainties can be significantly asymmetric. Values taken from runs before signals were subtracted. Stellar masses are taken from Gaia FLAME (Final Luminosity Age Mass Estimator) \citep{Gaia2020} except where previously published.}
	\label{tab:System_parameters}
	\begin{tabular}{llllllll}
		\hline
		\hline
		 & BD+03580 & DMPP-1 & DMPP-2 & DMPP-3 & DMPP-4 & HD28471 (DMPP-5) & HD103991 \\

		\hline
		{\it System properties} & & & & &\\
		TIC & - & 66560666 & 32432767 & 141765836 & 26884478 & 38698692 & 428657196 \\
		\textit{Gaia} DR2 & 3283466836680393472 & 3011005587574818432 & 5017319806652067072 & 5266148569447305600 & 2136027771930903808 & 4675222436007028608 & 3492834741329843072 \\
		$\alpha$ [deg]$^1$& 04:16:40.66 & 05:47:06.26 & 01:49:37.89 & 06:06:29.85 & 19:34:19.79 & 04:25:09.15 & 11:58:27.87 \\
		$\delta$ [deg]$^1$& 03:41:56.0 & -10:37:48.8 & -34:27:32.9 & -72:30:45.6 & +51:14:11.8 & -64:04:48.25 & -23:38:08.8 \\
		$V$ mag$^2$ & 9.83(04) & 7.98(01) & 8.57(01) & 9.09 & 5.703(30)$^3$ & 7.91 & 9.49 \\
		Distance [pc]$^1$ & 134.28(32) & 62.599(83) & 135.93(34) & 47.06(25) & 25.431(18) & 22.870(19) & 40.988(33) \\
		$M$ [$\rm M_\odot$]$^1$& 0.995(45) & 1.21(03) & 1.41(04) & 0.900(09) & 1.25(02)$^3$ & 0.980(44)$^4$ & 0.810(97)$^5$ \\
		\hline
		{\it System parameters} & & & & & & & \\
		$\gamma$ [$\rm{km\,s^{-1}}$] & 145401.4(1.9) & 28660.6(2.6) & 7544.2(2.9) & -667.3(2.6) & 0.2(1.3)$^3$ & 54126.93(55)$^{4}$ & 75172.40(48) \\
		$\sigma_{\rm jit}$ [$\rm m\,s^{-1}$] & 4.3$^{+1.7}_{-1.3}$ & 0.96$^{+0.11}_{-0.13}$,  5.4$^{+1.9}_{-1.6}$ & 4.14$^{+1.23}_{-1.15}$ & 8.24$^{+1.63}_{-3.28}$,  0.59$^{+0.21}_{-0.51}$, 3.68$^{+1.22}_{-0.67}$, 0.54$^{+0.22}_{-0.17}$ & 0.82$^{+0.79}_{-0.73}$ & 2.25$^{+0.65}_{-0.48}$, 0.48$^{+0.09}_{-0.09}$, 1.37$^{+0.54}_{-0.53}$ & 1.36$^{+0.62}_{-0.45}$ \\
        RV Offsets [$\rm m\,s^{-1}$] & - & -5.88$^{+2.42}_{-2.8}$ & - & -140.01$^{+1.83}_{-2.62}$,  12.35$^{+1.97}_{-2.82}$, 17.21$^{+4.93}_{-14.79}$ & - & 16.12$^{+0.22}_{-0.65}$, 21.11$^{+1.32}_{-0.92}$ & - \\
		\hline
	\multicolumn{5}{l}{	{\bf Notes:} 1 - \cite{Gaia2020}, 2 - \cite{Hog2000Tycho2}, 3 - \cite{barnes2023}, 4 - \cite{Stevenson2025-hd28471}, 5 - \cite{Paegert2021}}
	\end{tabular}

\end{table}

\begin{table}
    \centering
	\caption{Table~\ref{tab:System_parameters} continued.}
	\begin{tabular}{llllllll}
		\hline
		\hline
		 & HD118006 (DMPP-7) & HD122640 & HD135204 & HD143424 & HD144840 & HD149079 & HD149189 \\

		\hline
		{\it System properties} & & & & & & & \\
		TIC & 365158631 & 60737418 & 38838329 & 410724316 & 411083824 & 402834325 & 25428659 \\
		\textit{Gaia} DR2 & 3663051957490779264 & 6273522428481412224 & 4418203587193988480 & 6262841463297004672 & 4341712281239980032 & 5815781787804787456 & 5940913632776578432 \\
		$\alpha$ [deg]$^1$& 13:34:06.87 & 14:03:37.18 & 15:13:50.89 & 16:00:42.77 & 16:08:24.49 & 16:36:51.41 & 16:35:25.33 \\
		$\delta$ [deg]$^1$& 00:29:56.6 & -27:00:35.9 & -01:21:05.0 & -15:24:43.7 & -13:08:07.8 & -65:35:12.2 & -48:41:54.0 \\
		$V$ mag$^2$ & 8.80(02) & 9.26(02) & 6.60 & 9.11 & 9.15 & 7.68(01) & 9.448(63) \\
		Distance [pc]$^1$ & 102.28(19) & 63.148(69) & 17.2650(89) & 39.722(26) & 28.264(17) & 48.876(44) & 93.58(15) \\
		$M$ [$\rm M_\odot$]$^1$ & 1.09(04) & 0.90(11)$^3$ & 0.94(12)$^3$ & 0.84(10)$^3$ & 0.762(04) & 1.07(04) & 1.08(04) \\
		\hline
		{\it System parameters} & & & & & & & \\
		$\gamma$ [$\rm{km\,s^{-1}}$] & -36650.21(96) & -44533.3(5.2) & -74159.98(45) & -13208.73(55) & -37916.3(1.4) & 28400.53(87) & -17763.78(74) \\
		$\sigma_{\rm jit}$ [$\rm m\,s^{-1}$] & 3.51$^{+0.89}_{-0.71}$ & 0.25$^{+0.49}_{-0.22}$, 1.95$^{+0.59}_{-0.48}$ & 1.27$^{+0.55}_{-0.39}$ & 1.66$^{0.52}_{-0.48}$ & 1.76$^{1.11}_{-0.95}$ & 2.70$^{0.76}_{-0.59}$ & 1.16$^{+0.28}_{-0.28}$, 3.7$^{+1.9}_{-1.3}$, 0.91$^{+1.18}_{-0.78}$ \\
		RV Offsets [$\rm m\,s^{-1}$] & - & 1.38$^{+4.65}_{-4.63}$ & - & - & - & - & 14.73$^{+2.46}_{-0.89}$, 20.15$^{+2.93}_{-3.26}$ \\

		\hline
	\multicolumn{5}{l}{	{\bf Notes:} 1 - \cite{Gaia2020}, 2 - \cite{Hog2000Tycho2}, 3 - \cite{Paegert2021}}
	\end{tabular}

\end{table}
\end{landscape}

\begin{landscape}

\begin{table}
    \centering
	\caption{Table~\ref{tab:System_parameters} continued.}
	\begin{tabular}{lllllllllll}
		\hline
		\hline
		 & HD181433 & HD191122 (DMPP-8) & HD200133 (DMPP-9) & HD210975 & HD2134 & HD39194 & HD58489\\

		\hline
		{\it System properties} & & & & & & & \\
		TIC & 410399074 & 318042295 & 376777782 & 60111044 & 290575588 & 389506883 & 173079045 \\
		\textit{Gaia} DR2 & 6434153380720177664 & 6349054822860923008 & 6401028668786744960 & 6612306455196349056 & 4635564769579703936 & 4657193606465368704 & 5585594664826865152 \\
		$\alpha$ [deg]$^1$& 19:25:09.57 & 20:21:13.24 & 21:05:13.30 & 22:14:34.80 & 00:24:26.62 & 05:44:31.92 & 07:24:04.61 \\
		$\delta$ [deg]$^1$& -66:28:07.7 & -81:36:57.3 & -66:57:28.3 & -32:18:29.8 & -78:15:01.4 & -70:08:36.9 & -40:02:45.8 \\
		$V$ mag$^2$ & 8.38 & 8.63(01) & 9.20(01) & 9.55 & 8.82(01) & 8.075(12) & 9.74 \\
		Distance [pc]$^1$ & 26.990(15) & 76.036(79) & 110.09(18) & 31.665(18) & 71.443(61) & 26.439(14) & 43.101(19) \\
		$M$ [$\rm M_\odot$]$^1$& 0.86(06)$^{3,4}$ & 0.99(04) & 1.10(04) & 0.760(84)$^5$ & 0.92(04) & 0.86(04) & 0.770(94)$^5$ \\
		\hline
		{\it System parameters} & & & & & & & \\
		$\gamma$ [$\rm{km\,s^{-1}}$] & 39776.76(19) & 12534.29(62) & 11649.12(78) & 23886.53(61) & 4262.53(53) & 7688.02(09) & -14901.9(1.3) \\
		$\sigma_{\rm jit}$ [$\rm m\,s^{-1}$] & 0.93$^{+0.75}_{-0.44}$, 1.61$^{+0.93}_{-0.63}$, 0.84$^{+0.11}_{-0.11}$ & 2.9$^{+1.7}_{-1.3}$, 0.60$^{+0.84}_{-0.56}$ & 1.41$^{+0.32}_{-0.31}$ & 2.7$^{+0.51}_{-0.43}$ &  1.18$^{+0.16}_{-0.14}$ & 1.58$^{+0.59}_{-0.41}$, 3.4$^{+0.99}_{-0.69}$, 1.17$^{+0.09}_{-0.10}$ & 0.34$^{+3.20}_{-0.31}$, 12.6$^{+3.0}_{-2.4}$, 3.7$^{+1.1}_{-1.8}$ \\
		RV Offsets [$\rm m\,s^{-1}$] & 4.21$^{+0.64}_{-0.65}$, 21.28$^{+1.16}_{-1.04}$ & -10.18$^{+2.33}_{-2.01}$ & - & - &  - & 10.14$^{+0.61}_{-0.62}$, 15.82$^{+1.19}_{-1.19}$ & 1.75$^{+3.61}_{-3.59}$, 7.74$^{+3.90}_{-3.62}$ \\
		\hline
	\multicolumn{5}{l}{	{\bf Notes:} 1 - \cite{Gaia2020}, 2 - \cite{Hog2000Tycho2}, 3 - \cite{Trevisan2011}, 4 - \cite{Horner2019}, 5 - \cite{Paegert2021}}
	\end{tabular}

\end{table}

\begin{table}
    \centering
	\caption{Table~\ref{tab:System_parameters} continued.}
    \label{tab:System_parameters_4}
	\begin{tabular}{lllllllllll}
		\hline
		\hline
		  & HD67200 (DMPP-6) & HD85249 & HD89839\\

		\hline
		{\it System properties} & \\
		TIC  & 306668068 & 445976787 & 294028147\\
		\textit{Gaia} DR2  & 5270480576538756352 & 5308568655781883264 & 5355799586503538176\\
		$\alpha$ [deg]$^1$ & 08:01:32.54 & 09:48:52.01 &  10:20:40.54\\
		$\delta$ [deg]$^1$& -53:39:50.7 & -53:35:07.07 & -70:01:26.0\\
		$V$ mag$^2$ & 7.70(01) & 8.53 & 7.64(01)\\
		Distance [pc]$^1$ & 54.438(41) & 92.85(15) & 57.144(60)\\
		$M$ [$\rm M_\odot$]$^3$ & 1.079(45) & 1.207(040) & 1.21(03)\\
		\hline
		{\it System parameters} & & & \\
		$\gamma$ [$\rm{km\,s^{-1}}$] & 15044.3(1.1) & 8784.2(165.2) & 31755.60(58)\\
		$\sigma_{\rm jit}$ [$\rm m\,s^{-1}$] & 0.88$^{+0.09}_{-0.08}$, 1.42$^{+0.38}_{-0.27}$ & 2.79$^{+1.13}_{-1.37}$, 4.79$^{+1.05}_{-0.92}$ & 1.62$^{+0.59}_{-0.51}$, 2.31$^{+0.57}_{-0.49}$, 2.58$^{+0.82}_{-0.99}$ \\
		RV Offsets [$\rm m\,s^{-1}$] & -13.58$^{+1.08}_{-1.09}$ & 439.89$^{+69.7}_{-557.88}$ & 18.15$^{+1.24}_{-1.18}$, 32.24$^{+2.13}_{-2.33}$ \\
		\hline
	\multicolumn{4}{l}{	{\bf Notes:} 1 - \cite{Gaia2020}, 2 - \cite{Hog2000Tycho2}, 3 - \cite{Xiao2023}}
	\end{tabular}

\end{table}
\end{landscape}

\begin{table*}
    \centering
	\caption{Observation metrics for the targets analysed in this work. This table details the number of observations, mean SNR, standard deviation of the SNR, min/max SNR, date of first/last observation in BJD-2,450,000, and the standard deviation in the FWHM and BIS. All columns are combined values across all runs/instruments.}
	\label{tab:obs_summary}

	\begin{tabular}{lrrrrrrrrr}
		\hline
		\hline
		Star & $N_{\rm obs}$ & SNR & $\sigma_{\rm SNR}$ & minSNR & maxSNR & Start BJD & End BJD & $\sigma\rm FWHM$ [$\rm m\,s^{-1}$] & $\sigma\rm BIS$ [$\rm m\,s^{-1}$] \\
		\hline

BD+03580 & 10  & 60.787 & 10.529 & 45.953 & 76.280  & 59995.546 & 60004.533 & 4.890 & 4.907 \\
DMPP-1   & 158 & 154.715 & 27.795 & 63.740 & 220.522 & 57375.590 & 60004.519 & 3.831 & 2.308 \\
DMPP-2   & 48  & 98.103 & 26.363 & 49.161 & 153.552 & 57286.789 & 57762.650 & 74.491 & 38.860 \\
DMPP-3   & 135 & 89.614 & 17.621 & 43.095 & 119.012 & 54579.565 & 60290.595 & 5.970 & 5.855 \\
DMPP-4   & 71  & 215.420 & 35.680 & 137.700 & 302.810 & 57141.602 & 58698.714 & 26.530 & 7.700 \\
HD28471  & 122 & 153.739 & 31.047 & 75.559 & 217.029 & 52945.805 & 60004.641 & 5.734 & 1.515 \\
HD103991 & 17  & 80.149 & 8.068  & 69.218 & 94.284  & 59994.787 & 60004.893 & 8.423 & 3.503 \\
HD118006 & 19  & 81.948 & 9.718  & 61.741 & 98.909  & 59994.868 & 60004.840 & 8.619 & 5.499 \\
HD122640 & 32  & 83.012 & 9.524  & 62.568 & 104.322 & 57893.611 & 60004.759 & 6.657 & 2.687 \\
HD135204 & 15  & 230.594 & 27.648 & 181.027 & 269.509 & 59994.815 & 60004.825 & 5.973 & 1.427 \\
HD143424 & 15  & 81.231 & 7.300  & 67.085 & 90.509  & 59994.825 & 60004.811 & 7.208 & 3.397 \\
HD144840 & 14  & 92.008 & 12.501 & 57.086 & 105.698 & 59994.837 & 60004.875 & 5.770 & 4.687 \\
HD149079 & 14  & 129.245 & 15.061 & 100.647 & 167.660 & 59994.799 & 60004.861 & 7.184 & 2.965 \\
HD149189 & 55  & 98.767 & 20.148 & 52.793 & 147.166 & 54365.583 & 60004.783 & 4.928 & 2.639 \\
HD181433 & 10  & 113.370 & 31.806 & 34.019 & 192.008 & 52942.567 & 60004.847 & 8.330 & 2.246 \\
HD191122 & 25  & 65.524 & 17.570 & 29.677 & 99.046  & 54250.813 & 60004.902 & 7.982 & 4.728 \\
HD200133 & 54  & 71.162 & 18.277 & 28.498 & 100.113 & 57286.741 & 57895.875 & 7.563 & 3.932 \\
HD210975 & 26  & 74.725 & 14.569 & 51.037 & 107.556 & 52944.543 & 56604.553 & 9.061 & 3.360 \\
HD2134   & 87  & 86.225 & 21.352 & 29.958 & 127.848 & 57286.763 & 57895.858 & 6.172 & 3.430 \\
HD39194  & 12  & 133.034 & 28.030 & 50.371 & 200.140 & 52944.821 & 60004.631 & 10.245 & 2.029 \\
HD58489  & 40  & 63.636 & 17.480 & 31.334 & 95.421  & 53016.705 & 60004.703 & 26.896 & 3.963 \\
HD67200  & 114 & 170.410 & 32.068 & 61.829 & 230.789 & 57722.754 & 60004.686 & 4.868 & 1.582 \\
HD85249  & 31  & 108.161 & 22.707 & 66.746 & 149.240 & 57722.846 & 60004.675 & 24.262 & 11.970 \\
HD89839  & 22  & 88.764 & 51.271 & 30.198 & 223.655 & 53056.737 & 60004.732 & 10.979 & 7.650 \\

		\hline
	\end{tabular}
\end{table*}

\section{Detection Limits}\label{app:det-lims}

\sisetup{
  round-mode          = places,
  round-precision     = 2,
  separate-uncertainty = true,
  table-align-text-post = false
}

Here we present the individual detection limits for each of our systems as in Figure~\ref{fig:HD118006_Det_lim}. Green points on the plots are detected planetary signals with errorbars from the uncertainties in Table~\ref{tab:Prev_published} or Table~\ref{tab:Detected_planets}.

\begin{figure}
    \centering
	\includegraphics[width=0.48\textwidth]{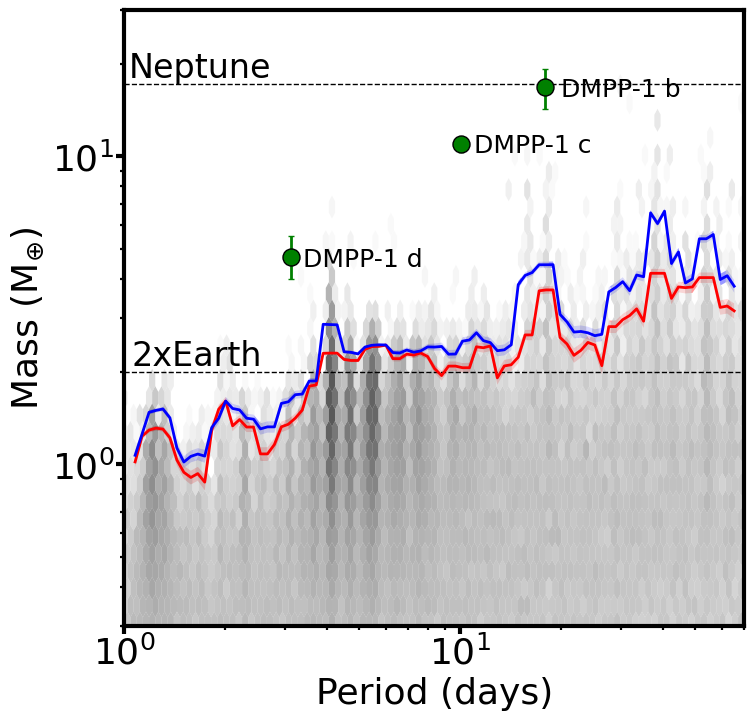}
    \caption{Detection limit plot for target DMPP-1. The standard $3\sigma$ (99\%) detection limit is plotted in blue and the corresponding $3\sigma$ limit calculated using only posterior samples with $e_{\rm{p}}<0.1$ is plotted in red. }
    \label{fig:Det-lim_DMPP-1}
\end{figure}

\begin{figure}
    \centering
	\includegraphics[width=0.48\textwidth]{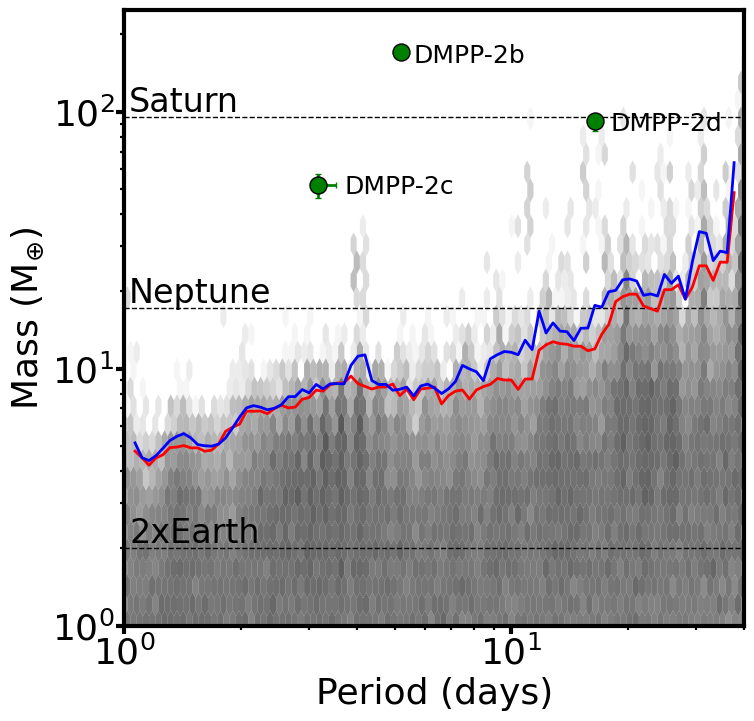}
    \caption{As for Fig~\ref{fig:Det-lim_DMPP-1}, but for DMPP-2.}
    \label{fig:Det-lim_DMPP-2}
\end{figure}

\begin{figure}
    \centering
	\includegraphics[width=0.48\textwidth]{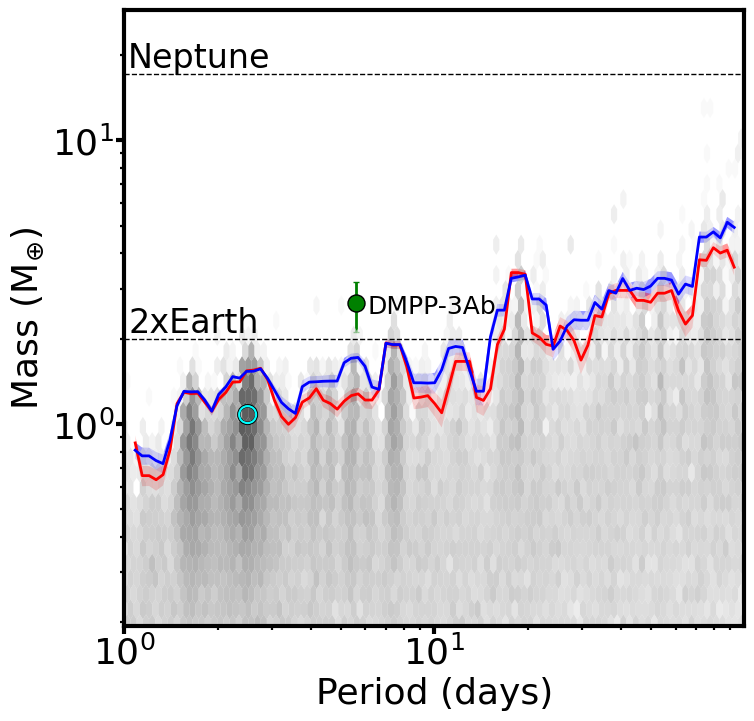}
    \caption{As for Fig~\ref{fig:Det-lim_DMPP-1}, but for DMPP-3. Open cyan circle represents candidate planetary signal.}
    \label{fig:Det-lim_DMPP-3}
\end{figure}

\begin{figure}
    \centering
	\includegraphics[width=0.48\textwidth]{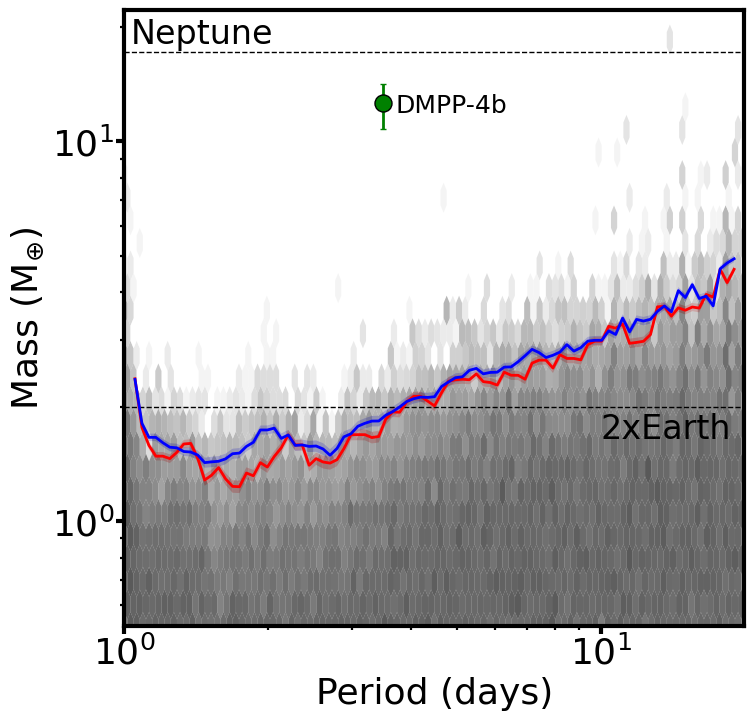}
    \caption{As for Fig~\ref{fig:Det-lim_DMPP-1}, but for DMPP-4.}
    \label{fig:Det-lim_DMPP-4}
\end{figure}

\begin{figure}
    \centering
	\includegraphics[width=0.48\textwidth]{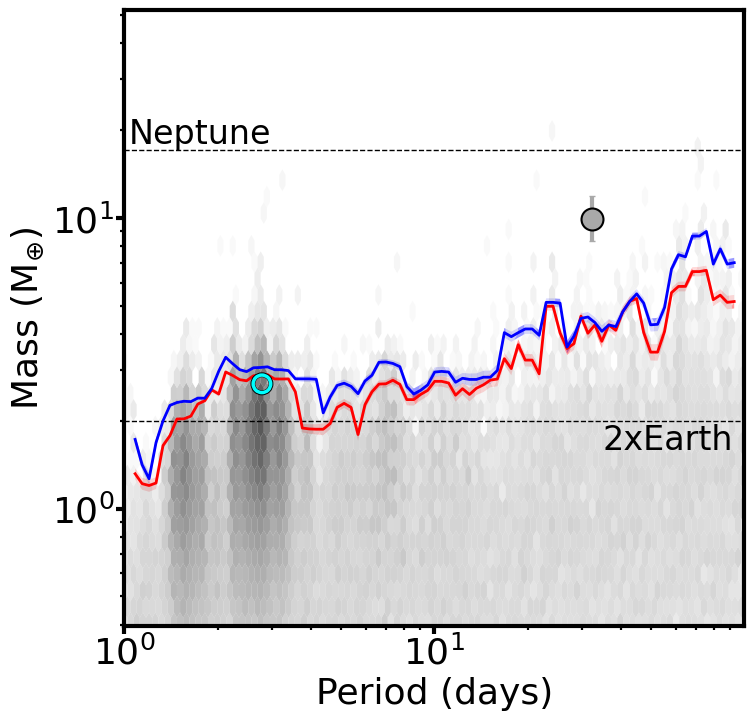}
    \caption{As for Fig~\ref{fig:Det-lim_DMPP-1}, but for HD\,2134. Grey circle represents the subtracted signal we attribute to activity. Open cyan circle represents candidate planetary signal.}
    \label{fig:Det-lim_HD2134}
\end{figure}

\begin{figure}
    \centering
	\includegraphics[width=0.48\textwidth]{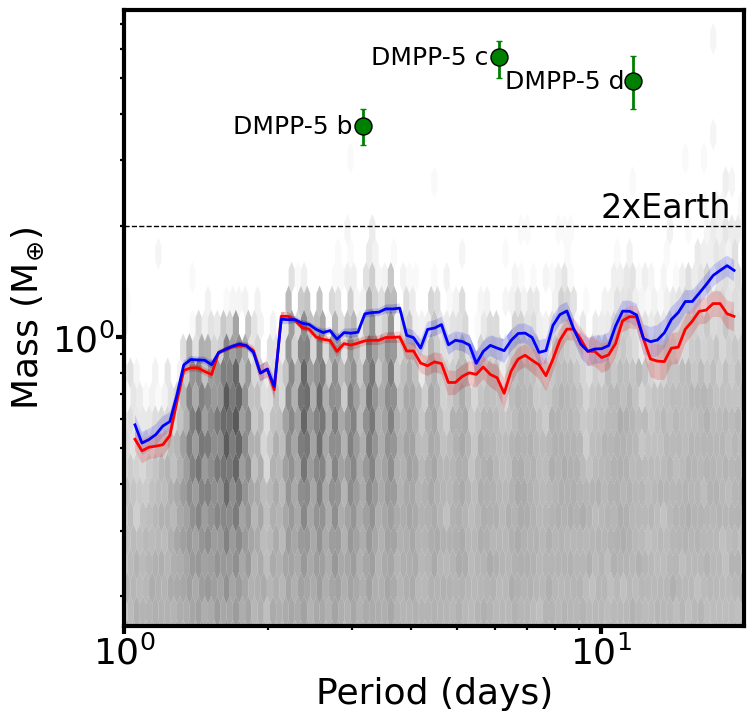}
    \caption{As for Fig~\ref{fig:Det-lim_DMPP-1}, but for HD\,28471 / DMPP-5.}
    \label{fig:Det-lim_HD28471}
\end{figure}

\begin{figure}
    \centering
	\includegraphics[width=0.48\textwidth]{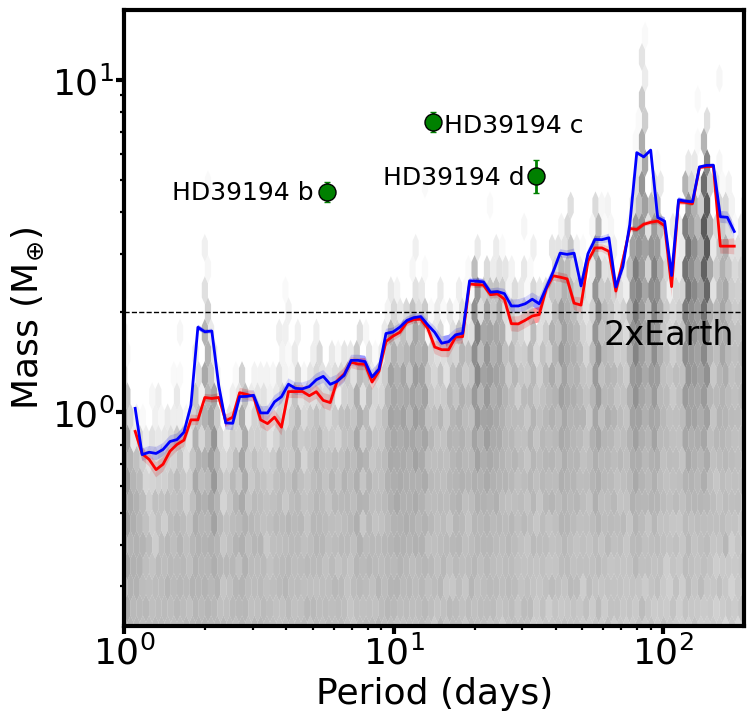}
    \caption{As for Fig~\ref{fig:Det-lim_DMPP-1}, but for HD\,39194.}
    \label{fig:Det-lim_HD39194}
\end{figure}

\begin{figure}
    \centering
	\includegraphics[width=0.48\textwidth]{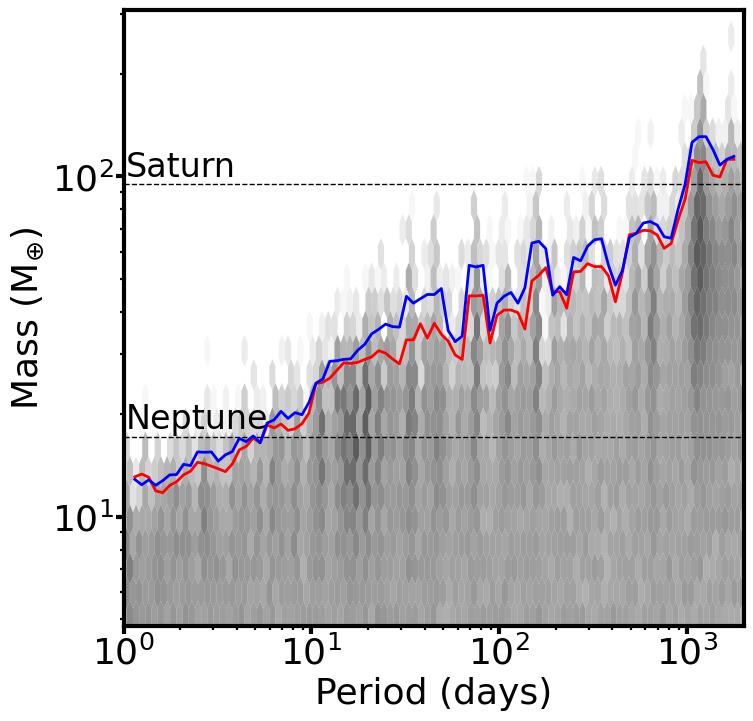}
    \caption{As for Fig~\ref{fig:Det-lim_DMPP-1}, but for HD\,58489.}
    \label{fig:Det-lim_HD58489}
\end{figure}

\begin{figure}
    \centering
	\includegraphics[width=0.48\textwidth]{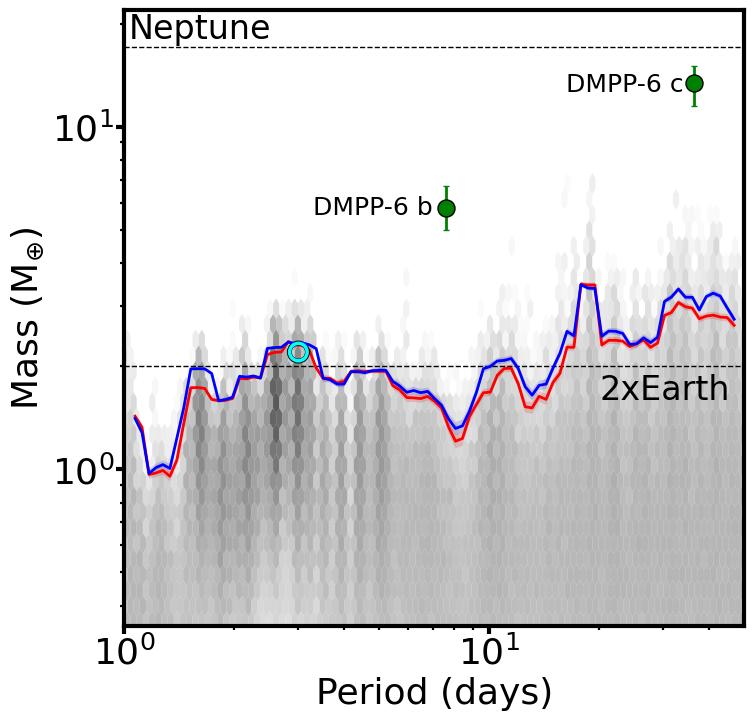}
    \caption{As for Fig~\ref{fig:Det-lim_DMPP-1}, but for HD\,67200 / DMPP-6. Open cyan circle represents candidate planetary signal.}
    \label{fig:Det-lim_HD67200}
\end{figure}

\begin{figure}
    \centering
	\includegraphics[width=0.48\textwidth]{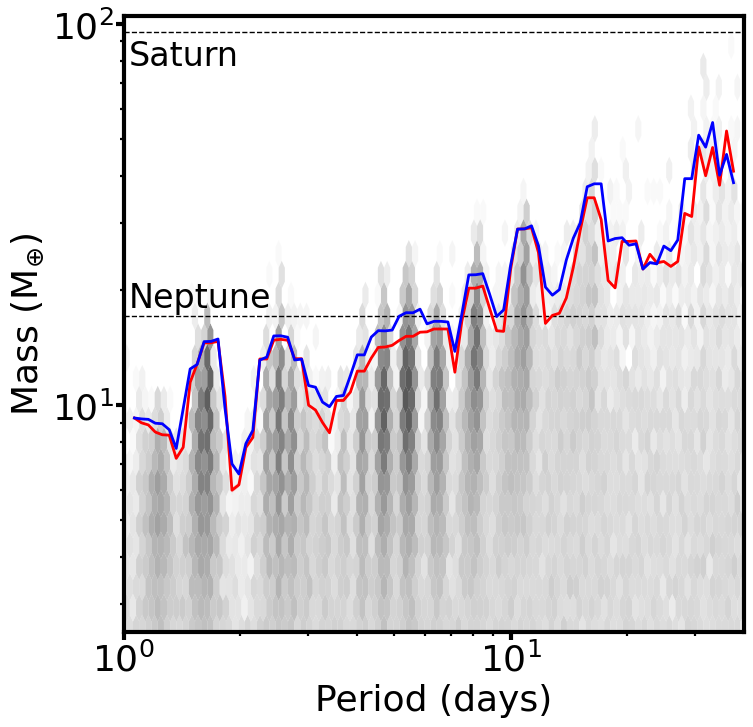}
    \caption{As for Fig~\ref{fig:Det-lim_DMPP-1}, but for HD\,85249.}
    \label{fig:Det-lim_HD85249}
\end{figure}

\begin{figure}
    \centering
	\includegraphics[width=0.48\textwidth]{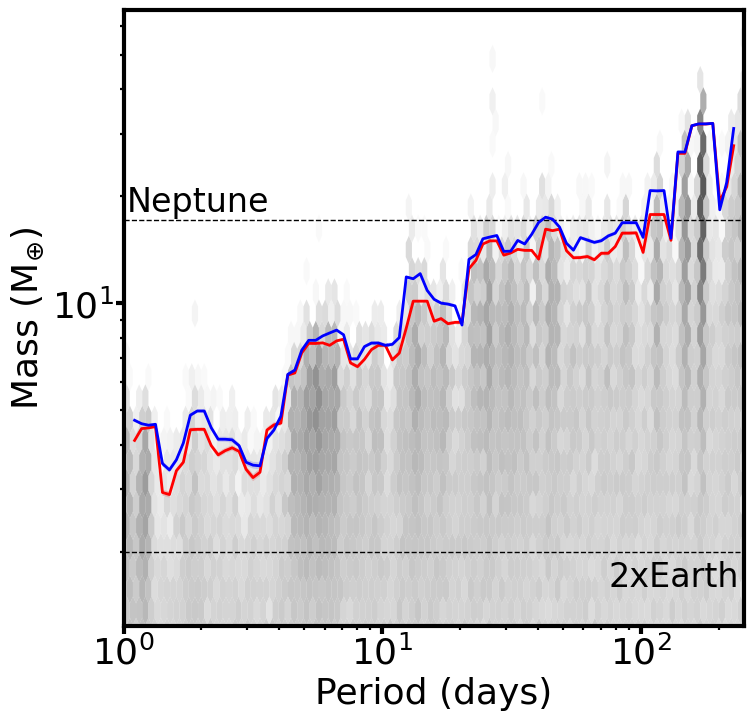}
    \caption{As for Fig~\ref{fig:Det-lim_DMPP-1}, but for HD\,89839.}
    \label{fig:Det-lim_HD89839}
\end{figure}

\begin{figure}
    \centering
	\includegraphics[width=0.48\textwidth]{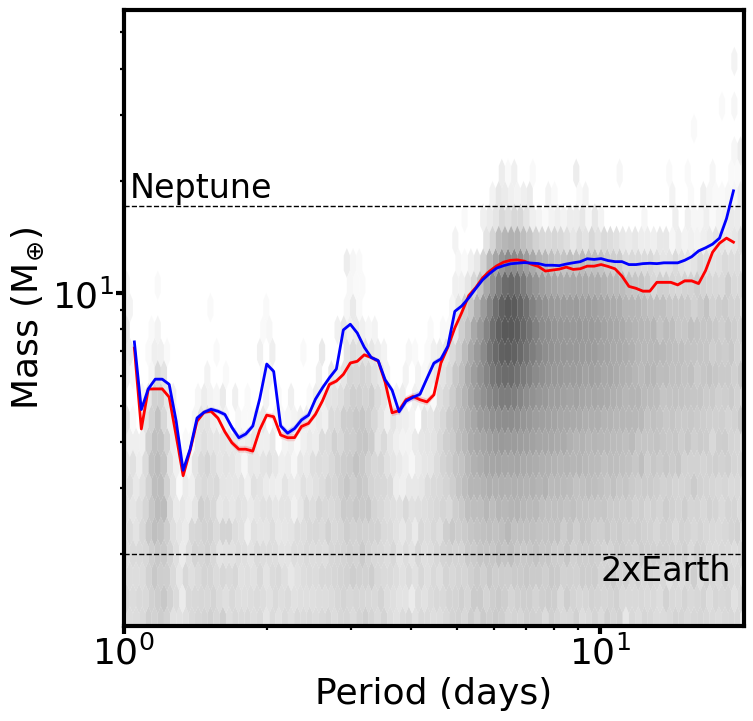}
    \caption{As for Fig~\ref{fig:Det-lim_DMPP-1}, but for HD\,103991.}
    \label{fig:Det-lim_HD103991}
\end{figure}

\begin{figure}
    \centering
	\includegraphics[width=0.48\textwidth]{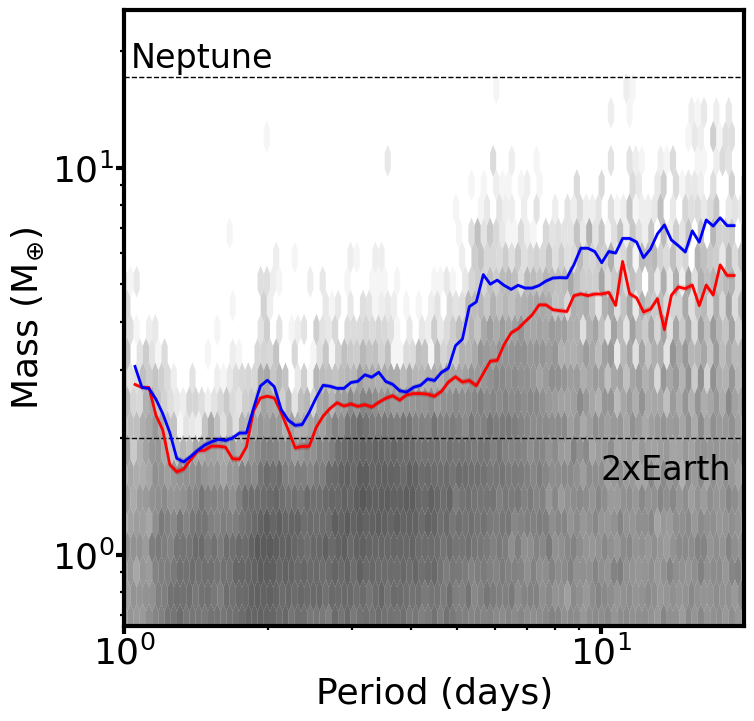}
    \caption{As for Fig~\ref{fig:Det-lim_DMPP-1}, but for HD\,122640. The removed signal from Table~\ref{tab:Subtracted_signals} has a mass of $\sim44$\mearth and is out of the scale of the plot.}
    \label{fig:Det-lim_122640}
\end{figure}

\begin{figure}
    \centering
	\includegraphics[width=0.48\textwidth]{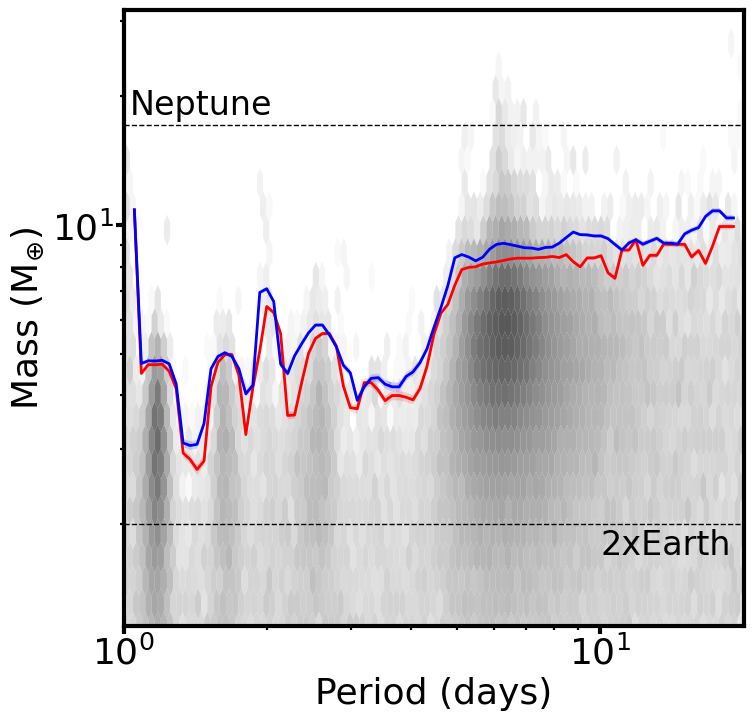}
    \caption{As for Fig~\ref{fig:Det-lim_DMPP-1}, but for HD\,135204.}
    \label{fig:Det-lim_135204}
\end{figure}

\begin{figure}
    \centering
	\includegraphics[width=0.48\textwidth]{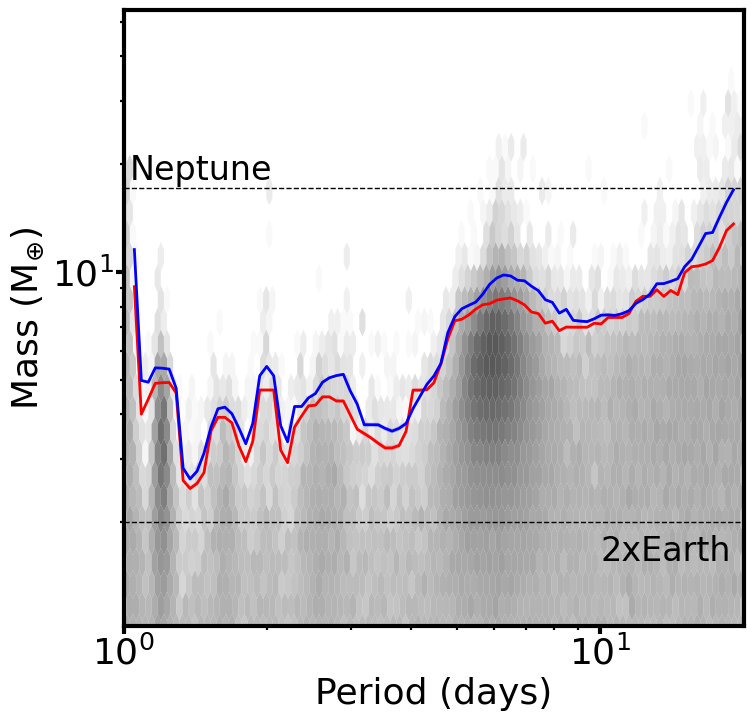}
    \caption{As for Fig~\ref{fig:Det-lim_DMPP-1}, but for HD\,143424.}
    \label{fig:Det-lim_HD143424}
\end{figure}

\begin{figure}
    \centering
	\includegraphics[width=0.48\textwidth]{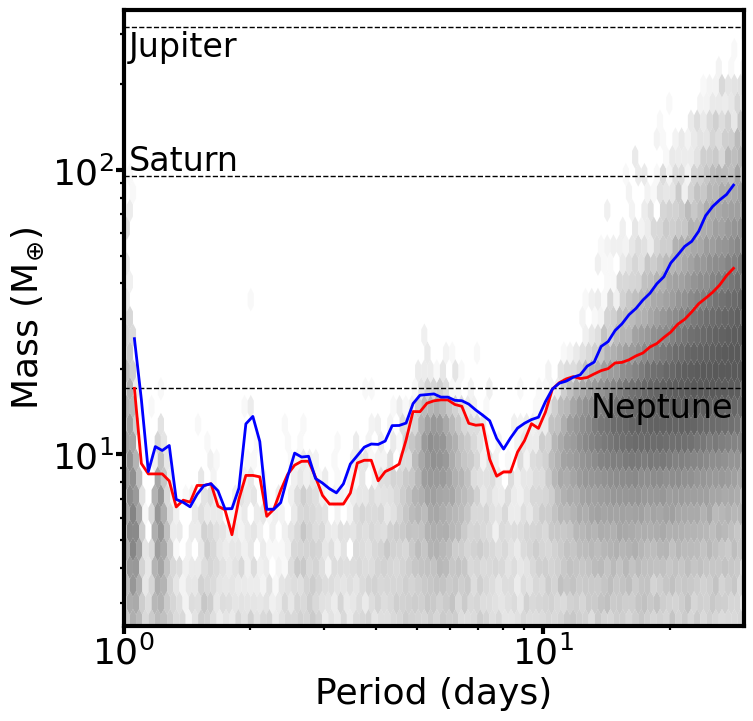}
    \caption{As for Fig~\ref{fig:Det-lim_DMPP-1}, but for HD\,144840.}
    \label{fig:Det-lim_HD144840}
\end{figure}

\begin{figure}
    \centering
	\includegraphics[width=0.48\textwidth]{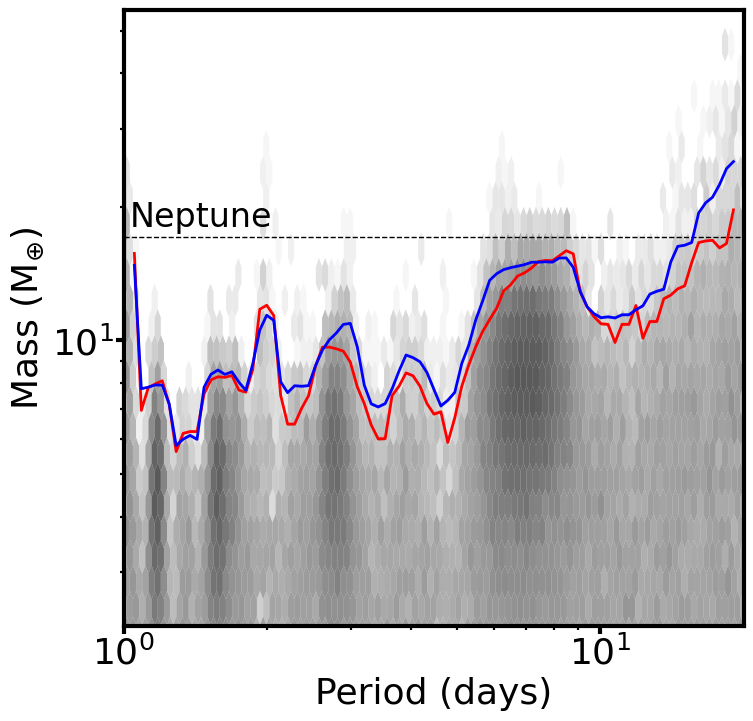}
    \caption{As for Fig~\ref{fig:Det-lim_DMPP-1}, but for HD\,149079.}
    \label{fig:Det-lim_149079}
\end{figure}

\begin{figure}
    \centering
	\includegraphics[width=0.48\textwidth]{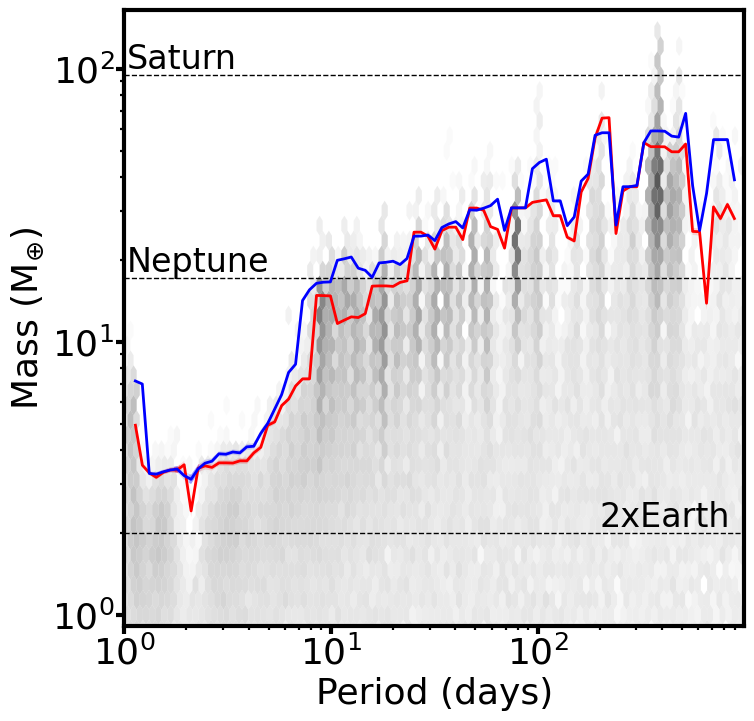}
    \caption{As for Fig~\ref{fig:Det-lim_DMPP-1}, but for HD\,149189.}
    \label{fig:Det-lim_HD149189}
\end{figure}

\begin{figure}
    \centering
	\includegraphics[width=0.48\textwidth]{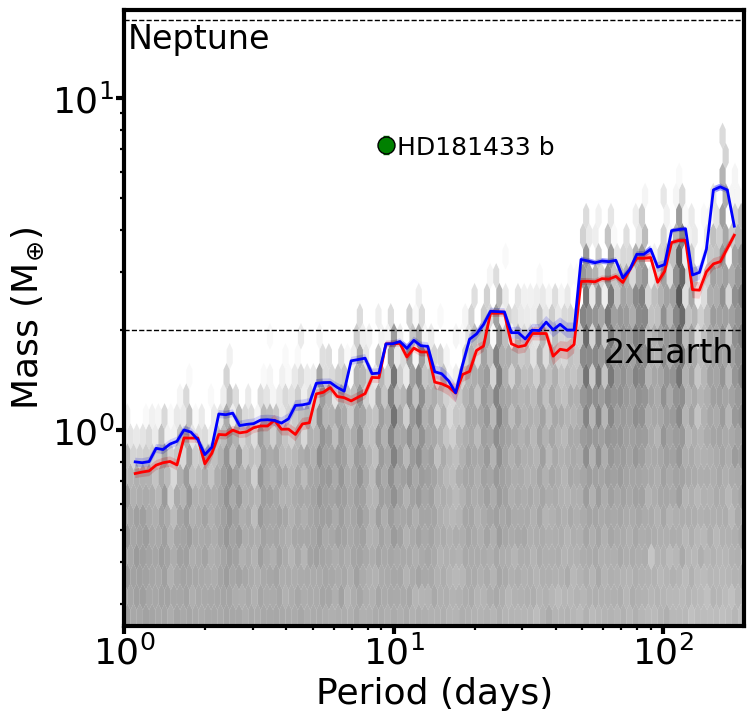}
    \caption{As for Fig~\ref{fig:Det-lim_DMPP-1}, but for HD\,181433.}
    \label{fig:Det-lim_HD181433}
\end{figure}

\begin{figure}
    \centering
	\includegraphics[width=0.48\textwidth]{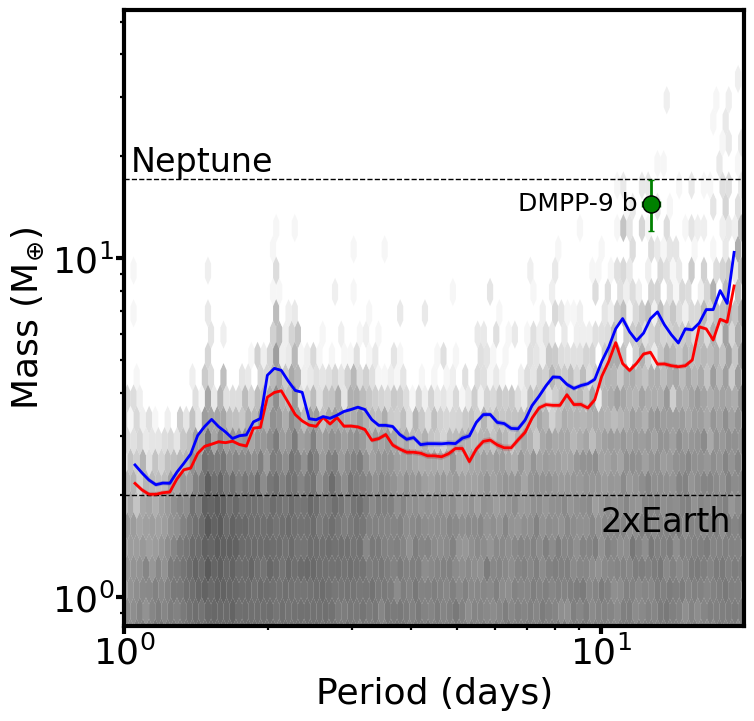}
    \caption{As for Fig~\ref{fig:Det-lim_DMPP-1}, but for HD\,200133 / DMPP-9.}
    \label{fig:Det-lim_HD200133}
\end{figure}

\begin{figure}
    \centering
	\includegraphics[width=0.48\textwidth]{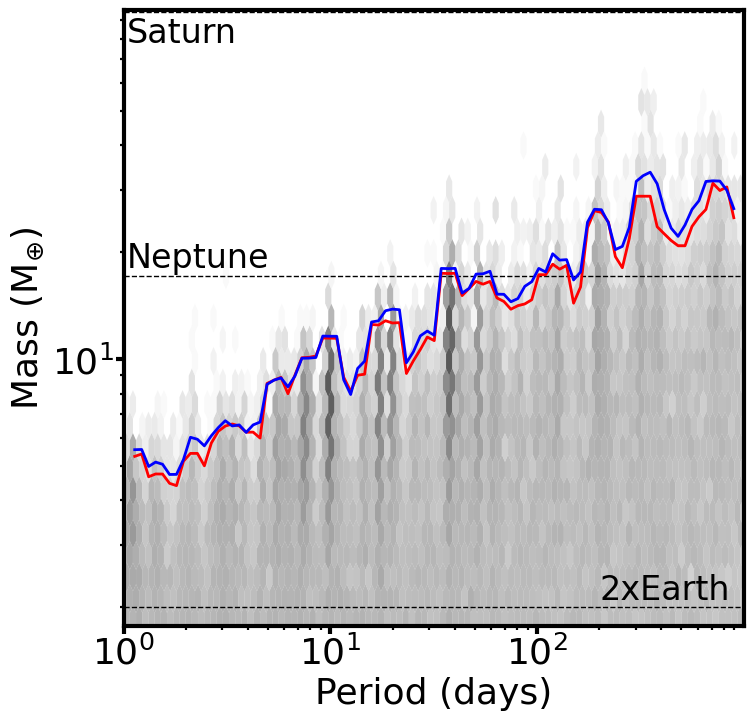}
    \caption{As for Fig~\ref{fig:Det-lim_DMPP-1}, but for HD\,210975.}
    \label{fig:Det-lim_HD210975}
\end{figure}

\section{Subtracted signals}

In this section, we present the signals subtracted from our datasets in order to calculate individual detection limits, and overall occurrence rates.

\begin{table*}
    \centering
	\caption{Parameters from highest likelihood posterior samples, subtracted from the dataset in order to calculate detection limits and subsequently survey completeness and occurrence rates.}
	\label{tab:Subtracted_signals}

	\begin{tabular}{lllllll}
		\hline
		\hline
		 System name & Period [days] & Semi-amplitude [m\,s$^{-1}$] & Eccentricity & $\omega$ & $\Phi$ & T$_0$ [BJD-2400000] \\

		\hline
		DMPP-1\,b & 17.8274043 & 4.10470933 & 0.14730062 & 6.11995822 & 5.29753014 & 57360.55902469\\
		  DMPP-1\,c & 9.7862891 & 3.03337298 & 0.18263751 & 2.9558394 & 3.70120839 & 57369.82504398\\
		  DMPP-1\,d & 3.14234479 & 1.79225768 & 0.12322107 & 4.80563562 & 2.58193113 & 57374.29853571\\
          
		DMPP-2\,b & 5.20333937 & 49.41240186 & 0.02721602 & 2.67882508 & 3.73374387 & 57283.69665724568\\
		  DMPP-2\,c & 3.16795946 & 18.48087325 & 0.24555166 & 1.86990002 & 0.75230228 & 57286.409401262965\\
		  DMPP-2\,d & 16.48508291 & 17.98622006 & 0.11366426 & 5.46727443 & 2.91853125 & 57279.1314102009\\
          
		DMPP-3\,b & 5.58964908 & 1.34811473 & 0.46766518 & 1.74460343 & 0.31757298 & 57434.797185626136\\
		DMPP-3\,B & 506.868256 & 2669.806406 & 0.59748 & 2.775321 & 6.178657 & 56936.64378758205\\
            DMPP-3+ & 809.034509 & 2.498941 & 0.0 & 1.689299 & 3.115262 & 57033.952889478205\\
    
		DMPP-4\,b & 3.49813174 & 5.01401372 & 0.0257469 & 4.64726898 & 6.13782424 & 57138.1843813\\

        HD\,28471 / DMPP-5\,b & 2.87573122 & 1.28665949 & 0.16181718 & 2.80163733 & 5.92074613 & 52943.095250861246\\
		HD\,28471 / DMPP-5\,c & 6.20937197 & 1.86215241 & 0.10760704 & 0.52039189 & 4.44622571 & 52945.0822669725\\
        HD\,28471 / DMPP-5\,c & 8.8655346 & 1.7087995 & 0.03719353 & 1.75942746 & 0.37183296 & 52945.28044449285\\

		HD\,181433\,b & 9.37467082 & 2.73659688 & 0.4107677 & 3.51456492 & 2.45042423 & 52938.91044288627\\
		HD\,181433\,c & 1018.37583726 & 17.27654281 & 0.24951593 & 0.16561441 & 0.73142406 & 52178.00334389883\\
		HD\,181433\,d & 6973.28325926 & 8.53881203 & 0.45943362 & 4.39295463 & 5.30304593 & 47057.07381393424\\

            HD\,39194\,b & 5.63687105 & 1.73455193 & 0.00822653 & 3.81388864 & 5.72521668 & 52939.6850636518\\
		HD\,39194\,c & 14.03121165 & 2.22851883 & 0.02362284 & 1.29771115 & 0.08663022 & 52944.627903897315\\
		HD\,39194\,d & 33.87735548 & 1.20269546 & 0.02740975 & 6.14602254 & 6.13660265 & 52911.734342374424\\

            HD\,118006 / DMPP-7\,b & 5.0162790 & 22.05399487 & 0.0573552 & 2.99923775 & 4.4909205 & 59991.28245223\\
            
		HD\,122640 & 138.49425273 & 5.89333551 & 0.2976916 & 4.16233507 & 1.70658196 & 57855.99452034\\
		
            HD\,191122 / DMPP-8\,b & 62.90705511 & 13.49436464 & 0.28556875 & 3.31680284 & 4.54104098 & 57082.39283831\\
        
		HD\,200133 / DMPP-9\,b & 12.73459852 & 4.24859186 & 0.1998315 & 2.526275345 & 1.53505777 & 57283.62956265\\
	
            HD\,2134 & 32.41548814 & 2.45926102 & 0.19273226 & 0.65488088 & 0.0808832 & 57286.34669756\\
	
            HD\,67200 / DMPP-6\,b & 7.60135752 & 1.89331578 & 0.38473448 & 3.29490636 & 0.16647627 & 57722.55283184016\\
		HD\,67200 / DMPP-6\,c & 36.40340479 & 2.69770204 & 0.21762472 & 3.07693338 & 4.44622571 & 57696.993772094065\\

\end{tabular}
\end{table*}

\section{Phase plots of previously published signals}\label{app:phase_plots}
Here we present the phase plots of the posterior samples with the highest likelihoods for each of the signals presented in Section~\ref{sec:prev_published}.

\begin{figure}
    \centering
    \begin{subfigure}[b]{0.48\textwidth}
        \centering
        \includegraphics[width=\textwidth]{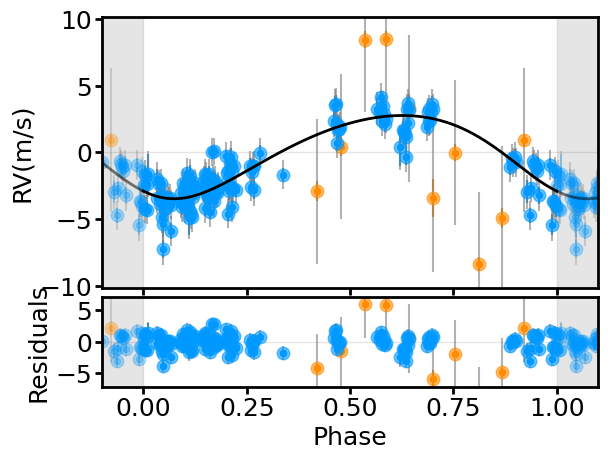}
    \end{subfigure}
    \hfill
    \begin{subfigure}[b]{0.48\textwidth}
        \centering
        \includegraphics[width=\textwidth]{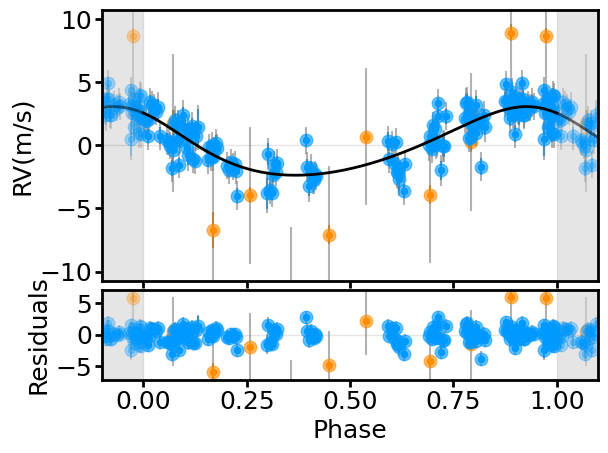}
    \end{subfigure}
    \begin{subfigure}[b]{0.48\textwidth}
        \centering
        \includegraphics[width=\textwidth]{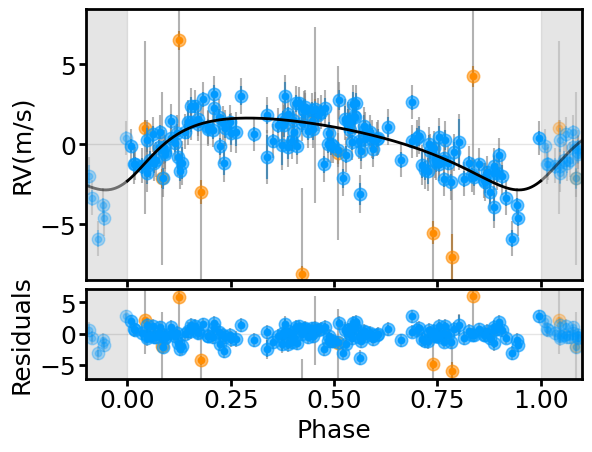}
    \end{subfigure}    
    \caption{Phased Keplerian RV models of DMPP-1\,b (top), DMPP-1\,c (middle), and DMPP-1\,d (bottom) with HARPS data post-fibre upgrade (blue) and post-COVID warmp-up (orange) along with associated residuals after removing the planetary signals. The additional RV jitter term has been added to the plotted uncertainties and is shown by the grey error bars. No random samples are shown for this signal as the eccentricity is unconstrained and consistent with 0. The shaded regions display the repeating signal.}
    \label{fig:DMPP-1_combined_phase}
\end{figure}

\begin{figure}
    \centering
    
    \begin{subfigure}[b]{0.48\textwidth}
        \centering
        \includegraphics[width=\textwidth]{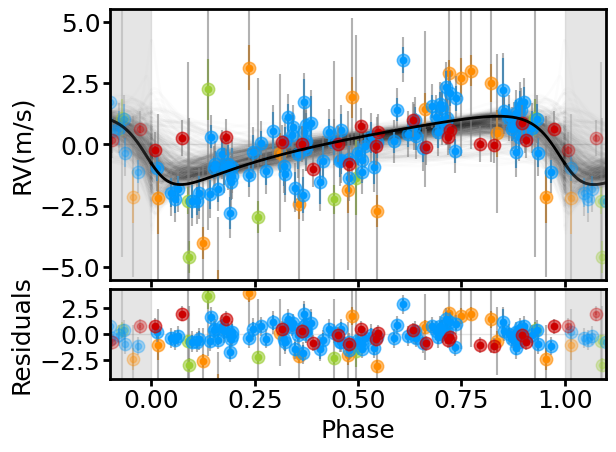}
    \end{subfigure}
    \hfill
    \begin{subfigure}[b]{0.48\textwidth}
        \centering
        \includegraphics[width=\textwidth]{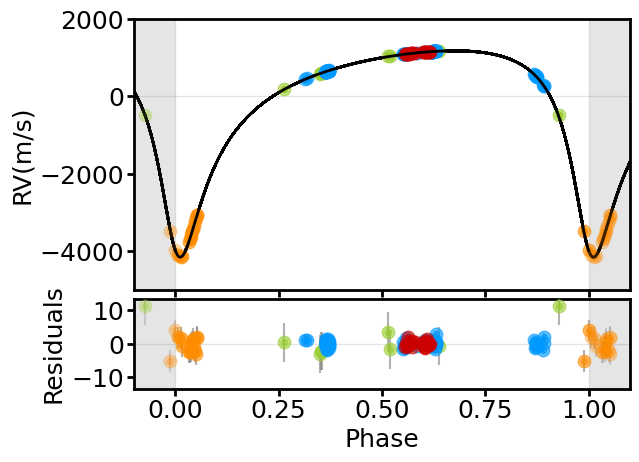}
    \end{subfigure} 
    \caption{Phased Keplerian RV models of DMPP-3A\,b (top), DMPP-3B (bottom) with HARPS data pre-fibre upgrade (green), HARPS data post-fibre upgrade (blue), HARPS data post-COVID warm-up (orange) and ESPRESSO data (red) along with associated residuals after removing the planetary signal. The additional RV jitter term has been added to the plotted uncertainties and is shown by the grey error bars. No random samples are shown for this signal as the eccentricity is unconstrained and consistent with 0. The shaded regions display the repeating signal.}
    \label{fig:DMPP-3_combined_phase}
\end{figure}

\begin{figure}
    \centering
    \begin{subfigure}[b]{0.48\textwidth}
        \centering
        \includegraphics[width=\textwidth]{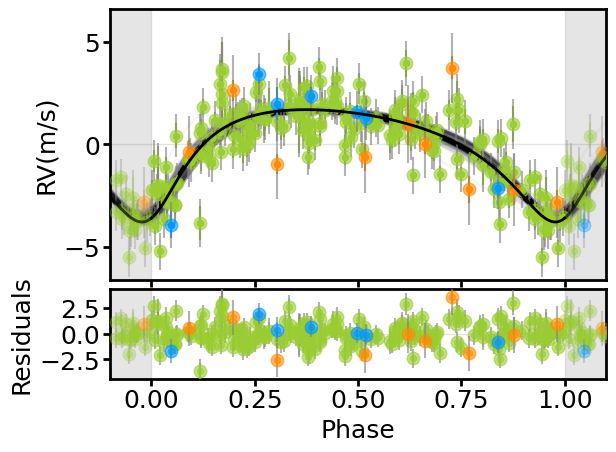}
    \end{subfigure}
    \hfill
    \begin{subfigure}[b]{0.48\textwidth}
        \centering
        \includegraphics[width=\textwidth]{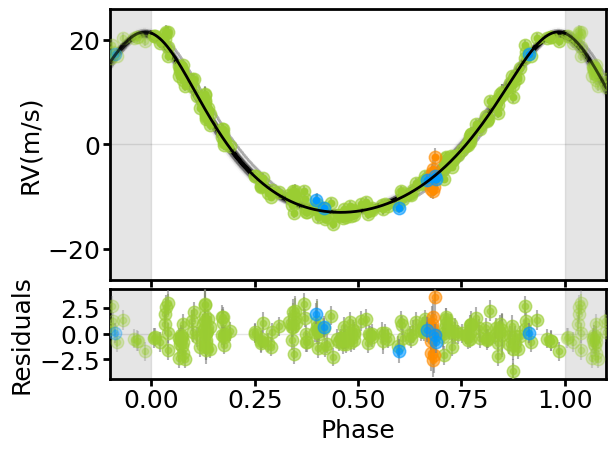}
    \end{subfigure}
    \begin{subfigure}[b]{0.48\textwidth}
        \centering
        \includegraphics[width=\textwidth]{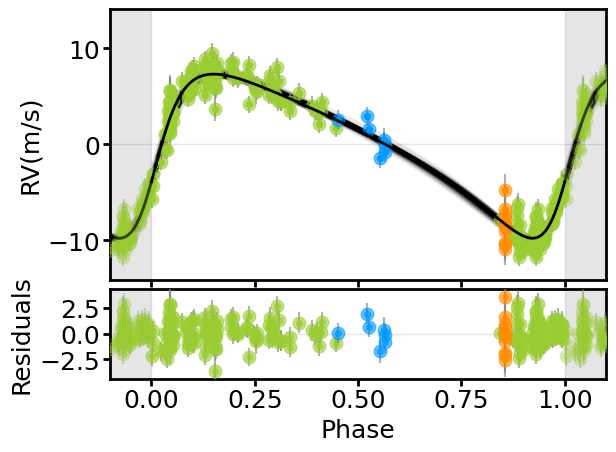}
    \end{subfigure}    
    \caption{Phased Keplerian RV models of HD\,181433\,b (top), HD\,181433\,c (middle), and HD\,181433\,d (bottom) with HARPS data pre-fibre upgrade (green), HARPS data post-fibre upgrade (blue), HARPS data post-COVID warm-up (orange) and their respective residuals after removing the planetary signals. The additional RV jitter term has been added to the plotted uncertainties and is shown by the grey error bars. Faded Keplerian models are based on 500 randomly drawn posterior samples from a \kima run. The shaded regions display the repeating signal.}
    \label{fig:HD181433_combined_phase}
\end{figure}

\begin{figure}
    \centering
    \begin{subfigure}[b]{0.48\textwidth}
        \centering
        \includegraphics[width=\textwidth]{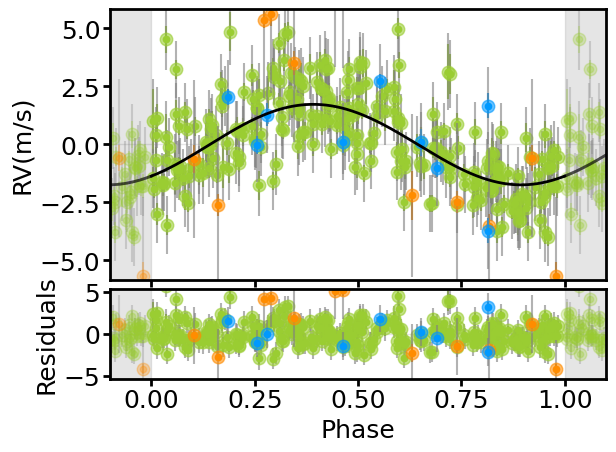}
    \end{subfigure}
    \hfill
    \begin{subfigure}[b]{0.48\textwidth}
        \centering
        \includegraphics[width=\textwidth]{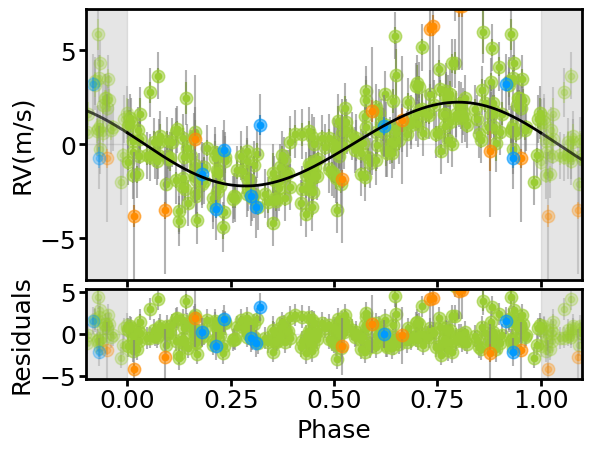}
    \end{subfigure}
    \begin{subfigure}[b]{0.48\textwidth}
        \centering
        \includegraphics[width=\textwidth]{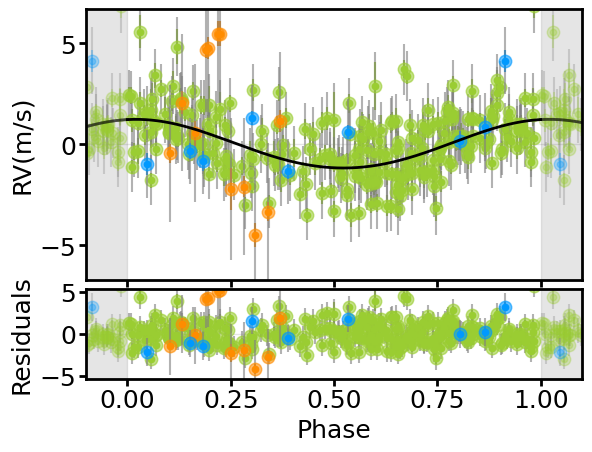}
    \end{subfigure}    
 
    \caption{Phased Keplerian RV models of HD\,39194\,b (top), HD\,39194\,c (middle), and HD\,39194\,d (bottom) with HARPS data pre-fibre upgrade (green), HARPS data post-fibre upgrade (blue), and HARPS data post-COVID warm-up (orange) and their respective residuals after removing the planetary signals. The additional RV jitter term has been added to the plotted uncertainties and is shown by the grey error bars. No random samples are shown for this signal as the eccentricity is unconstrained and consistent with 0.. The shaded regions display the repeating signal.}
    \label{fig:HD39194_combined_phase}
\end{figure}

\begin{figure}
    \centering
	\includegraphics[width=0.48\textwidth]{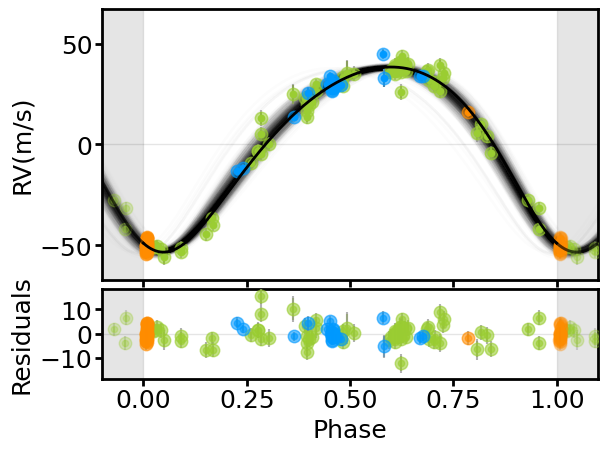}
    \caption{Phased Keplerian Radial-Velocity (RV) models of HD\,89839\,b with  HARPS data pre-fibre upgrade (green), HARPS data post-fibre upgrade (blue), and HARPS data post-COVID warm-up (orange) and their respective residuals after removing the planetary signal. The additional RV jitter term has been added to the plotted uncertainties and is shown by the grey error bars. Faded Keplerian models are based on 500 randomly drawn posterior samples from a \kima run. The shaded regions display the repeating signal.}
    \label{fig:HD89839b_phase}
\end{figure}

\section{Signal corner plots}\label{app:corners}
Here we present the corner plots of posterior sample distributions for the orbital parameters of each of the signals presented in Tables~\ref{tab:Prev_published} and \ref{tab:Detected_planets}.

\begin{figure*}
    \centering
	\includegraphics[width=\textwidth]{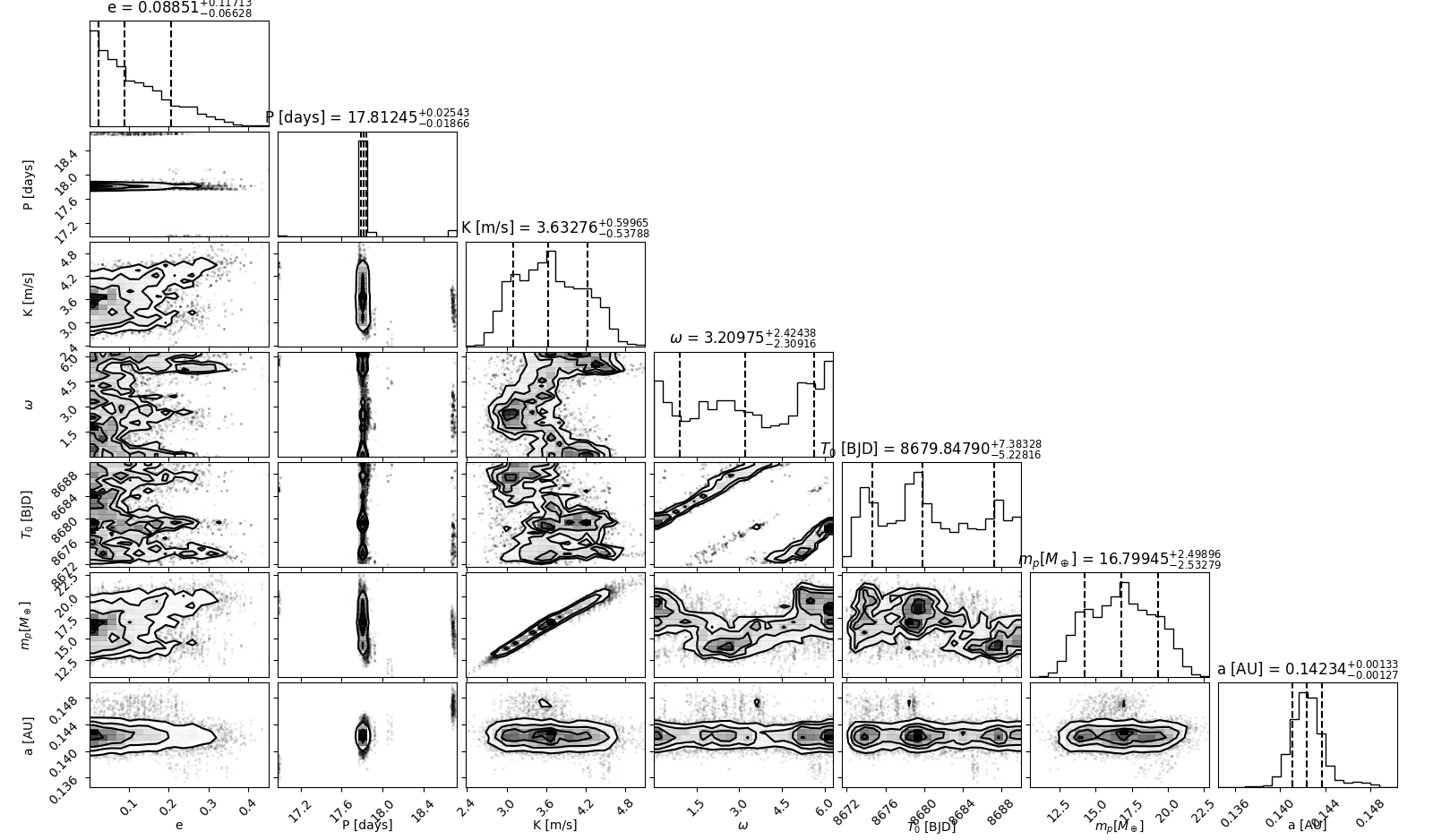}
    \caption{Corner plot for DMPP-1\,b.}
    \label{fig:corner_DMPP-1b}
\end{figure*}

\begin{figure*}
    \centering
	\includegraphics[width=\textwidth]{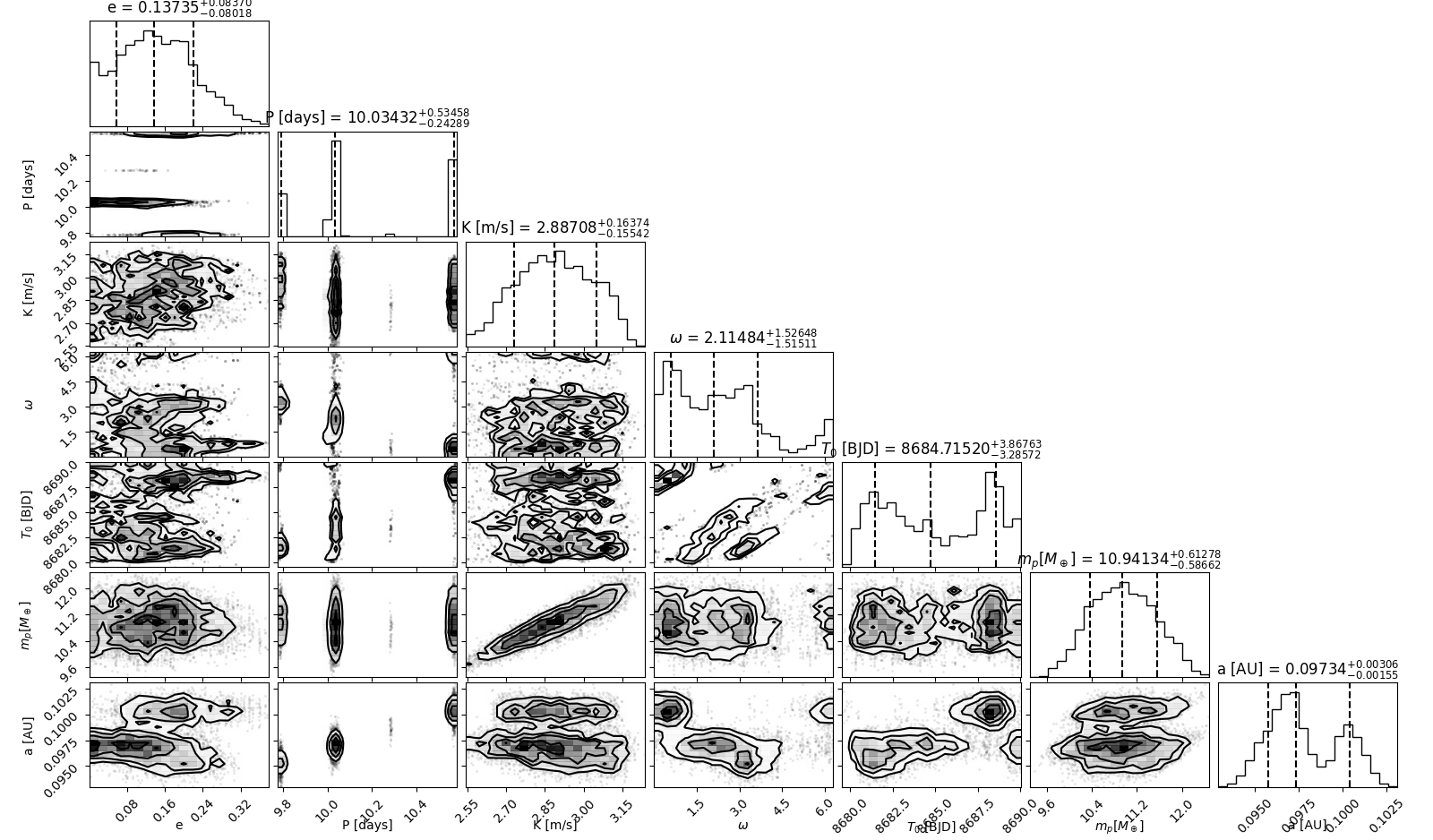}
    \caption{Corner plot for DMPP-1\,c.}
    \label{fig:corner_DMPP-1c}
\end{figure*}

\begin{figure*}
    \centering
	\includegraphics[width=\textwidth]{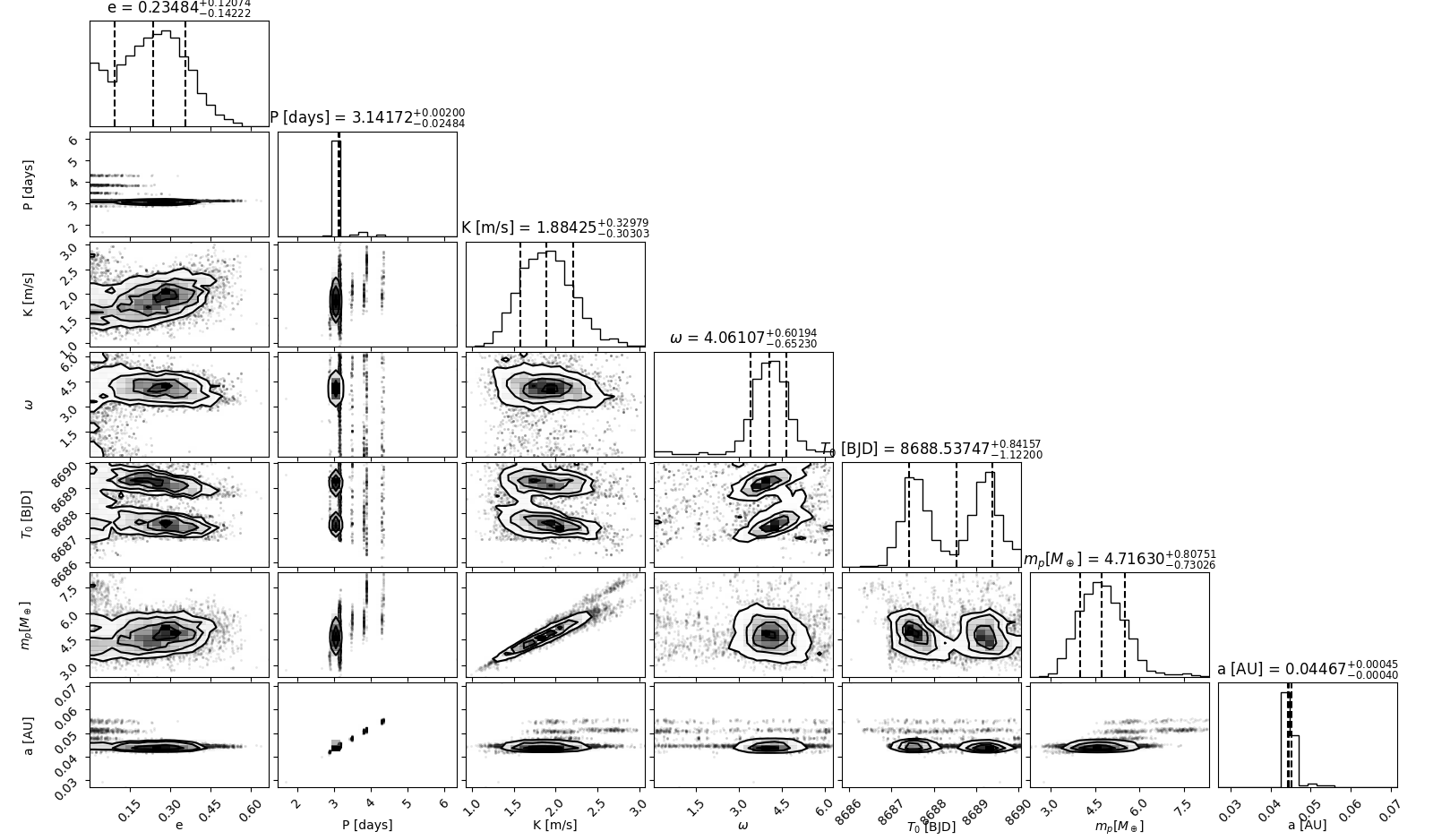}
    \caption{Corner plot for DMPP-1\,d.}
    \label{fig:corner_DMPP-1d}
\end{figure*}

\begin{figure*}
    \centering
	\includegraphics[width=\textwidth]{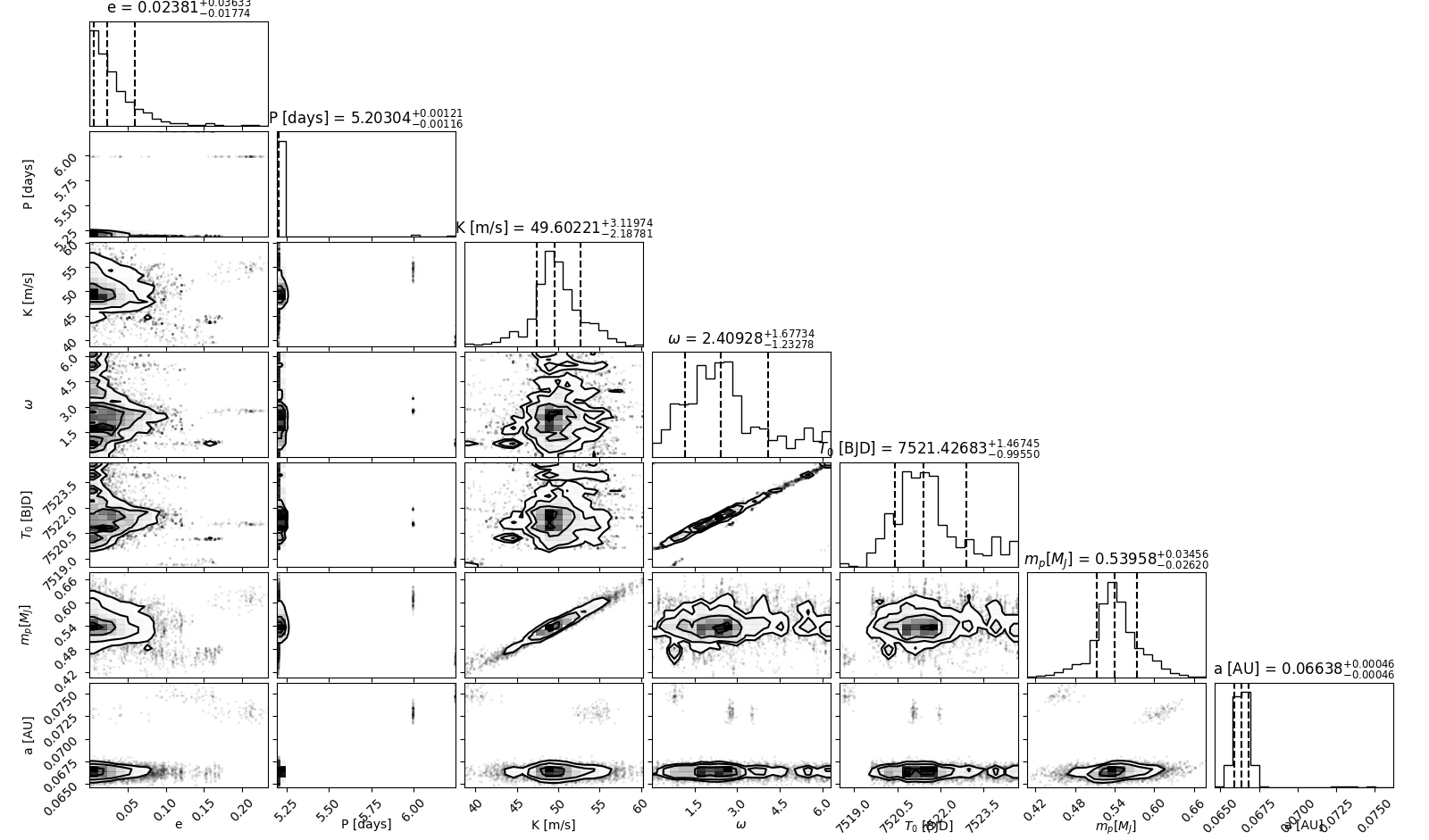}
    \caption{Corner plot for DMPP-2\,b.}
    \label{fig:corner_DMPP-2b}
\end{figure*}

\begin{figure*}
    \centering
	\includegraphics[width=\textwidth]{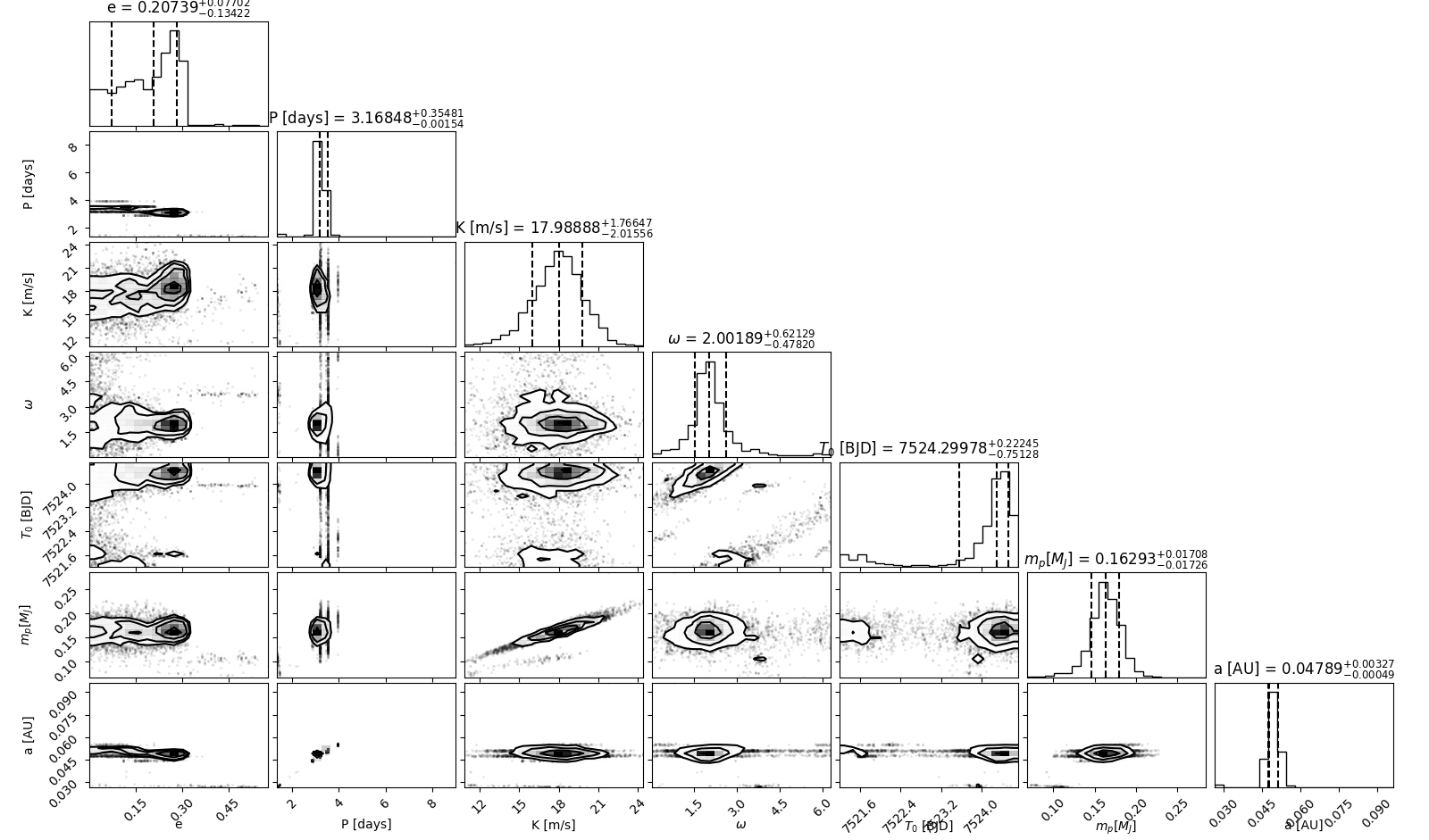}
    \caption{Corner plot for DMPP-2\,c.}
    \label{fig:corner_DMPP-2c}
\end{figure*}

\begin{figure*}
    \centering
	\includegraphics[width=\textwidth]{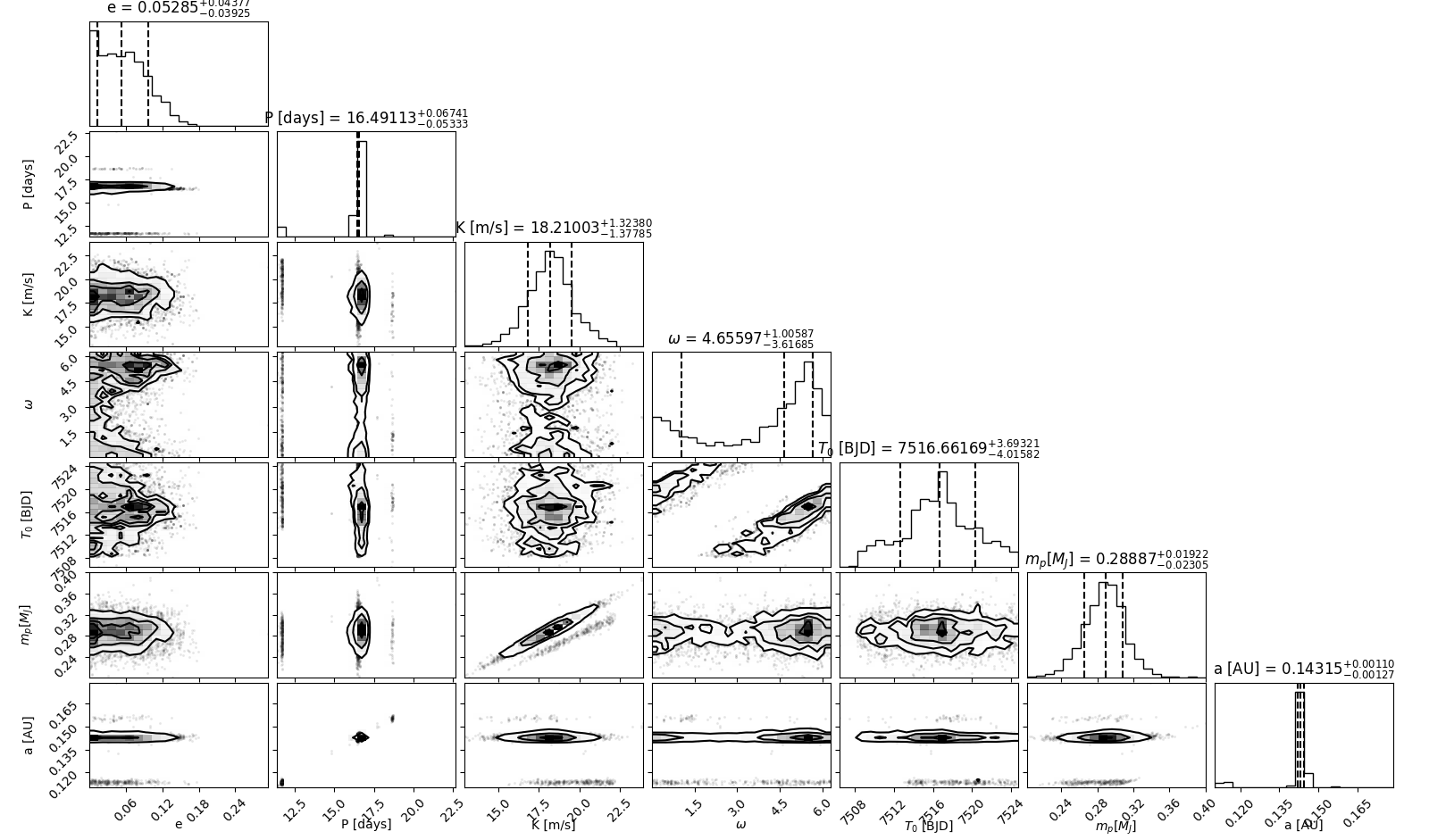}
    \caption{Corner plot for DMPP-2\,d.}
    \label{fig:corner_DMPP-2d}
\end{figure*}

\begin{figure*}
    \centering
	\includegraphics[width=\textwidth]{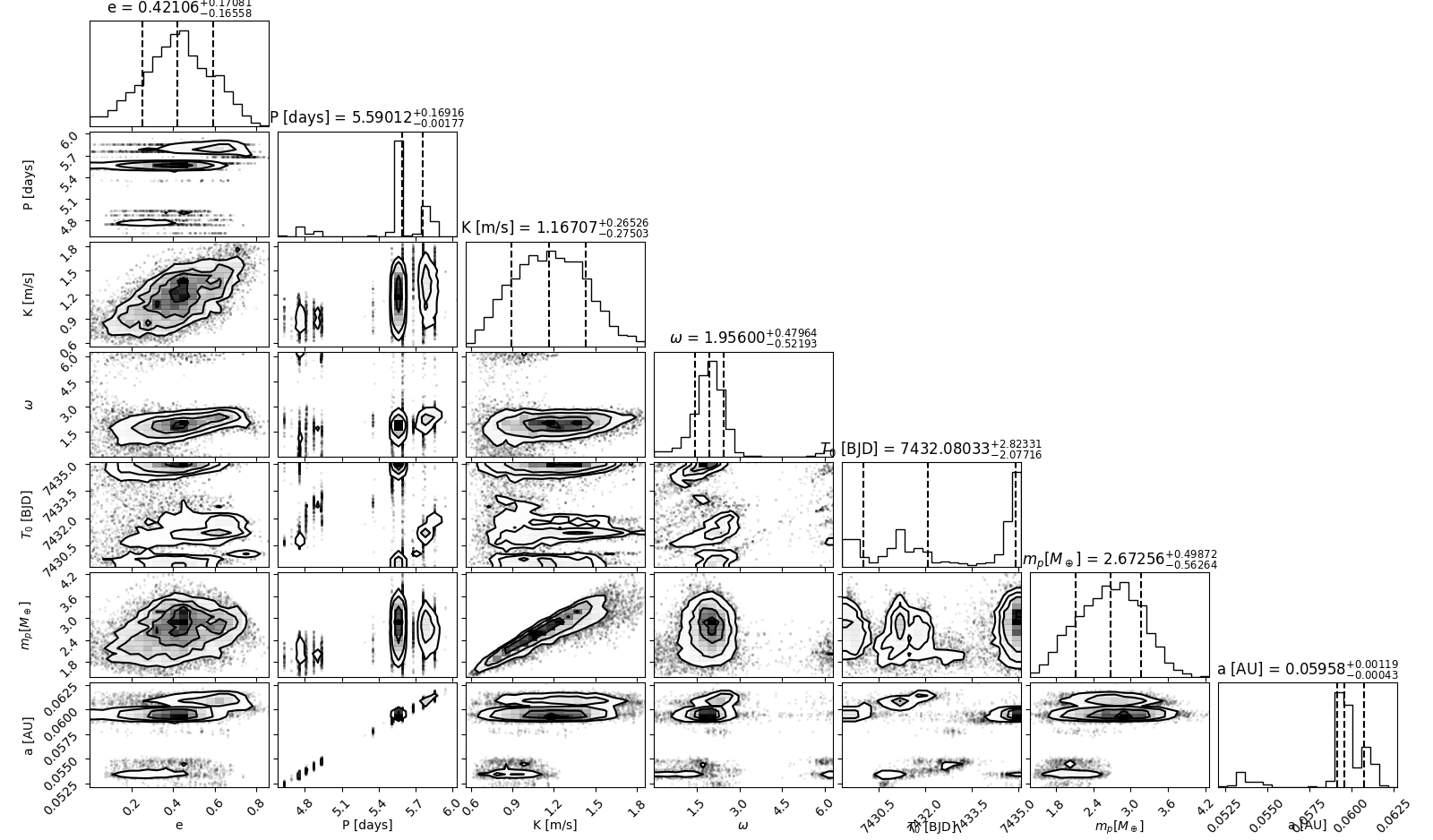}
    \caption{Corner plot for DMPP-3\,b.}
    \label{fig:corner_DMPP-3b}
\end{figure*}

\begin{figure*}
    \centering
	\includegraphics[width=\textwidth]{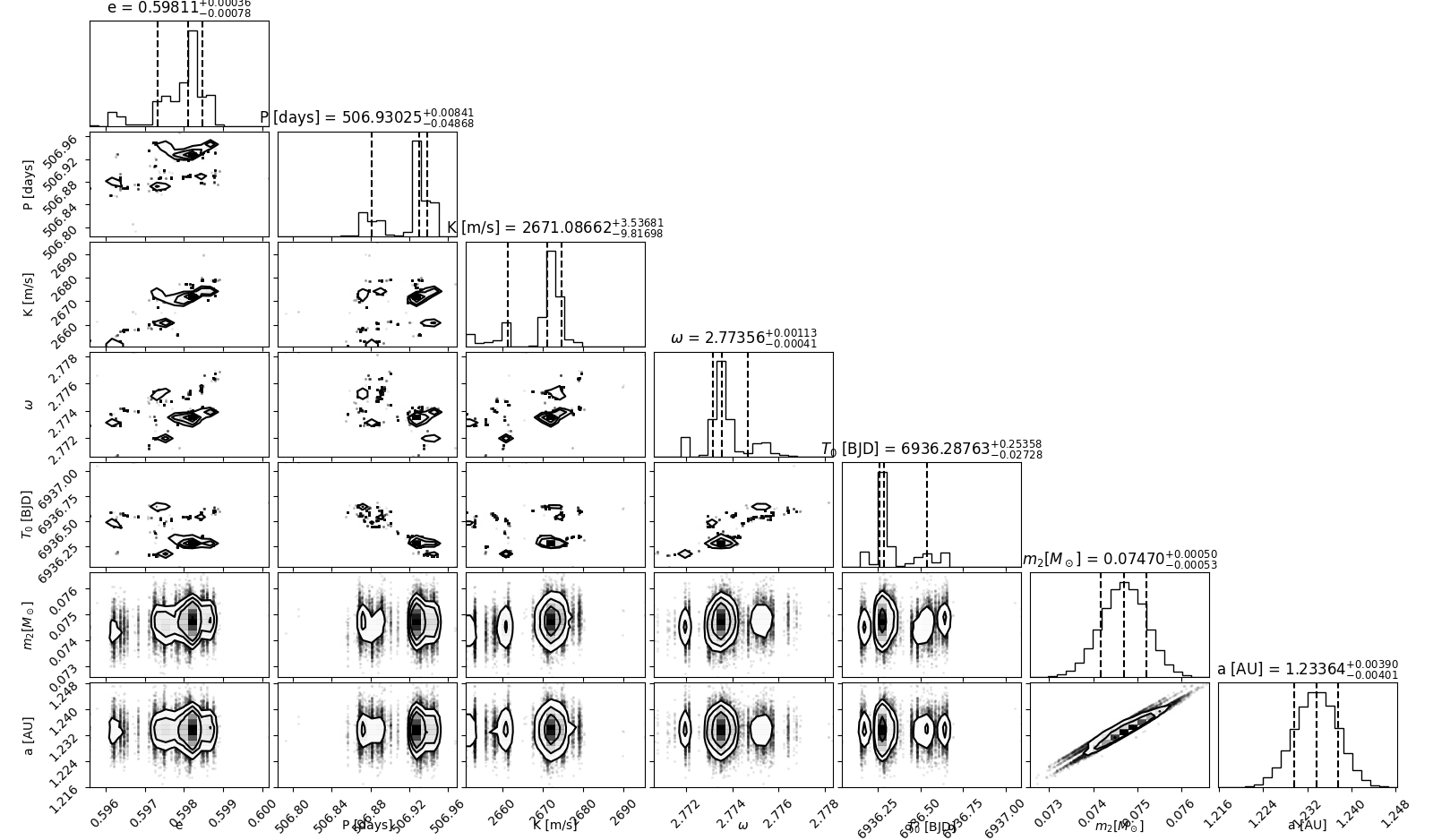}
    \caption{Corner plot for DMPP-3\,AB.}
    \label{fig:corner_DMPP-3AB}
\end{figure*}

\begin{figure*}
    \centering
	\includegraphics[width=\textwidth]{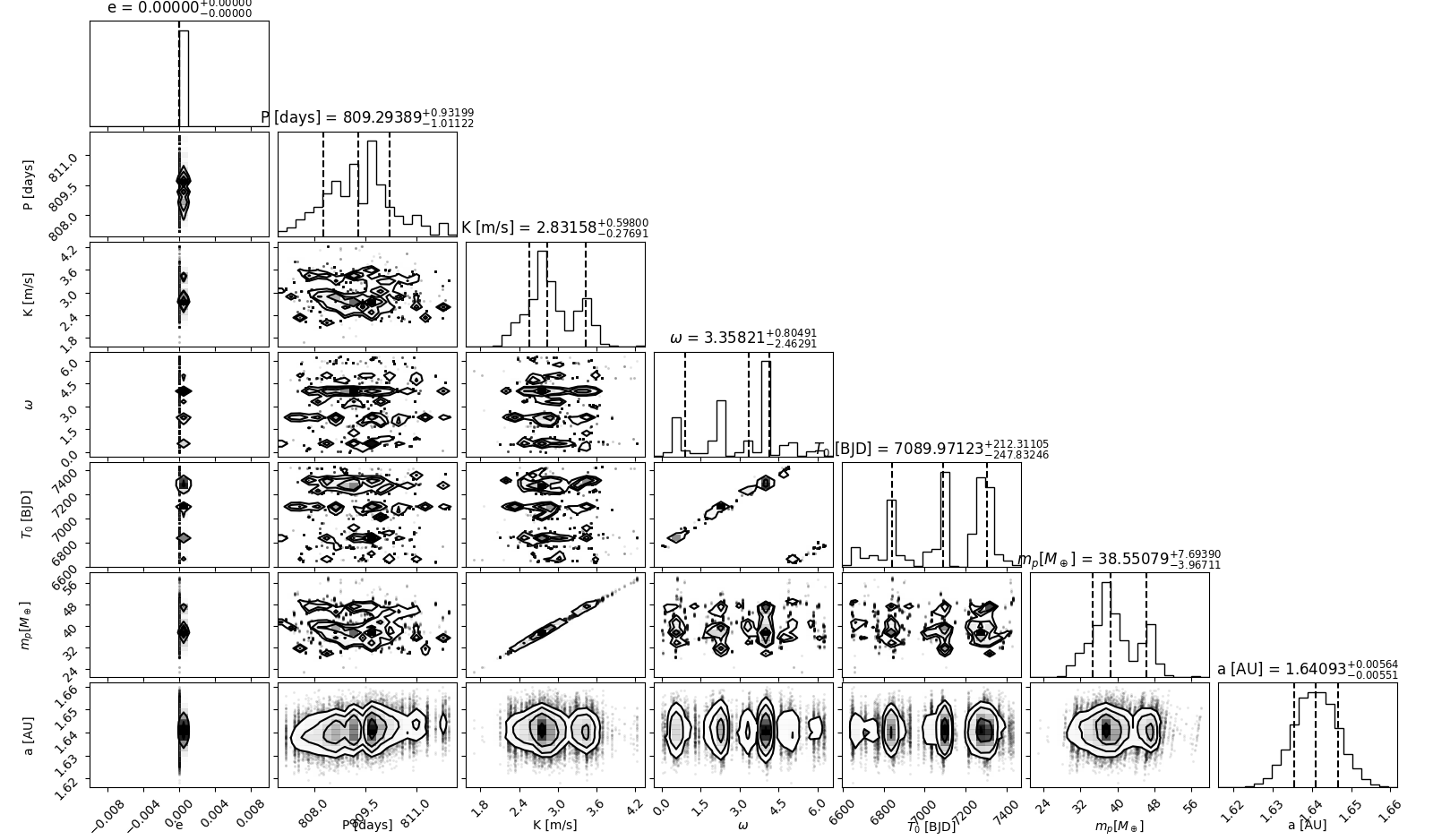}
    \caption{Corner plot for DMPP-3 800~d signal.}
    \label{fig:corner_DMPP-3_800d}
\end{figure*}

\begin{figure*}
    \centering
	\includegraphics[width=\textwidth]{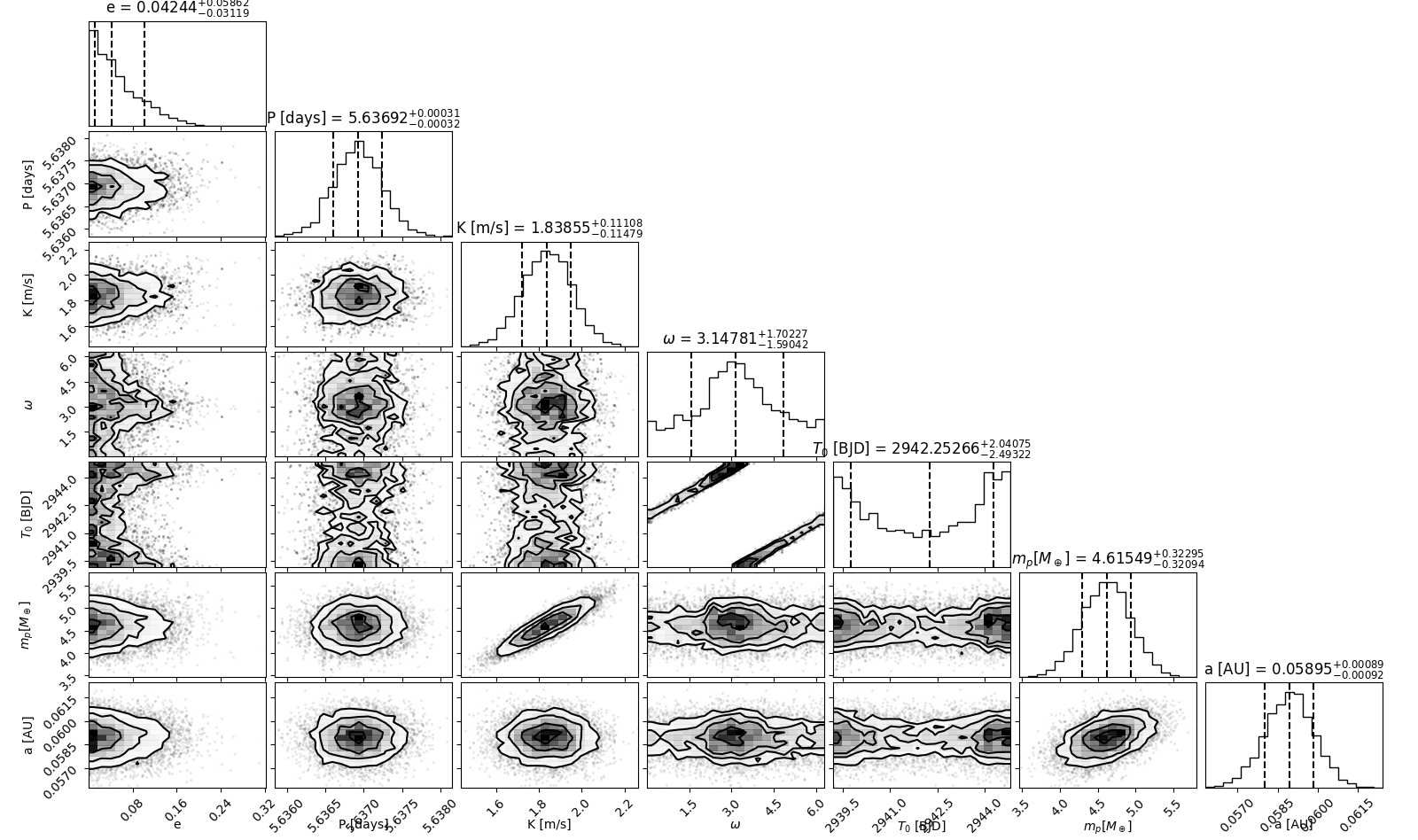}
    \caption{Corner plot for HD39194\,b.}
    \label{fig:corner_HD39194b}
\end{figure*}

\begin{figure*}
    \centering
	\includegraphics[width=\textwidth]{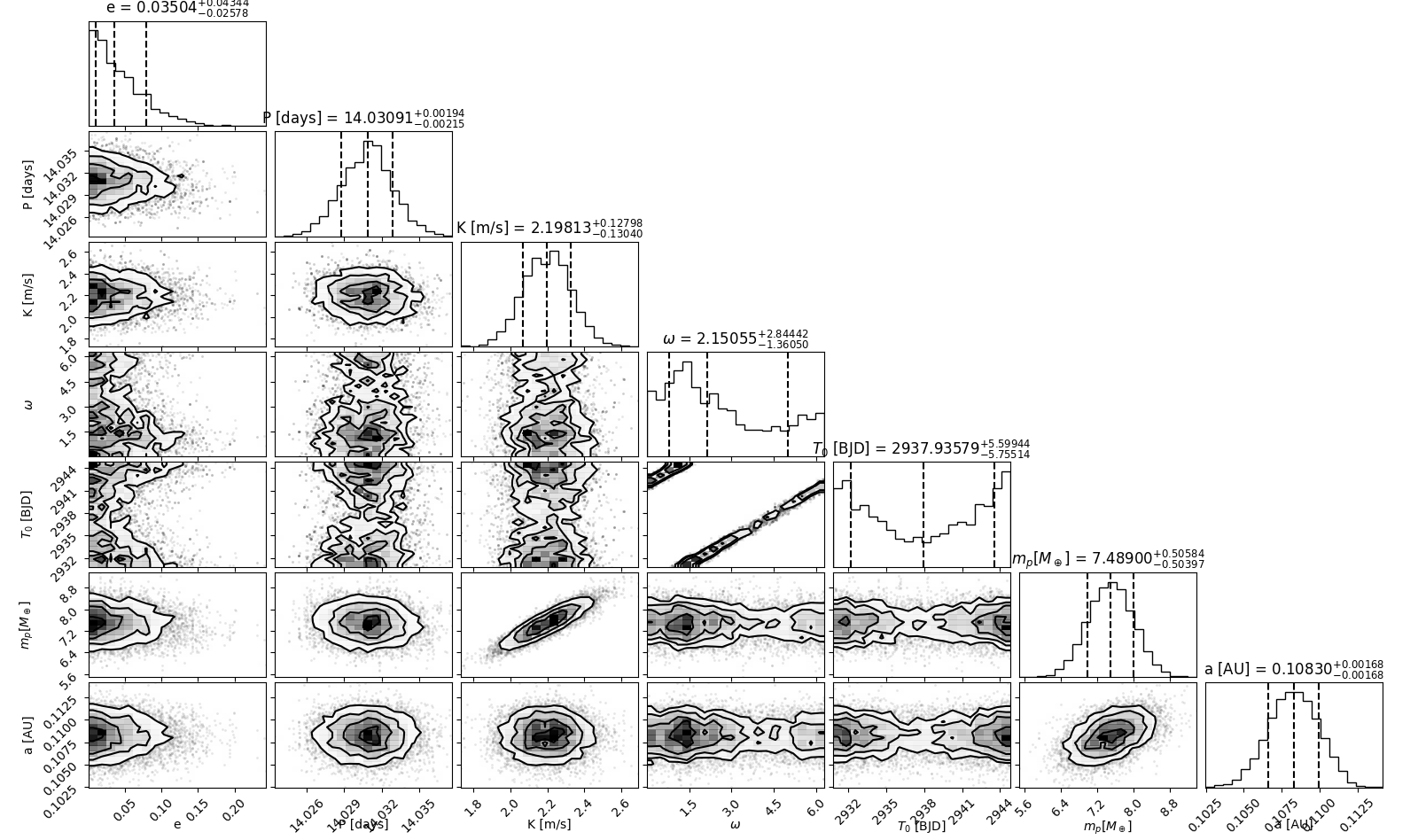}
    \caption{Corner plot for HD39194\,c.}
    \label{fig:corner_HD39194c}
\end{figure*}

\begin{figure*}
    \centering
	\includegraphics[width=\textwidth]{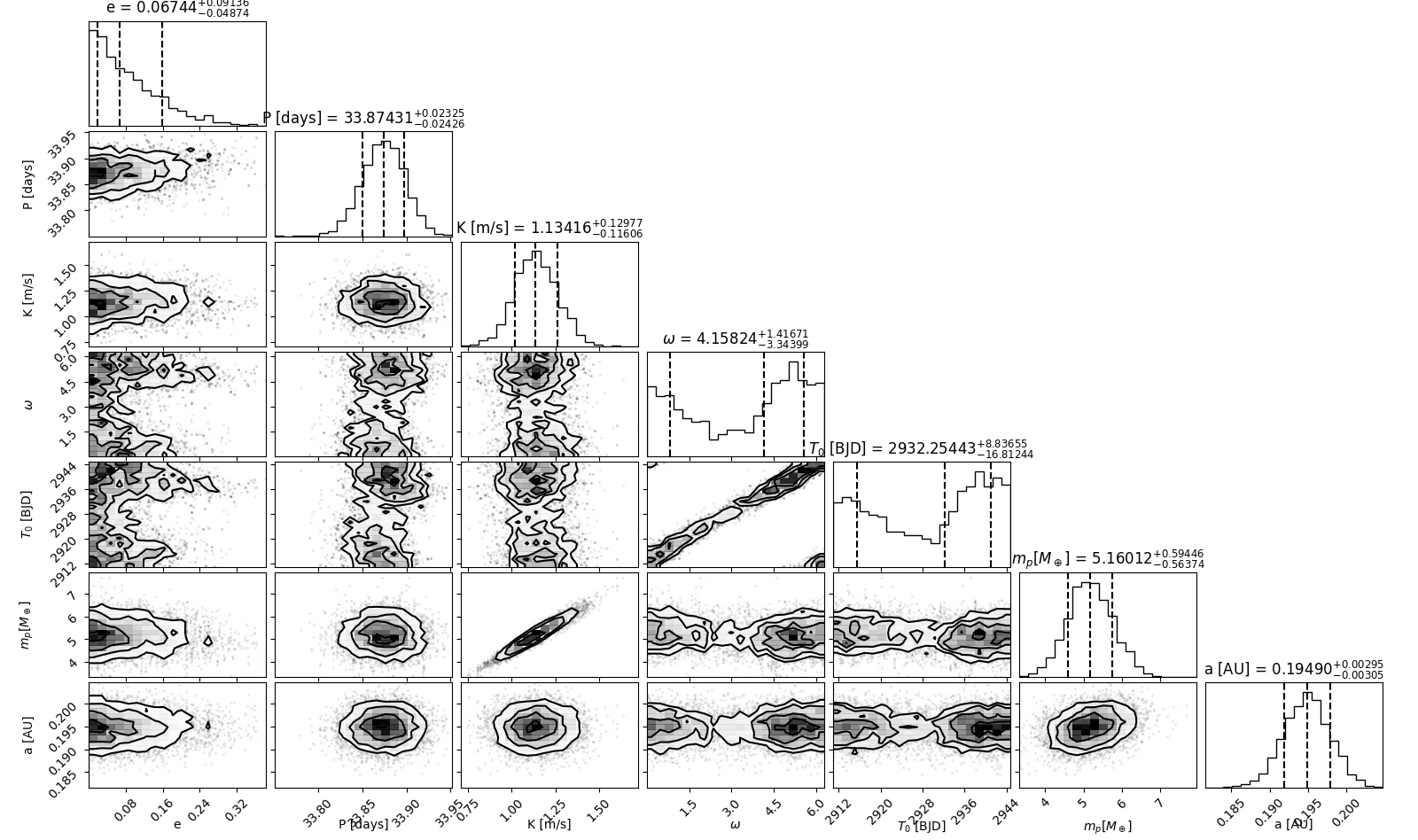}
    \caption{Corner plot for HD39194\,d.}
    \label{fig:corner_HD39194d}
\end{figure*}

\begin{figure*}
    \centering
	\includegraphics[width=\textwidth]{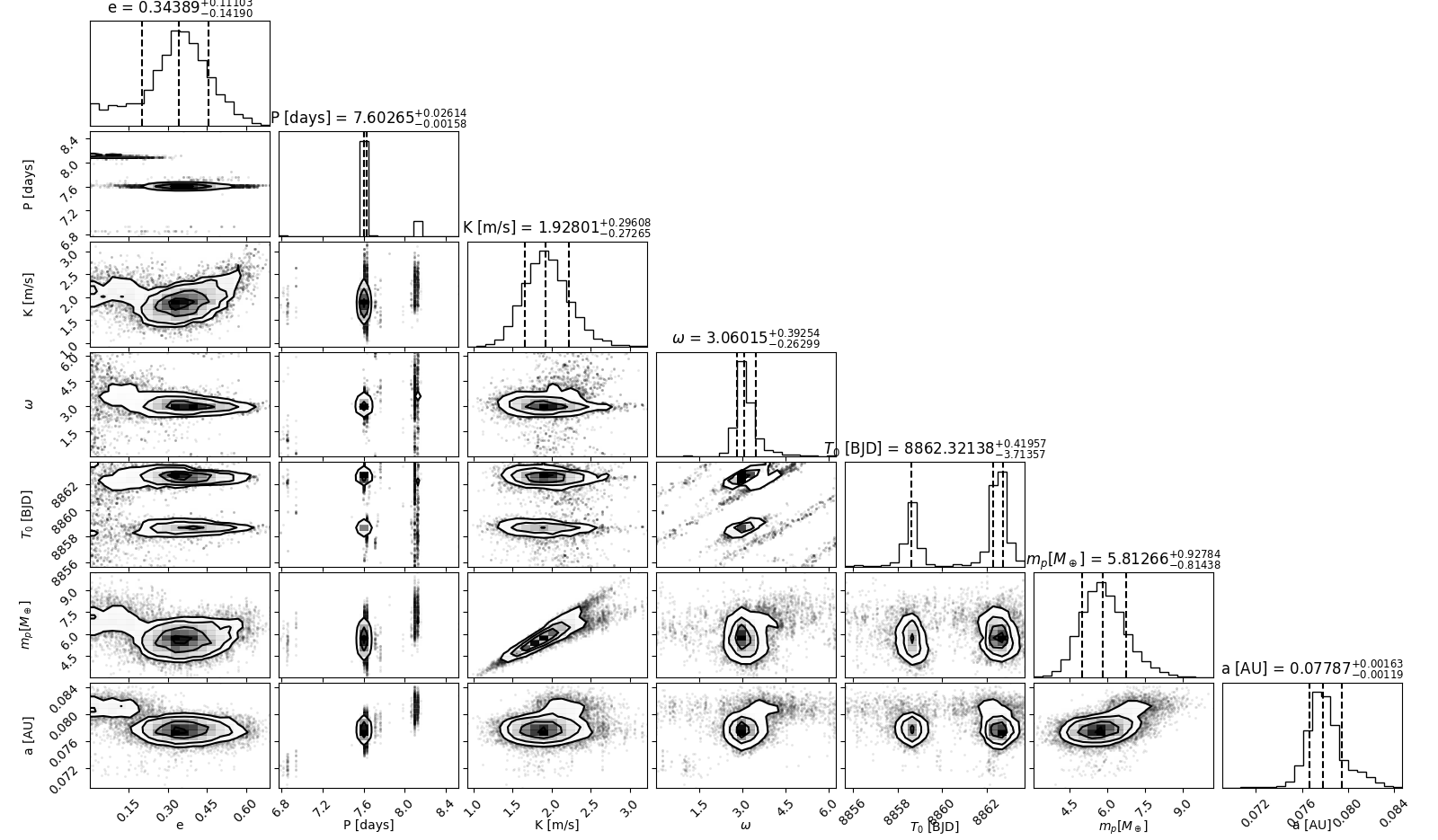}
    \caption{Corner plot for HD67200 / DMPP-6\,b.}
    \label{fig:corner_HD67200b}
\end{figure*}

\begin{figure*}
    \centering
	\includegraphics[width=\textwidth]{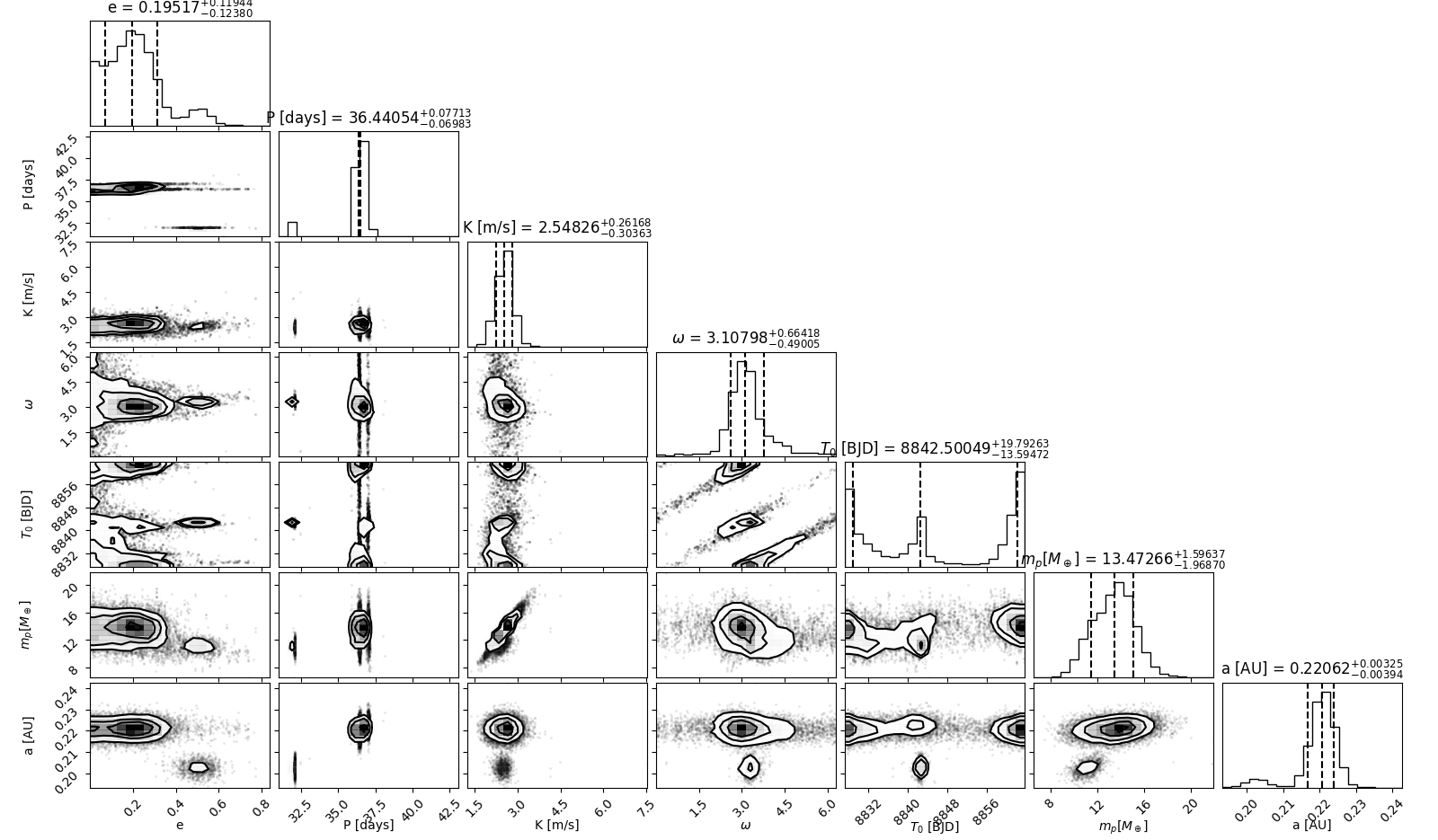}
    \caption{Corner plot for HD67200 / DMPP-6\,c.}
    \label{fig:corner_HD67200c}
\end{figure*}

\begin{figure*}
    \centering
	\includegraphics[width=\textwidth]{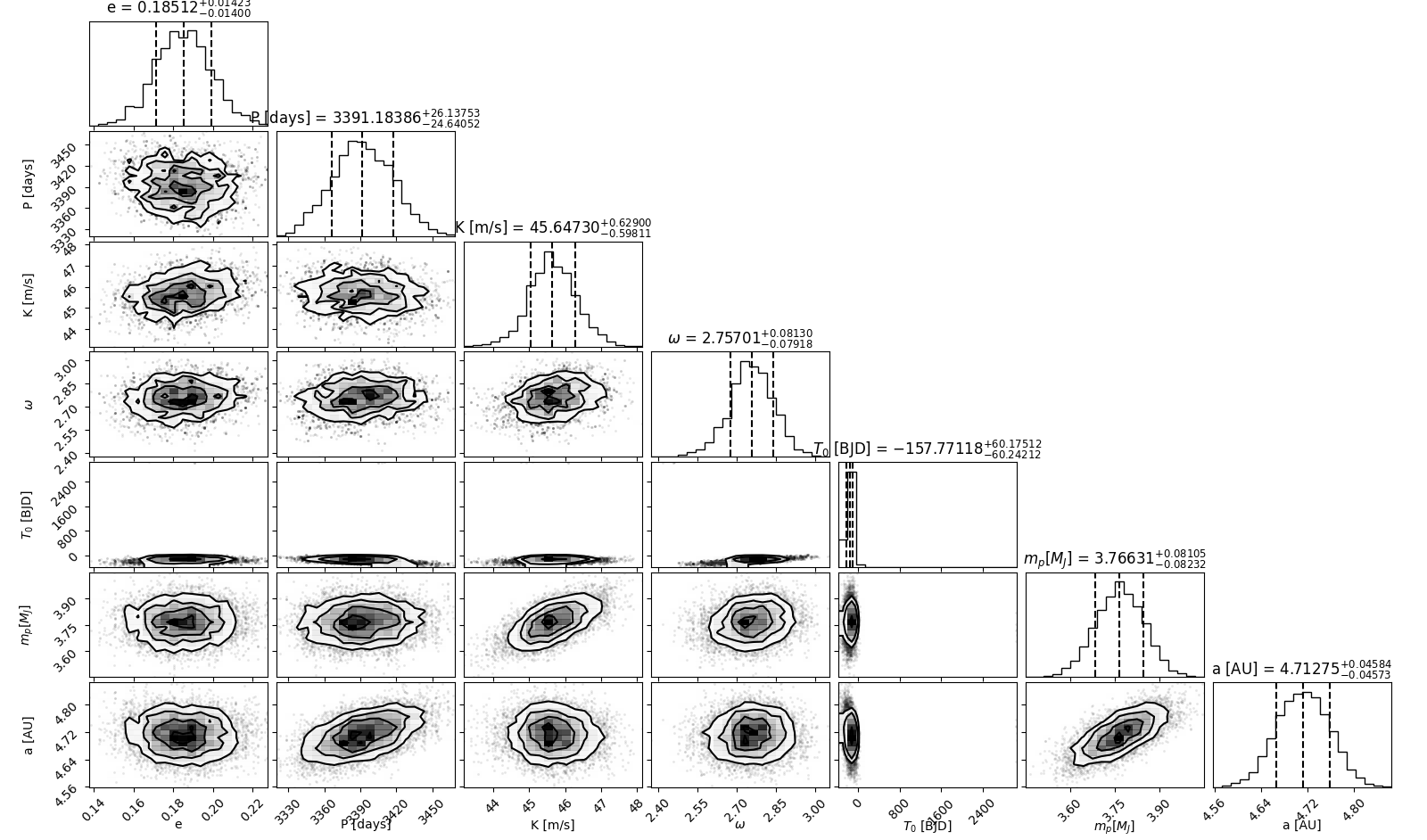}
    \caption{Corner plot for HD89839\,b.}
    \label{fig:corner_HD89839b}
\end{figure*}

\begin{figure*}
    \centering
	\includegraphics[width=\textwidth]{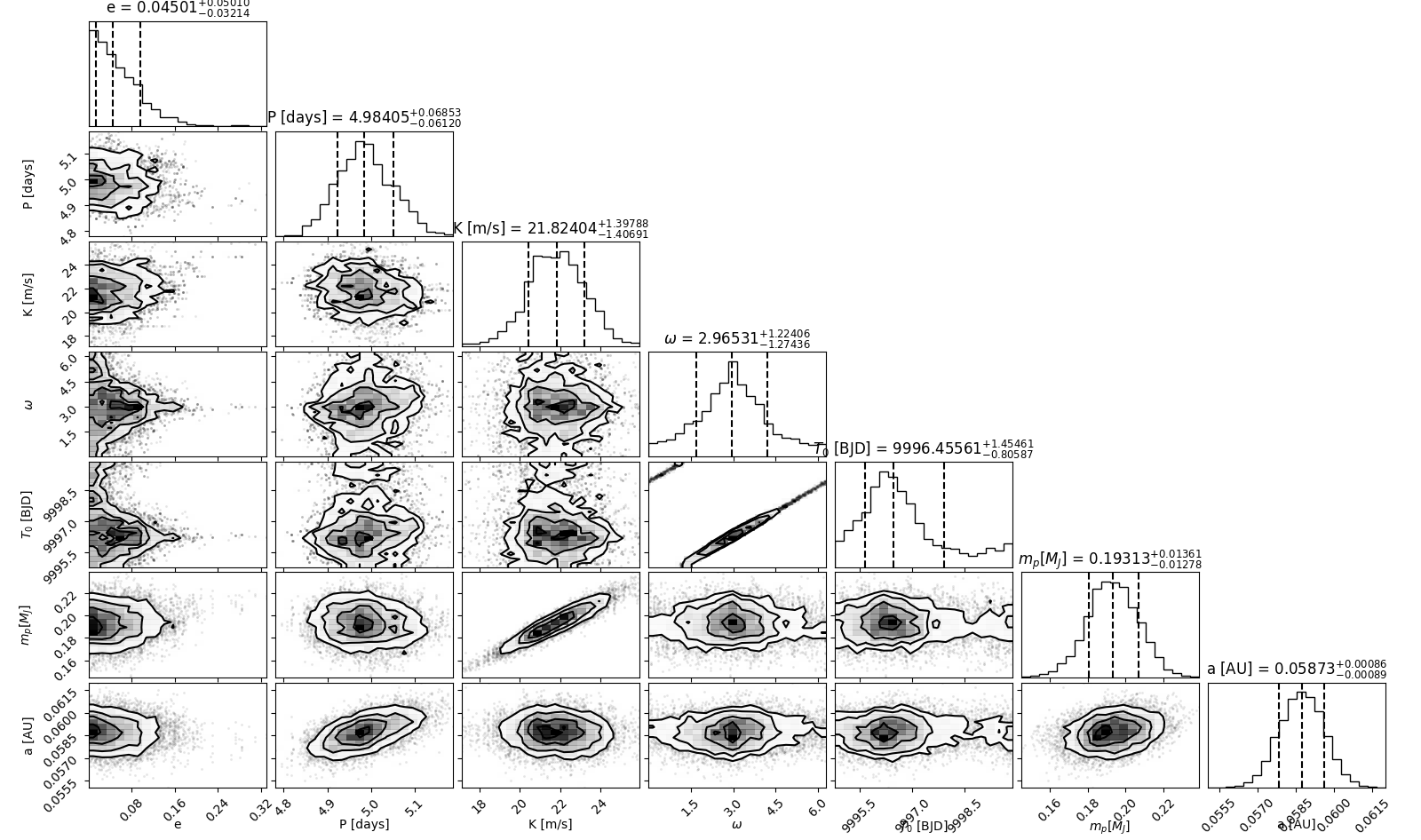}
    \caption{Corner plot for HD118006 / DMPP-7\,b.}
    \label{fig:corner_HD118006b}
\end{figure*}

\begin{figure*}
    \centering
	\includegraphics[width=\textwidth]{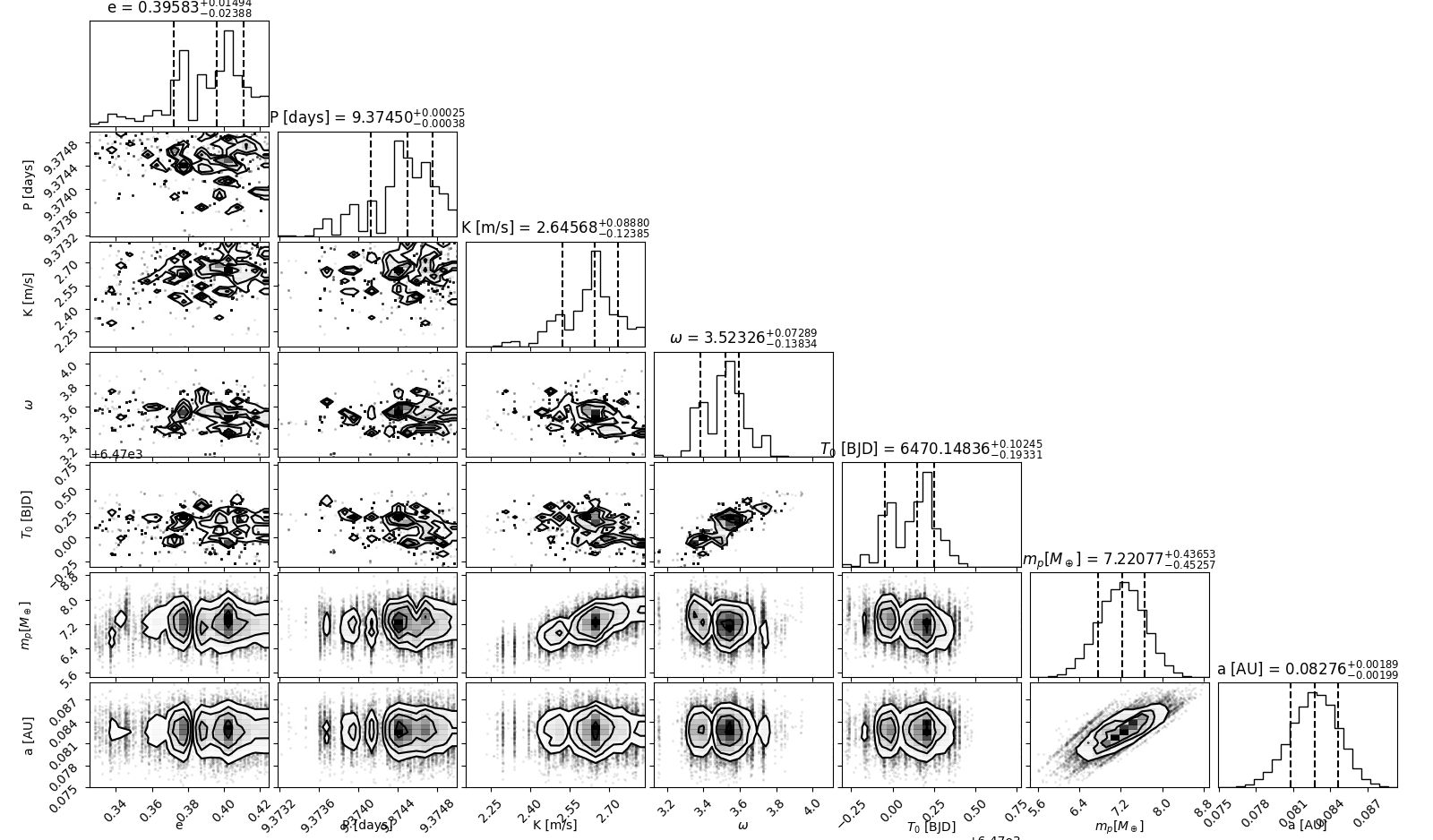}
    \caption{Corner plot for HD181433\,b.}
    \label{fig:corner_HD181433b}
\end{figure*}

\begin{figure*}
    \centering
	\includegraphics[width=\textwidth]{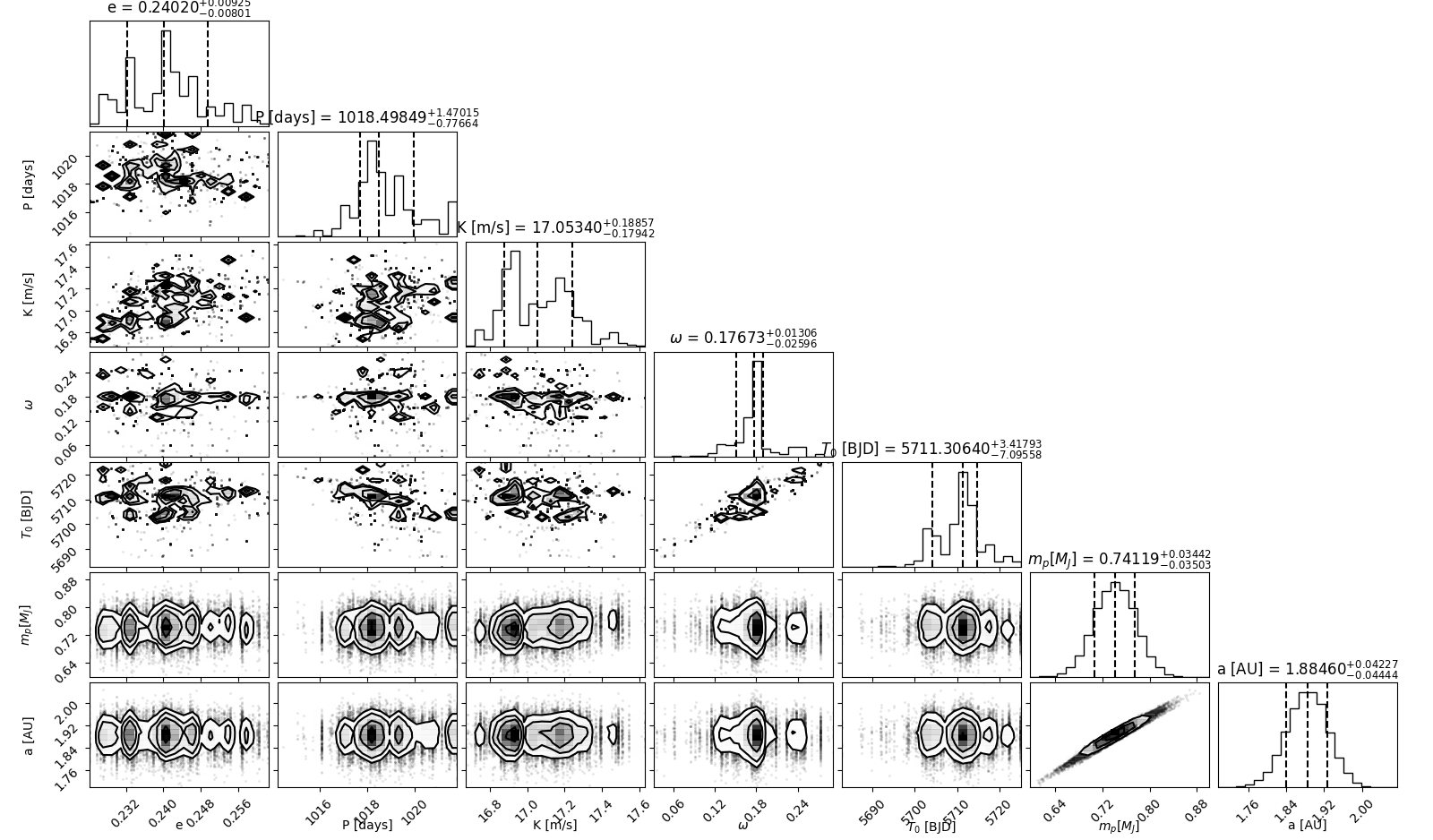}
    \caption{Corner plot for HD181433\,c.}
    \label{fig:corner_HD181433c}
\end{figure*}

\begin{figure*}
    \centering
	\includegraphics[width=\textwidth]{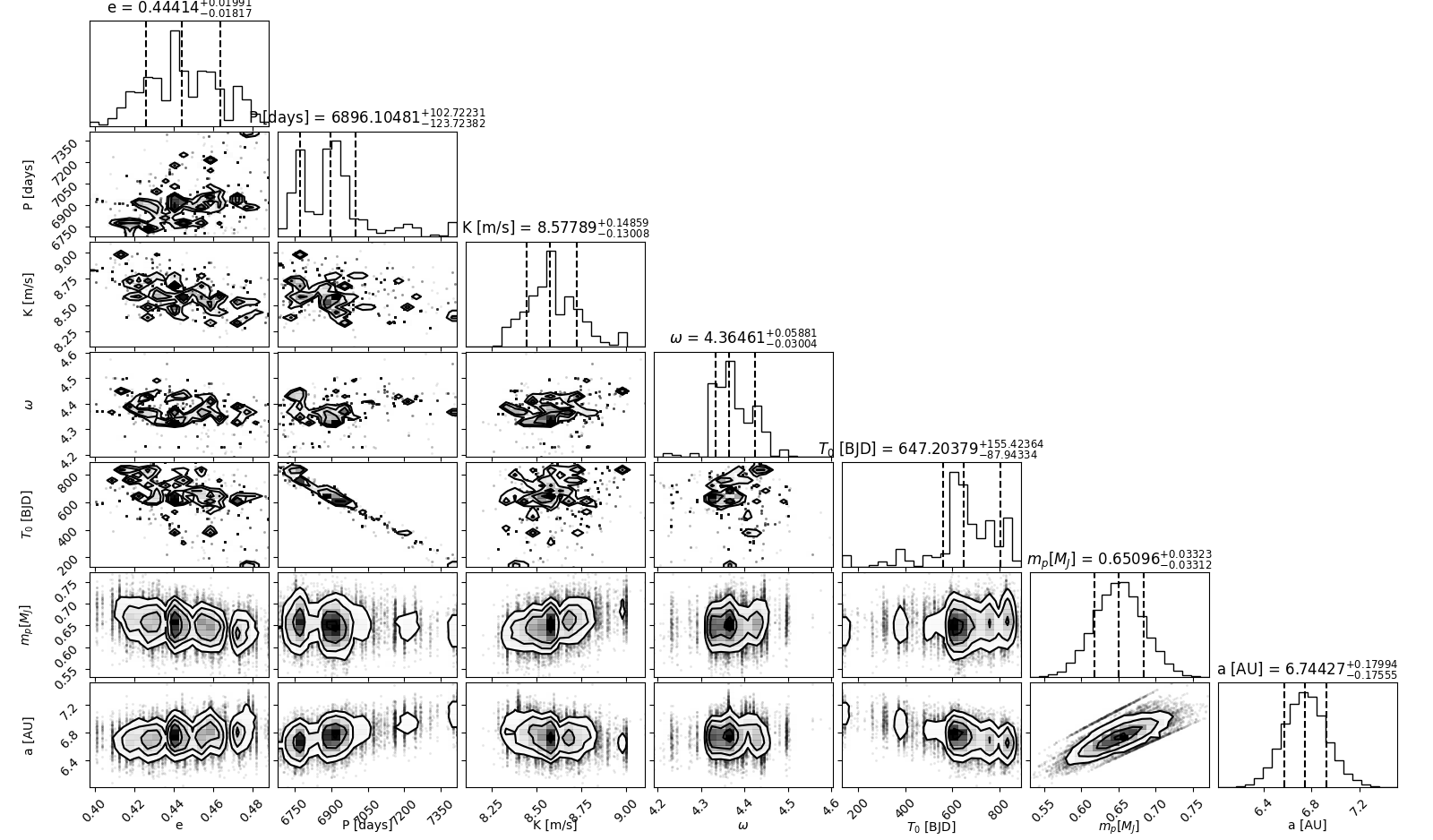}
    \caption{Corner plot for HD181433\,d.}
    \label{fig:corner_HD181433d}
\end{figure*}

\begin{figure*}
    \centering
	\includegraphics[width=\textwidth]{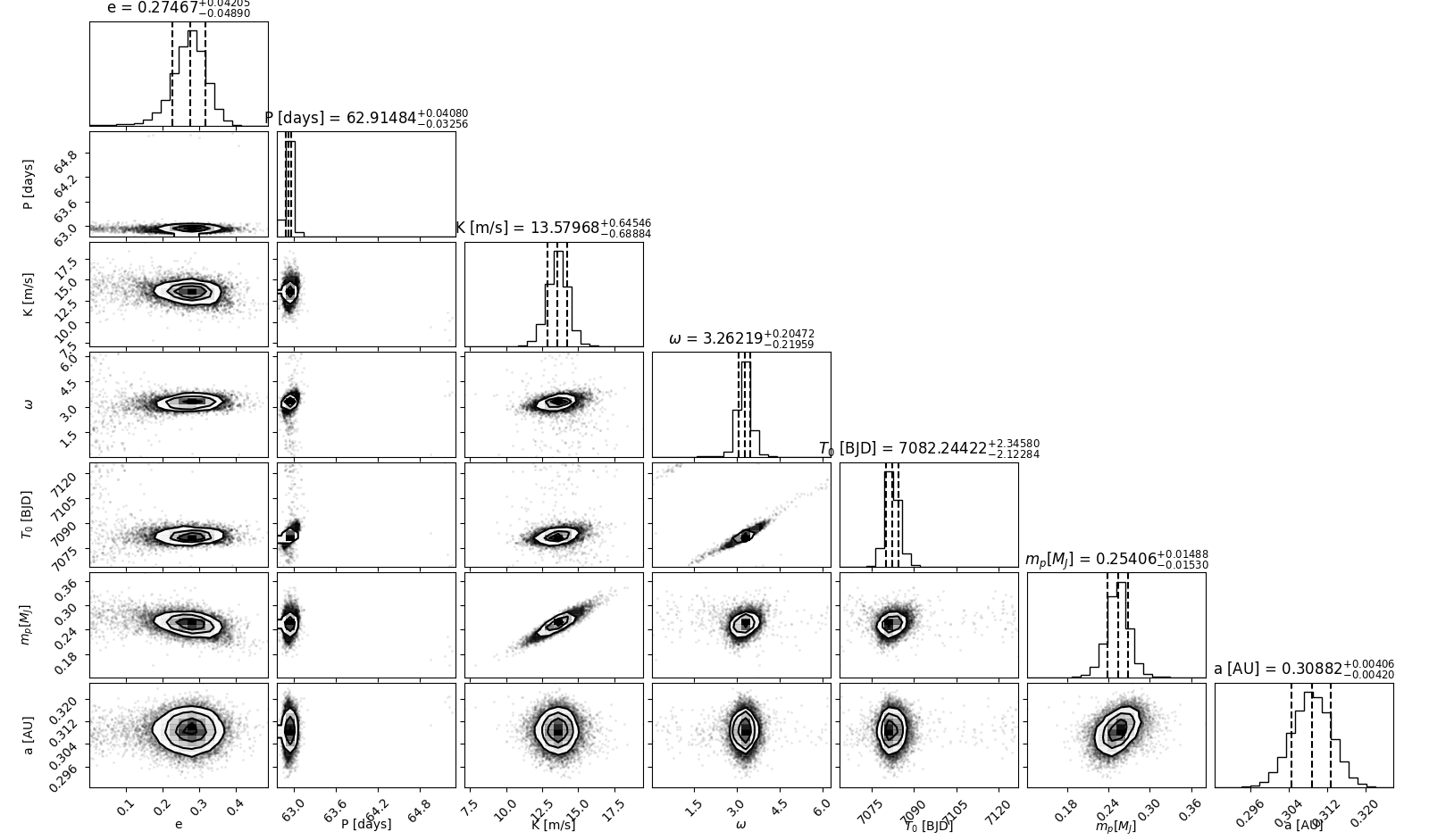}
    \caption{Corner plot for HD191122 / DMPP-8\,b.}
    \label{fig:corner_HD191122b}
\end{figure*}

\begin{figure*}
    \centering
	\includegraphics[width=\textwidth]{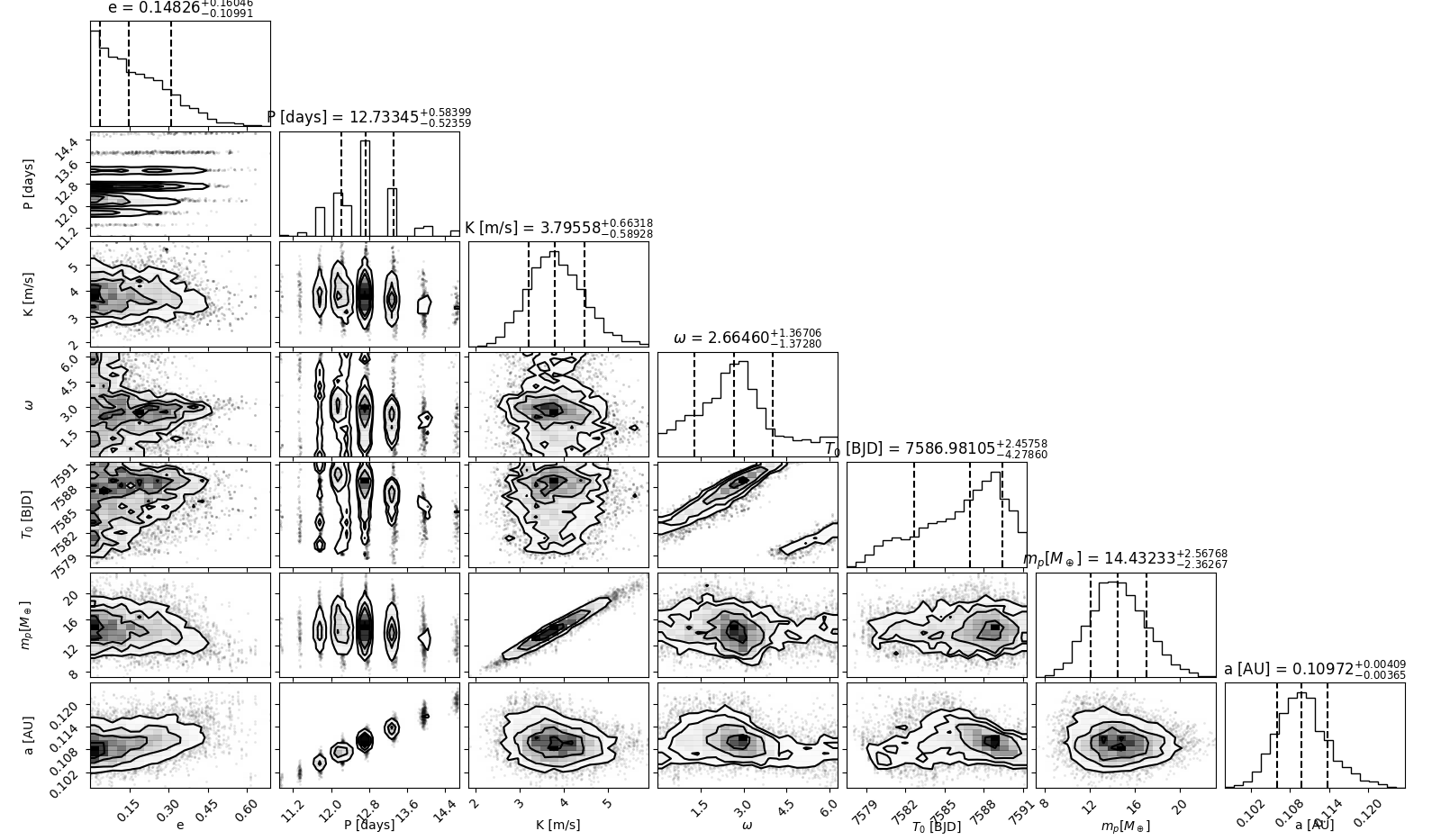}
    \caption{Corner plot for HD200133 / DMPP-9\,b.}
    \label{fig:corner_HD200133b}
\end{figure*}

%%%%%%%%%%%%%%%%%%%%%%%%%%%%%%%%%%%%%%%%%%%%%%%%%%

% Don't change these lines
\bsp	% typesetting comment
\label{lastpage}
\FloatBarrier
\end{document}